\documentclass[12pt]{iopart}

\usepackage{iopams}  
\expandafter\let\csname equation*\endcsname\relax
\expandafter\let\csname endequation*\endcsname\relax

\usepackage{bm}				% bold math symbols
\usepackage{amsmath}
\usepackage{amssymb}
\usepackage{latexsym}
\usepackage{amsfonts}
\usepackage{graphicx}
\usepackage[usenames,dvipsnames]{xcolor}
\usepackage{dsfont}
\usepackage{wasysym}
\usepackage{cite}

\usepackage{color}
\definecolor{myblue}{RGB}{65,105,225}
\definecolor{mygreen}{RGB}{34,139,34}
\definecolor{myorange}{RGB}{255,69,0}
\usepackage[colorlinks,linkcolor=myorange,urlcolor=myblue,citecolor=myblue]{hyperref}

\newcommand{\ket}[1]{\vert{#1}\rangle} 
\newcommand{\bra}[1]{\langle{#1}\vert} 
\newcommand{\bracket}[2]{\langle{#1}\vert{#2}\rangle} 
\newcommand{\bbraket}[2]{\langle\langle{#1}\vert{#2}\rangle\rangle} 
\newcommand{\op}[2]{\ket{#1}\!\bra{#2}}
\newcommand{\mean}[1]{\langle #1 \rangle}
\newcommand{\abs}[1]{\left|#1\right|} 
\newcommand{\pare}[1]{\left( #1 \right)}
\newcommand{\be}{\begin{equation}}
\newcommand{\ee}{\end{equation}}
\newcommand{\key}[1]{\left\{ #1 \right\}}
\newcommand{\cor}[1]{\left[ #1 \right]}
\newcommand{\acor}[1]{\left\{ #1 \right\}}
\newcommand{\la}{\langle}
\newcommand{\ra}{\rangle}
\newcommand{\bc}{\begin{center}}
\newcommand{\ec}{\end{center}}

\newcommand{\ben}{\begin{eqnarray}}
\newcommand{\een}{\end{eqnarray}}

\newcommand{\cHt}{{\cal H}_T}
\newcommand{\cH}{{\cal H}}
\newcommand{\cHe}{{\cal H}_E}
\newcommand{\cL}{{\cal L}}

\newcommand{\cU}{{\cal U}}
\newcommand{\cW}{{\cal W}}

\newcommand{\vlamb}{{\bm \lambda}}

\newcommand{\tO}{\tilde{O}}
\newcommand{\trhoT}{\tilde{\rho}_T}
\newcommand{\trho}{\tilde{\rho}}
\newcommand{\tV}{\tilde{V}}
\newcommand{\tE}{\tilde{E}}

\newcommand{\cB}{{\cal B}}
\newcommand{\cBH}{{\cal B}({\cal H})}
\newcommand{\cBHt}{{\cal B}({\cal H}_T)}
\newcommand{\cBHe}{{\cal B}({\cal H}_E)}
\newcommand{\cBBHt}{{\cal B}({\cal B}({\cal H}_T))}
\newcommand{\cBBH}{{\cal B}({\cal B}({\cal H}))}
\newcommand{\cBa}{\cB_{\alpha\beta}}
\newcommand{\cBaa}{\cB_{\alpha\alpha}}
\newcommand{\rhoa}{\rho_{\alpha\beta}}
\newcommand{\rhoaa}{\rho_{\alpha}}
\newcommand{\id}{\mathds{1}}
\newcommand{\ide}{\mathds{1}_E}

\newcommand{\PP}{\textrm{P}}
\newcommand{\ii}{\textrm{i}}
\newcommand{\te}[1]{\textrm{#1}}

\newcommand{\cLl}{{\cal L}_{\lambda}}

\newcommand{\oa}{\omega_{\alpha\beta\nu}(\lambda)}
\newcommand{\loa}{\hat{\omega}_{\alpha\beta\nu}(\lambda)}

\begin{document}

\title{Harnessing symmetry to control quantum transport}

\author{D. Manzano}
\address{Departamento de Electromagnetismo y F\'{\i}sica de la Materia. Universidad de Granada. E-18071 Granada. Spain.}
\ead{manzano@onsager.ugr.es}

\author{P.I. Hurtado}
\address{Departamento de Electromagnetismo y F\'{\i}sica de la Materia, and Instituto Carlos I de F{\'\i}sica Te{\'o}rica y Computacional. Universidad de Granada. E-18071 Granada. Spain }
\ead{phurtado@onsager.ugr.es}

\begin{abstract}
Controlling transport in quantum systems holds the key to many promising quantum technologies. Here we review the power of symmetry as a resource to manipulate quantum transport, and apply these ideas to engineer novel quantum devices. Using tools from open quantum systems and large deviation theory, we show that symmetry-mediated control of transport is enabled by a pair of twin dynamic phase transitions in current statistics, accompanied by a coexistence of different transport channels. By playing with the symmetry decomposition of the initial state, one can modulate the importance of the different transport channels and hence control the flowing current. Motivated by the problem of energy harvesting we illustrate these ideas in open quantum networks, an analysis which leads to the design of a symmetry-controlled quantum thermal switch. We review an experimental setup recently proposed for symmetry-mediated quantum control in the lab based on a linear array of atom-doped optical cavities, and the possibility of using transport as a probe to uncover hidden symmetries, as recently demonstrated in molecular junctions, is also discussed. Other symmetry-mediated control mechanisms are also described. Overall, these results demonstrate the importance of symmetry not only as a organizing principle in physics but also as a tool to control quantum systems. 
\end{abstract}

\vspace{1pc}
\noindent{\it PACS}: 5.60.Gg, 03.65.Yz, 44.10.+i.

\vspace{1pc}

\newpage

\tableofcontents

\newpage

\section{Introduction}
\label{sec:intro}

The control of transport and dynamical response in quantum systems is nowadays of fundamental technological interest \cite{wiseman09a,brandes10a}. Such interest is fueled by the remarkable advances of modern nanotechnologies and the possibility to manipulate with high precision systems in the quantum realm, ranging from ultracold atoms to trapped ions or molecular junctions, to mention just a few. The possibility to control transport at these scales opens the door to the design of e.g. programmable molecular circuits \cite{seminario05a,aradhya13a} and molecular junctions \cite{reddy07a,cui17a}, quantum machines \cite{linden10a,seifert12a,devoret14a}, high-performance energy harvesting devices \cite{wu10a,scholak11a,sothmann15a,thierschmann15a}, or quantum thermal switches and transistors \cite{manzano14a,manzano16a,joulain16a}. 

Importantly, most devices of interest to prevailing quantum technologies are \emph{open} and subject to dissipative interactions with an environment. This dissipation has been usually considered negative for the emerging quantum technologies as it destroys quantum coherence, a key resource at the heart of this second quantum revolution. However, in a recent series of breakthroughs \cite{barreiro11a,verstraete09a,diehl08a,kraus08a,lin13a,pastawski11a}, it has been shown that a careful engineering of the dissipative interactions with the environment may favor the quantum nature of the associated process. This idea has been recently used for instance in order to devise optimal quantum control strategies \cite{barreiro11a}, to implement universal quantum computation \cite{verstraete09a}, to drive the system to desired target states (maximally entangled, matrix-product, etc.) \cite{diehl08a,kraus08a,lin13a}, or to protect quantum states by prolonging their lifetime \cite{pastawski11a}.

In all cases, the natural framework to investigate the physics of systems in contact with a decohering environment is the theory of open quantum systems \cite{davies72a,breuer02a,gardiner00a,plenio98a,daley17}. This set of techniques has been applied to a myriad of problems in diverse fields, including quantum optics \cite{kay70a,paul04a}, atomic physics \cite{bransden03a,cohen11a} and quantum information \cite{briegel12a,schuld14b}. More recently, the open quantum systems approach has been applied to the study of quantum effects in biological systems \cite{plenio08a,cai10b,cai10a,scholes11,mohseni08a,pelzer14a,levi15a} and quantum transport in condensed matter \cite{prosen11a,znidaric13a,walschaers13a,manzano16b}, the latter being the focus of this paper. A complete characterization of transport in open quantum systems requires the understanding of their current statistics, and this is achieved by employing the tools of full counting statistics and large deviation theory \cite{touchette09a,esposito09a,flindt09a,garrahan10a,garrahan11a,ates12a,hickey12a,genway12a,flindt13a,lesanovsky13a,maisi14a,buca14a,zannetti14a,buca15a,znidaric14a,znidaric14b,znidaric14c,buca17a,carollo17a}. The central observable of this theory is the current large deviation function (LDF), which measures the likelihood of different current fluctuations, typical or rare. Large deviation functions are of fundamental importance in nonequilibrium statistical mechanics, in addition to their practical relevance expressed above. Indeed LDFs play in nonequilibrium physics a role equivalent to the equilibrium free energy and related potentials, and govern the \emph{thermodynamics of currents} out of equilibrium \cite{bertini01a,bertini02a,bertini05a,bertini06a,bertini15a,bodineau04a,derrida07a,hurtado09a,hurtado09b,hurtado10a,prados11a,hurtado11a,hurtado13a,hurtado14a,hurtado11b}.

An important lesson of modern theoretical physics is the importance of symmetries as a tool to uncover unifying principles and regularities in otherwise complex physical situations \cite{wigner64a,gross96a}. As we will see repeatedly in this paper, analyzing the consequences of symmetries on quantum transport and dynamics allows to gain deep insights into the physics of open systems, even though the associated dynamical problems are too complex to be solved analytically. A prominent example of the importance and many uses of symmetry in physics is Noether's theorem \cite{noether1918a}\footnote[1]{An english translation of the original Noether's paper can be found at \cite{noether71a}.}. Noether 
%understood for the first time the deep connection between symmetries and conservation laws in classical  physics,
originally proved that in classical systems every symmetry leads to a conserved quantity, though her result applies also to quantum systems and it constitutes a key result in quantum field theory \cite{ward50a,zee10a,mandl10a}. In this way, by analyzing the symmetries of a given (isolated) system one may deduce the associated conservation laws, which in turn define the slowly-varying fields which control the system long-time and large-scale relaxation. Interestingly, the situation in open quantum systems is more complex, and the relation between symmetries and conservation laws is not as clear-cut as for isolated systems where Noether's theorem applies, giving rise to a richer phenomenology \cite{buca12a,albert14a,albert16a}. For instance, it has been recently shown \cite{albert14a} that open quantum systems described by a Lindblad-type master equation may exhibit conservation laws which do not correspond to symmetries (as found also in some classical integrable systems), even though every symmetry yields a conserved quantity in these systems.

Another example of the importance of symmetries to obtain insights into complex physics concerns the different fluctuation theorems derived for nonequilibrium systems in the classical and quantum realm \cite{evans93a,evans94a,gallavotti95a,gallavotti95b,gallavotti96a,jarzynski97a,crooks98a,kurchan98a,lebowitz99a,hatano01a,seifert12a,esposito09a,andrieux09a,chetrite12a,hurtado11b}. These theorems, which strongly constraint the probability distributions of fluctuations far from equilibrium, are different expressions of the time-reversal symmetry of microscopic dynamics at the mesoscopic, irreversible level. Remarkably, by demanding invariance of the optimal paths responsible of a given fluctuation under additional symmetry transformations (beyond time-reversibility), further fluctuation theorems can be obtained  which remain valid arbitrarily far from equilibrium \cite{hurtado11b,villavicencio14a,lacoste14a,lacoste15a,kumar15a}. A particular example is the recently unveiled isometric fluctuation theorem for current statistics in diffusive transport problems, which links the probability of different but isometric vector flux fluctuations in a way that extends and generalizes the Gallavotti-Cohen relation in this context \cite{hurtado11b}. At the quantum transport level, symmetry ideas have also proven useful in past years. A first example is related to anomalous collective effects, as e.g. superradiance (enhanced relaxation rate) \cite{dicke54a,palacios02a} and supertransfer (enhanced exciton transfer rate and diffusion length) \cite{strek77a,scholes02a,lloyd10a,abasto12a}, which result from geometric symmetries in the system Hamiltonian. Moreover, symmetries have been recently shown to constraint strongly the reduced density matrix associated to nonequilibrium steady states in open quantum systems \cite{popkov13a}, while violations of time-reversal symmetry lead to an enhancement of quantum transport in continuous-time quantum walks \cite{zimboras13a}. Finally, optimal quantum control schemes have been recently proposed using symmetry as a guiding principle \cite{zeier11a}.

All these results suggest that symmetry plays a relevant role in transport, both at the classical and quantum level. Indeed, we will demonstrate here the power of symmetry as a resource for quantum transport. In particular, the purpose of this paper is to review recent advances in the use of symmetry ideas to control energy transport and fluctuations in open quantum systems. With this aim in mind, we introduce first the mathematical tools that we will need in our endeavor. In particular, Section \ref{sec:subspaces} is devoted to a brief but self-consistent introduction to the physics of open quantum systems and their description in terms of master equations for the reduced density matrix, including Redfield- and Lindblad-type master equations. The spectral properties of the resulting Lindblad-Liouville evolution superoperator will be analyzed in Section \ref{sec:sym}, with particular emphasis on the steady state behavior and the first relaxation modes \cite{albert14a}. This will lead naturally to the question of the uniqueness of the steady state. Building on Refs. \cite{buca12a,buca15a} (see also \cite{baumgartner08a}), we will show how the presence of a strong symmetry (to be defined below) in an otherwise dissipative quantum system leads to a degenerate steady state with multiplicity linked to the symmetry spectrum. Secion \ref{sec:symej} then describes a few examples of driven spin chains \cite{buca12a} and ladders \cite{znidaric13b} where symmetries and their effect on transport become apparent. To understand how and why symmetry affects transport properties, including both average behavior and fluctuations, we introduce in Section \ref{sec:LDF} the full counting statistics for the current and the associated large deviation theory. Equipped with this tool, we next show in Section \ref{sec:LDF2.5} that the symmetry-induced steady-state degeneracy is nothing but a coexistence of different transport channels stemming from a general first-order-type dynamic phase transition (DPT) in current statistics \cite{manzano14a}. This DPT shows off as a non-analyticity in the cumulant generating function of the current or equivalently as a non-convex regime in the current large deviation function, and separates two (fluctuating) transport phases characterized by a maximal and minimal current, respectively. Moreover, the time-reversibility of the microscopic dynamics results in the appearance of a twin DPT for rare, reversed current fluctuations, such that the Gallavotti-Cohen fluctuation theorem holds across the whole spectrum of current fluctuations. The symmetry-induced degenerate steady state preserves part of the information of the initial state due to the lack of mixing between the different symmetry sectors, and we show how this opens the door to a complete control of transport properties (as e.g. the average current) by tailoring this information via initial-state preparation techniques.

Motivated by the problem of energy harvesting in (natural and artificial) photosynthetic complexes, and with the aim of validating our general results, we study in Section \ref{sec:networks} transport and current fluctuations in open quantum networks \cite{manzano14a}. These models exhibit exchange symmetries linked to their network topology, and therefore are expected to display the phenomenology described above. This is confirmed in detailed numerical analyses. Our results also suggest novel design strategies based on symmetry ideas for quantum devices with controllable transport properties. Indeed, using this approach we describe a novel design for a \emph{symmetry-controlled quantum thermal switch}, i.e. a quantum qubit device where the heat current can be completely blocked, modulated or turned on by preparing the symmetry of the initial state. This schematic idea is further developed in Section \ref{sec:switch}, where an experimental setup for symmetry-enabled quantum control in the lab is described \cite{manzano16a}. The setup consists in a linear array of three optical cavities coupled to terminal reservoirs, with the central cavity doped with two identical $\Lambda$-atoms driven by laser fields. This system is symmetric under the exchange of the two atoms, and we describe how, by switching on and off one of the lasers, the photon current across the optical cavities can be controlled at will. This symmetry-controlled atomic switch can be realized in current laboratory experiments, and interestingly this device can be also used to store maximally-entangled states for long periods of time due to its symmetry properties. The previous results show the power of symmetry as a tool to control transport. Conversely, we can also use transport to probe unknown symmetries of open quantum systems. This idea is explored in Section \ref{sec:transient}, where we describe a dynamical method to detect hidden symmetries in molecular junctions by analyzing the time evolution of the exciton current under a temperature gradient \cite{thingna16a}. The detection scheme includes a probe acting on the molecular complex which serves as a possible symmetry-breaking element. We explain the dynamical signatures of the underlying symmetry in terms of the spectral properties of the evolution superoperator, and show that these signatures remain robust in the presence of weak conformational disorder and/or environmental noise. Finally we review in Section \ref{sec:other} other symmetry-mediated mechanisms to control quantum transport, based either on weak symmetries of the steady-state density matrix \cite{popkov13a} or the violation of time-reversal symmetry to enhance transport \cite{zimboras13a}. We end this paper offering a summary of these results and a glance at future developments in Section \ref{sec:conclusions}.

\newpage

\section{Open quantum systems and master equations}
\label{sec:subspaces}

\begin{figure}
\bc
\includegraphics[scale=0.25]{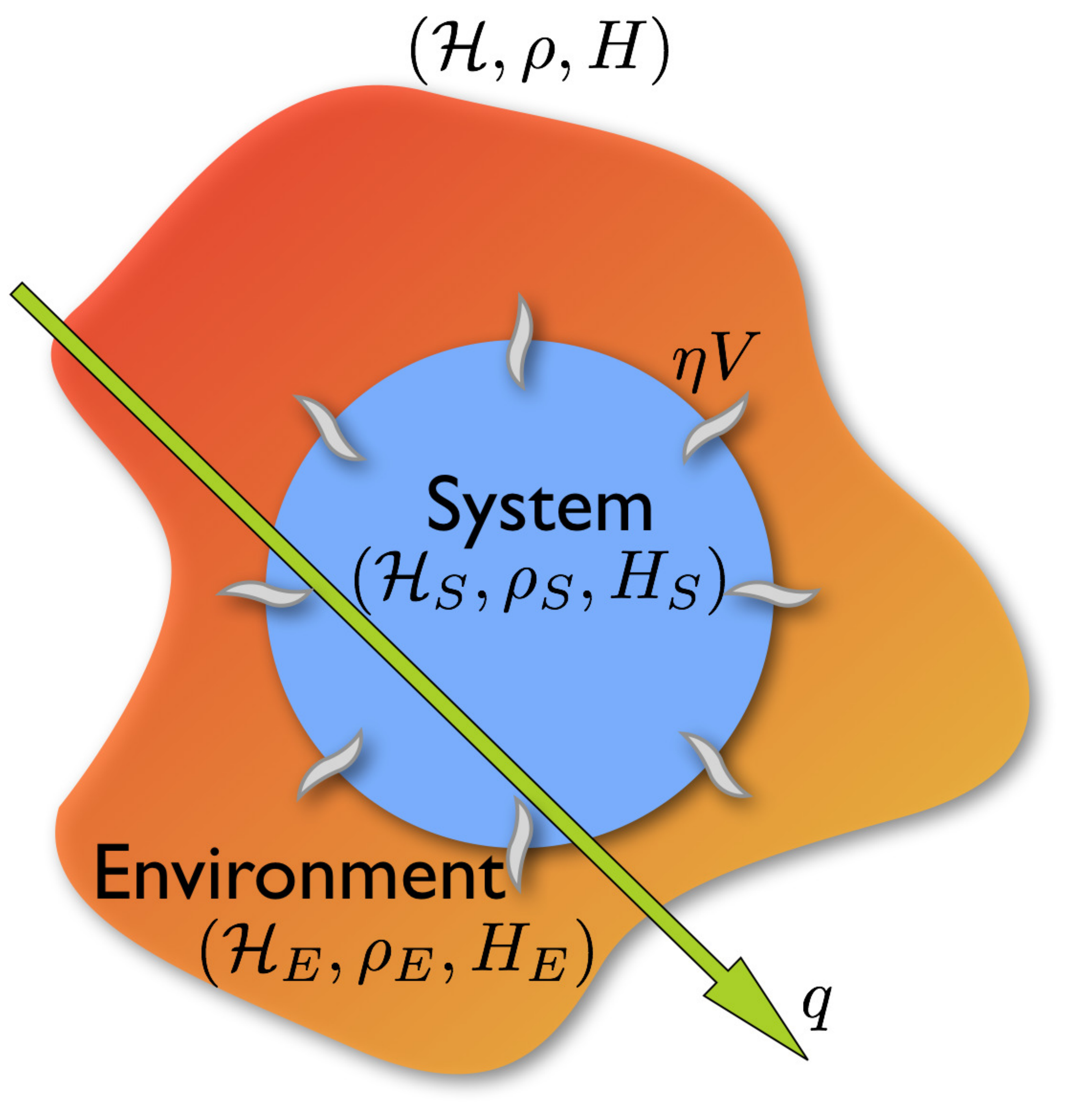}
\caption{Sketch of a quantum system in contact with an environment. The system-environment interaction typically gives rise to a nonequilibrium situation characterized by a net heat current $q$ flowing through the system.}
\label{fig1}
\ec
\end{figure}

We consider a general quantum system in contact with an environment, as the one displayed in Figure \ref{fig1}. The Hilbert space $\cHt$ of the total system can be decomposed in the tensor product of the Hilbert spaces of the system $\cH$ (which we assume of finite dimension $D$\footnote[1]{We make this assumption for the sake of simplicity, though later on we will generalize some of our results to bosonic systems with infinite-dimensional Hilbert spaces and unbounded operators.}) and the environment $\cHe$, i.e. $\cHt=\cH\otimes \cHe$. Note that now and hereafter we avoid using the subscript $_S$ for the system quantities in order not to clutter our notation (we instead use $_T$ for the total system and $_E$ for the environment). The state of the total system is described at any time by a density matrix $\rho_T(t)$, a unit trace operator in the space $\cBHt$ of bounded operators acting on the Hilbert space $\cHt$. The space $\cBHt$ is itself a Hilbert space once equipped with the Hilbert-Schmidt inner product \cite{breuer02a}, 
\be
\bbraket{\sigma}{\rho} = \Tr(\sigma^\dagger \rho) \, , \qquad \forall \sigma,\rho \in \cBHt \, ,
\label{inner}
\ee
where $\Tr(\omega)$ is the trace of the operator $\omega\in\cBHt$. The dynamics of the joint system and environment is determined by a time-independent hermitian Hamiltonian $H_T\in\cBHt$, with
\be
H_T = H\otimes\ide + \id\otimes H_{E} + \eta V \, ,
\label{eq:hamil}
\ee
where $H\in \cBH$ is the system Hamiltonian, $H_{E}\in \cBHe$ is the environment Hamiltonian, $\id_a$ denotes the identity operator in the $\cH_a$ subspace, and $V\in \cBHt$ describes the system-environment interaction. For the sake of clarity, we will not make explicit from now on the identity operators in the Hamiltonian (\ref{eq:hamil}) whenever clear from the context. The dimensionless parameter $\eta$ above defines the interaction strength, and will be assumed small below, leading to the so-called weak coupling approximation \cite{breuer02a,rivas10a,kryszewski08a}. Without loss of generality, the interaction term can be written in a direct product, bilinear form as
\be
V= \sum_{l}S_l \otimes E_{l} = \sum_{l}S_l^\dagger \otimes E_{l}^\dagger \, ,
\label{eq:inter}
\ee
with $S_l\in \cBH$ and $E_l\in \cBHe$ operators acting on the system and environment, respectively. The second equality reflects the Hermitian character of the interaction term (although the individual operators $S_l$ and $E_l$ need not be Hermitian, just $V$) \cite{kryszewski08a}. The environment can be any quantum system but for most applications it is useful to consider thermal baths at fixed temperatures. In this case, if there are more than one bath and their temperatures are different, then one should expect net currents flowing through the system, even after reaching a steady state. 

The time evolution of the total system is determined by the {\it Liouville-von Neumann equation} \cite{messiah99a} 
\be
\dot{\rho}_T(t)=-\ii \cor{H_T,\rho_T(t)} \equiv \cL_T \rho_T(t) \, , 
\label{eq:LvN}
\ee
where the dot represents time derivative, $\cor{A,B}=AB-BA$, and we fix our units so $\hbar=1$ throughout the paper. Moreover, $\cL_T\in \cBBHt$ defines the total Liouville superoperator. 

The system state at any time is captured by the \emph{reduced} density matrix $\rho(t)$ obtained by tracing over the environment degrees of freedom,
\be
\rho(t)=\Tr_E \pare{\rho_T(t)} = \sum_k \bra{e_k} \rho_T(t) \ket{e_k} \, ,
\label{eq:trE}
\ee
with $\{ \ket{e_k},k\in[1,\textrm{dim}(\cHe)]\}$ a suitable basis of the environment Hilbert space. To better describe now the system evolution, it is most convenient to work in the interaction (or Dirac) picture, where both the states and the operators carry part of the time dependence (as opposed to the standard Schr\"odinger and Heisenberg pictures, where either the states or the operators carry the whole time dependence, respectively \cite{messiah99a}). In the interaction picture operators carry the \emph{known} part of the time dependence, and evolve solely due to an \emph{additional} interaction term. This is most useful when dealing with the effect of perturbations on the known dynamics of an unperturbed system \cite{breuer02a}. In our case, denoting as $\tO(t)$ the interaction-picture representation of an arbitrary operator $O\in\cBHt$, we have that 
\be
\tO(t)= \text{e}^{\ii(H+H_E)t} \, O \, \text{e}^{-\ii(H+H_E)t} \, ,
\label{eq:intP}
\ee
where we did not make explicit the identity operators accompanying the system and environment Hamiltonians, see Eq. (\ref{eq:hamil}) and the accompanying discussion. Taking now the time derivative of $\trhoT(t)$, defined as in Eq. (\ref{eq:intP}), the Liouville-von Neumann equation (\ref{eq:LvN}) in the interaction picture reduces to
\be
\dot{\trho}_T(t)=-\ii \eta \cor{\tV(t),\trhoT(t)} \, . 
\label{eq:vn_int}
\ee
This equation can be formally integrated to yield
\be
\trhoT(t)=\trhoT(0)-\ii \eta \int_0^t ds \cor{\tV(s),\trhoT(s)} \, ,
\ee
and introducing this result in the right-hand side (rhs) of Eq. (\ref{eq:vn_int}) we obtain a Dyson-type expansion
\ben
\dot{\trho}_T(t)&=&-\ii \eta \cor{\tV(t),\trhoT(0)}  -\eta^2 \int_{0}^t ds \cor{\tV(t), \cor{ \tV(s),\trhoT(s)}} \nonumber \\ 
&=&-\ii \eta \cor{\tV(t),\trhoT(0)}  -\eta^2 \int_{0}^t ds \cor{\tV(t), \cor{ \tV(s),\trhoT(t)}} + {\cal O}(\eta^3) \, ,
\label{eq:red_vn0}
\een
where we have iterated the expansion to obtain the second equality. In the weak coupling limit, $\eta\to 0$, we can neglect higher-order terms in the coupling constant to obtain a local-in-time master equation
\be
\dot{\trho}_T(t)=-\ii \eta \cor{\tV(t),\trhoT(0)}  -\eta^2 \int_{0}^t ds \cor{\tV(t), \cor{ \tV(s),\trhoT(t)}} \, .
\label{eq:red_vn}
\ee
The evolution equation for the system reduced density matrix can be obtained now by tracing over the environment degrees of freedom, see Eq. (\ref{eq:trE}), arriving at
\be
\dot{\trho}(t)=-\ii \eta \Tr_E \cor{\tV(t),\trhoT(0)}  -\eta^2 \int_{0}^t ds \Tr_E \cor{\tV(t), \cor{ \tV(s),\trhoT(t)}} \, .
\label{eq:red_vn2}
\ee
Note that this is still not a closed evolution equation for $\trho$, as the rhs of the previous equation still depends on the full density matrix $\trhoT$.

In order to proceed, we now assume that initially the system and the environment are uncorrelated, $\trhoT(0)=\trho(0)\otimes \trho_E(0)$. Moreover, we also assume that the environment is in a thermal state at the initial time, $\trho_E(0)=\rho_{th}\equiv \exp(-H_E/T)/\Tr(\exp(-H_E/T))$, with $T$ some temperature and Boltzmann's constant $k_B=1$, so $\trhoT(0)=\trho(0)\otimes \rho_{th}$ \cite{breuer02a,rivas10a,kryszewski08a}. In this case it is easy to see that $\Tr_E \cor{\tV(t),\trho(0)\otimes \rho_{th}}=0$, so the first term in the rhs of Eq. (\ref{eq:red_vn2}) disappears. Indeed, using the form (\ref{eq:inter}) of the interaction Hamiltonian,
\be
\Tr_E \cor{\tV(t),\trho(0)\otimes \rho_{th}}=\sum_l \left( \tilde{S}_l(t)\trho(0) \Tr_E\big(\tilde{E}_l(t)\rho_{th}\big) - \trho(0) \tilde{S}_l(t) \Tr_E\big(\rho_{th} \tilde{E}_l(t)\big) \right) = 0 \, ,
\label{eq:zero}
\ee 
where we have used that $\la E_l\ra\equiv \Tr(E_l\rho_{th}) =0$, which in turn implies that $\Tr_E(\tilde{E}_l(t)\rho_{th})=0=\Tr_E(\rho_{th} \tilde{E}_l(t))$ due to the cyclic property of the trace and the fact that $\cor{H_E,\rho_{th}}=0$. The condition $\la E_l\ra =0$ does not need any extra assumption about the system \cite{kryszewski08a}, because any Hamiltonian $H_T=H+H_E+\eta V$ with an interaction term $V=\sum_l S_l\otimes E_l$ such that $\la E_l\ra \neq 0$ can be rewritten as $H_T=(H+\eta \sum_l \la E_l\ra S_l) + H_E + \eta V'$, with $V'=\sum_l S_l\otimes (E_l-\la E_l\ra)$ such that now $\Tr_E \cor{\tV'(t),\trho(0)\otimes \rho_{th}}=0$, the only difference being a global shift in the system energy with no physical effect. We hence assume without loss of generality an environment such that $\la E_l\ra =0$ $\forall l$. In this way, Eq. (\ref{eq:red_vn2}) reduces to
\be
\dot{\trho}(t)=  -\eta^2 \int_{0}^t ds \Tr_E \cor{\tV(t), \cor{ \tV(s),\trhoT(t)}} \, .
\label{eq:red_vn3}
\ee
For arbitrary times we can always write $\trhoT(t) = \trho(t)\otimes \trho_E(t) + \trho_\textrm{corr}(t)$, where $\trho_\textrm{corr}(t)$ represents the entangled part of the total density matrix induced by the system-environment interaction. Our next step consists in assuming a strong separation of timescales between the system and the environment. In particular, we assume that the environment correlation and relaxation timescales ($\tau_\text{corr}$ and $\tau_\text{rel}$, respectively) are much faster than the typical time-scale $\tau_0$ for the system to change due to its interaction with the environment \cite{kryszewski08a}. This assumption is physically motivated in the weak coupling limit $\eta \to 0$, where the system evolution due to its interaction with the environment slows down proportionally to $\eta^2$, an observation made clear in the interaction picture, see Eq. (\ref{eq:red_vn3}). In this way, the assumption $\tau_\text{corr} \ll\tau_0$ allows us to neglect the system-bath correlations at any time, so we can write $\trhoT(t) = \trho(t)\otimes \trho_E(t)$. Moreover, because $\tau_\text{rel} \ll\tau_0$ the environment relaxes to thermodynamic equilibrium before any appreciable change in the system state happens, so effectively the system always interact with a \emph{thermal} environment, $\trho_E(t)=\trho_{th}$. Therefore, under this strong time-scales separation hypothesis (well-motivated in the weak coupling limit \cite{kryszewski08a}), we can always write $\trhoT(t) = \trho(t)\otimes \trho_{th}$ to obtain 
\be
\dot{\trho}(t)=  -\eta^2 \int_{0}^t ds \Tr_E \cor{\tV(t), \cor{ \tV(s),\trho(t)\otimes \trho_{th}}} \, .
\label{eq:red_vn4}
\ee
This is a local-in-time, autonomous equation for the system evolution which is still non-Markovian due to its dependence of the initial system preparation \cite{breuer02a}. However, the memory kernel in the time integral of Eq. (\ref{eq:red_vn3}) typically decays fast enough so the system effectively \emph{forgets} about this initial state and we can switch the upper limit in the time integral to infinity. In this case, and changing variables in the integral, $s\to t-s$, we obtain a Markovian quantum master equation known as \emph{Redfield equation} \cite{redfield57a}
\be
\dot{\trho}(t)=  -\eta^2 \int_{0}^\infty ds \Tr_E \cor{\tV(t), \cor{ \tV(t-s),\trho(t)\otimes \trho_{th}}} \, .
\label{eq:red_vn5}
\ee

The last equation does not yet warrant a completely positive evolution for the system density matrix, so we still need one further approximation to obtain a generator of a CPTP dynamical map: the \emph{secular} or \emph{rotating wave} approximation \cite{rivas10a}. In order to do so, we now perform a spectral decomposition of the system operators $S_l\in\cBH$ entering the definition of the system-environment bilinear interaction in Eq. (\ref{eq:inter}) in terms of the eigenoperators of the superoperator $\cor{H,\bullet}\in\cBBH$, which form a complete basis of the Hilbert space of bounded operators $\cBH$. In particular, we can now write
\be
S_l= \sum_\nu S_l(\nu) \, ,
\label{eq:Sexpand}
\ee
where the eigenoperators $S_l(\nu)$ are defined via
\be
\cor{H,S_l(\nu)} = -\nu S_l(\nu) \, ,
\label{eq:Seigen1}
\ee
with $\nu$ the associated eigenvalue. Taking now the Hermitian conjugate, it is easy to see that
\be
\cor{H,S_l^\dagger(\nu)} = \nu S_l^\dagger(\nu) \, .
\label{eq:Seigen2}
\ee
In the interaction picture, the system-environment interaction operator can be written as $\tV(t)=\textrm{e}^{\ii t\cor{H+H_E,\bullet}} V = \textrm{e}^{\ii t(H+H_E)} V \textrm{e}^{-\ii t(H+H_E)}$ and therefore
\be
\tV(t)= \sum_{l,\nu} \textrm{e}^{-i\nu t} S_l(\nu)\otimes \tE_l(t) = \sum_{l,\nu} \textrm{e}^{+i\nu t} S_l^\dagger(\nu)\otimes \tE_l^\dagger(t) \, ,
\label{eq:Veigen}
\ee
where we have used the hermiticity of $V$ in the last equality. Note that $\tV(t-s)$ can be decomposed in similar terms. Expanding the commutators in Eq. (\ref{eq:red_vn5}) we trivially find
\ben
\hspace{-2cm}\dot{\trho}(t) =  -\eta^2 \Tr_E \Big\{ \int_{0}^\infty ds  \tV(t) \tV(t-s) \trho(t)\otimes \trho_{th}  - \int_{0}^\infty ds \tV(t) \trho(t)\otimes \trho_{th} \tV(t-s) \nonumber \\ 
\hspace{-0.75cm} -  \int_{0}^\infty ds  \tV(t-s) \trho(t)\otimes \trho_{th}\tV(t) + \int_{0}^\infty ds  \trho(t)\otimes \trho_{th} \tV(t-s)\tV(t) \Big\} \, ,
\label{eq:red_vn6}
\een
and applying the spectral decomposition (\ref{eq:Veigen}) in terms of $S_l(\nu)$ for $\tV(t-s)$ and $S_k^\dagger(\nu')$ for $\tV(t)$ in the first and third term of the rhs of the previous equation (and the complementary decomposition for the other two terms), we arrive after some algebra at
\be
\dot{\trho}(t) = \sum_{\substack{\nu, \nu'  \\ k,l}} \left(\textrm{e}^{\ii(\nu'-\nu)t} \Gamma_{kl}(\nu) \cor{S_l(\nu)\trho(t),S_k^\dagger(\nu')} + \textrm{e}^{\ii(\nu-\nu')t} \Gamma_{lk}^*(\nu) \cor{S_l(\nu'),\trho(t) S_k^\dagger(\nu)} \right)\, ,
\label{eq:red_vn7}
\ee
where we have defined
\be
\Gamma_{kl}(\nu) \equiv \int_0^\infty ds \, \textrm{e}^{\ii \nu s} \Tr_E \left(\tE_k^\dagger(t) \tE_l(t-s) \trho_{th}\right) = \int_0^\infty ds \, \textrm{e}^{\ii \nu s} \Tr_E  \left(\tE_k^\dagger(s) E_l \trho_{th}\right) \, ,
\label{eq:defGamma}
\ee
with $\tE_k(t)= \textrm{e}^{\ii H_E t} E_k \textrm{e}^{-\ii H_E t}$ in the interaction picture. Note that we have used the cyclic property of the trace and the commutator $\cor{H_E,\trho_{th}}=0$ for the second equality above. In Eq. (\ref{eq:red_vn7}), all terms with $|\nu'-\nu|\gg\eta^2$ will oscillate rapidly around zero before the system evolves appreciably due to its interaction with the environment (recall that $\tau_0\propto \eta^{-2}$, see above). Therefore in the weak coupling limit, where the strong time-scales separation hypothesis holds, only the terms with $\nu=\nu'$ contribute (secular or rotating-wave approximation) and we find
\be
\dot{\trho}(t) = \sum_{\substack{\nu \\ k,l}} \left( \Gamma_{kl}(\nu) \cor{S_l(\nu)\trho(t),S_k^\dagger(\nu)} +  \Gamma_{lk}^*(\nu) \cor{S_l(\nu),\trho(t) S_k^\dagger(\nu)} \right) \, .
\label{eq:red_vn8}
\ee
Decomposing now the coefficients $\Gamma_{kl}(\nu)$ into \emph{Hermitian} and \emph{anti-Hermitian} parts \cite{breuer02a,rivas10a,buca15a}, $\Gamma_{kl}(\nu)=\frac{1}{2}\gamma_{kl}(\nu) + \ii \pi_{kl}(\nu)$, with $\pi_{kl}(\nu) \equiv \frac{1}{2\ii}(\Gamma_{kl}(\nu) - \Gamma_{lk}^*(\nu))$ and\footnote[1]{Note that both $\gamma_{kl}(\nu)$ and $\pi_{kl}(\nu)$ are Hermitian.}
\be
\gamma_{kl}(\nu) \equiv \Gamma_{kl}(\nu) + \Gamma_{lk}^*(\nu) = \int_{-\infty}^\infty ds \, \textrm{e}^{\ii \nu s} \Tr_E  \left(\tE_k^\dagger(s) E_l \trho_{th}\right) \, ,
\label{eq:gamma}
\ee
and going back to the original Schr\"odinger picture, we arrive at
\be
\dot{\rho}(t) = -\ii \cor{H+H_\textrm{Ls},\rho(t)} + {\cal D}[\rho(t)] \, , 
\label{eq:red_vn9}
\ee
with the dissipator superoperator ${\cal D}[\bullet]\in \cBBH$ defined as
\be
{\cal D}[\bullet] \equiv \sum_{\substack{\nu \\ k,l}} \gamma_{kl}(\nu) \left( S_l(\nu)\bullet S_k^\dagger(\nu) - \frac{1}{2}\acor{\bullet,S_k^\dagger(\nu) S_l(\nu)} \right)\, ,
\label{eq:dissip}
\ee
with $\acor{A,B}=AB+BA$ the anti-commutator and $H_\text{Ls}\equiv \sum_{\nu,k,l} \pi_{kl}(\nu) S_k^\dagger(\nu) S_l(\nu) \in \cBH$ a \emph{Lamb shift} Hamiltonian \cite{breuer02a} which amounts to a renormalization of the system energy levels due to its interaction with the environment. Note that $\cor{H,H_\text{Ls}}=0$, see Eqs. (\ref{eq:Seigen1})-(\ref{eq:Seigen2}) above. Eqs. (\ref{eq:red_vn9})-(\ref{eq:dissip}) are the \emph{first standard form} of the quantum master equation.

The coefficients $\gamma_{kl}(\nu)$ are positive semidefinite in all cases as they are the Fourier transform of a positive function, the bath correlation function $\Tr_E  \left(\tE_k^\dagger(s) E_l \trho_{th}\right)$ \cite{breuer02a,rivas10a,buca15a}. Therefore the matrix $\gamma$ formed by these coefficients may be diagonalized by an appropriate unitary transformation $\tau$ such that
\be
\tau \, \gamma \, \tau^\dagger= \left ( 
\begin{array}{ccc}
d_1(\nu) & 0 & \cdots \\
0 & d_2(\nu) & \cdots \\
\vdots &  \vdots &  \ddots \end{array}
\right) \, ,
\label{eq:utrans}
\ee
so $d_i(\nu)=\sum_{kl} \tau_{ik}\gamma_{kl}(\nu)\tau_{il}^*$. This transformation allows now to write the master equation (\ref{eq:red_vn9})-(\ref{eq:dissip}) into Lindblad form
\be
\dot{\rho}(t) = -\ii \cor{H+H_\textrm{Ls},\rho(t)} + \sum_{i,\nu} \left( L_i(\nu)\rho(t) L_i^\dagger(\nu) - \frac{1}{2}\acor{\rho(t),L_i^\dagger(\nu) L_i(\nu)} \right) \equiv \cL \rho(t)\, , 
\label{eq:red_vn10}
\ee
with $L_i(\nu)\in\cBH$ Lindblad operators acting on the system defined via
\be
L_i(\nu) \equiv \sqrt{d_i(\nu)} \sum_l \tau_{il} S_l(\nu) \, . 
\label{eq:lindop}
\ee
Eq. (\ref{eq:red_vn10}) defines the Lindblad-Liouville superoperator $\cL$. This superoperator generates a completely positive and trace preserving (CPTP) map, and it is the most general Markovian generator that preserves the physical requirements of the density matrix \cite{breuer02a,gardiner00a}.

Lindblad (or rather Lindblad-Gorini-Kossakowski-Sudarshan) master equations similar to (\ref{eq:red_vn10}) were originally derived and used in the realm of quantum optics \cite{briegel93a,cohen11a,gardiner00a}. More recently it has been applied to a plethora to problems including quantum effects in photosynthetic complexes \cite{caruso09a,mohseni08a, manzano13a,witt13a}, transport in one-dimensional \cite{saito03a,znidaric10a, manzano12a,bermudez13a}  and multi-dimensional quantum systems \cite{znidaric13b,znidaric13a,asadian13a,manzano16b}, trapped ions \cite{barreiro11a} and optomechanical systems \cite{tomadin12a}, or quantum information and computation \cite{verstraete09a}, to mention just a few. The validity of this equation in a system with local coupling to the environment was analitically and numerically analized in \cite{rivas10a}.

\newpage

\section{Steady states, symmetries, and invariant subspaces}
\label{sec:sym}

\subsection{Some general properties of the spectrum of $\cL$}
\label{sec:sym1}

The main feature of Lindblad equation (\ref{eq:red_vn10}) is its dissipative character, a property associated to the non-Hermiticity of the generator $\cL$ and in stark contrast with the coherent evolution induced by standard Hamiltonian dynamics. As a non-Hermitian superoperator, $\cL$ may not be diagonalizable. However, if diagonalizable, $\cL$ will exhibit in general distinct sets of right and left eigenoperators, and its eigenvalues will typically come in complex conjugate pairs \footnote[1]{The spectral analysis of the superoperator $\cL$ is better understood in the Fock-Liouville space associated to the operator Hilbert space $\cBH$ \cite{prosen08a,albert16a,thingna16a}. In Fock-Liouville space, operators (and density matrices in particular) can be mapped onto complex \emph{vectors} of dimension $D^2$, while the Lindblad-Liouville superoperator $\cL$ is mapped onto a complex $D^2\times D^2$ matrix (with $D$ the finite dimension of the system Hilbert space $\cH$).}.
Let $\phi_k, \, \hat{\phi}_k \in\cBH$ be right and left eigenoperators of $\cL$, respectively, with (common) eigenvalue $\Lambda_k\in\mathbb{C}$ such that
\be
\cL \phi_k = \Lambda_k \phi_k \qquad , \qquad \hat{\phi}_k \cL \equiv \cL^\dagger \hat{\phi}_k = \Lambda_k \hat{\phi}_k \, ,
%\qquad \cL^\dagger \hat{\phi}_k \equiv \hat{\phi}_k \cL  = \Lambda_k \hat{\phi}_k \, ,
%\qquad \hat{\phi}_k \cL = \Lambda_k \hat{\phi}_k \, .
\label{eq:eig}
\ee
%where the action of the generator to the left is defined by the adjoint generator $\hat{\phi}_k \cL = \cL^\dagger \hat{\phi}_k$. For a generator defined by Eq.  (\ref{eq:red_vn10}) the adjoint generator is defined as
with the adjoint generator $\cL^\dagger$ defined as 
%see Eq. (\ref{eq:red_vn10}),
\be
 \cL^\dagger \rho =  \ii \cor{H+H_\textrm{Ls},\rho(t)} + \sum_{i,\nu} \left( L_i^\dagger(\nu)\rho(t) L_i(\nu) - \frac{1}{2}\acor{\rho(t),L_i(\nu) L_i^\dagger(\nu)} \right) \, ,
\ee
see Eq.~(\ref{eq:red_vn10}). This is written in the Schr\"odinger picture, meaning that $\cL$ is the generator of the dynamics of states while $\cL^{\dagger}$ is the generator of the dynamics of operators. The set of right and left eigenoperators of $\cL$ form a biorthogonal basis of $\cBH$, such that $\bbraket{\hat{\phi}_i}{\phi_j}=\delta_{i,j}$, with the Hilbert-Schmidt inner product $\bbraket{\hat{\phi}_k}{\rho(0)}=\Tr\left( {\hat{\phi}_k}^\dagger \rho(0) \right)$ as defined in Eq.~(\ref{inner}). In this way,
%so 
any arbitrary density matrix $\rho(0)$ can be decomposed into this basis, and its time evolution $\rho(t)=\exp(+t\cL) \rho(0)$ can be written as
\be
\rho(t)=\sum_k e^{+\Lambda_k t} \bbraket{\hat{\phi}_k}{\rho(0)}\; \phi_k \, .
\label{eq:LF-evol}
\ee
%with the Hilbert-Schmidt inner product $\bbraket{\hat{\phi}_k}{\rho(0)}=\Tr\left( {\hat{\phi}_k}^\dagger \rho(0) \right)$, as defined in Eq. (\ref{inner}). The biorthogonality of the basis means that $\bbraket{\hat{\phi}_i}{\phi_j}=\delta_{i,j}$. 
Since the Lindblad-Liouville superoperator is trace preserving, we conclude that all its eigenvalues $\Lambda_k$ must have a non-positive real part, $\text{Re}(\Lambda_k)\le 0$, with at least one eigenvalue such that $\text{Re}(\Lambda_k)= 0$ \cite{albert14a,evans77a,albert16a}. Moreover, steady states now correspond to the null fixed points of $\cL$, i.e. to the eigenmatrices corresponding to zero eigenvalue, $\cL \rho_\text{st}=0$. Note that we will be interested below in Lindblad operators $L_i$ describing most common physical situations, namely (i) coupling to different reservoirs (of energy, spin, etc.) which locally inject and extract excitations at constant rate, or (ii) the effect of environmental dephasing noise which causes local decoherence and thus classical behavior \cite{breuer02a,schlosshauer07a}. In this way, Eq.~(\ref{eq:red_vn10}) will describe all sorts of nonequilibrium situations driven by external gradients and noise sources, giving rise in general to non-zero currents flowing between the system and the baths. We will refer to these fixed points of the dynamics as nonequilibrium steady states (NESS), and denote the associated density matrix as $\rho^\textrm{NESS}$.

\begin{figure}
\centerline{\includegraphics[scale=0.2]{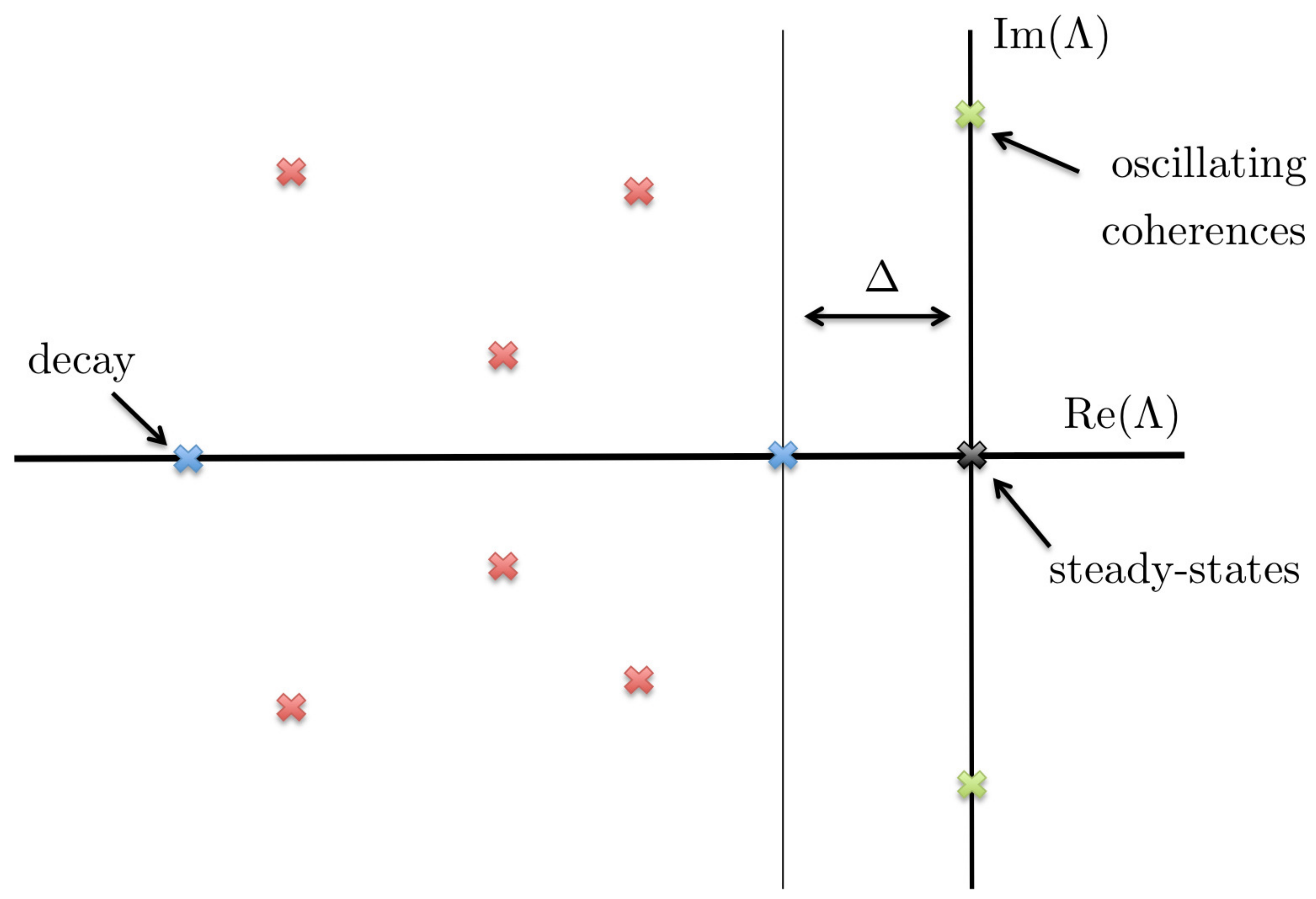}}
\caption{Sketch inspired by Fig. 1 of  Ref. \cite{albert14a}. The black cross represents the steady-states (eigenmatrices of the Lindblad-Liouville superoperator with zero eigenvalue), blue crosses represent pure exponential decay (zero imaginary value but negative real part),  green crosses are oscillating coherences (zero real part and non-zero imaginary part), and red crosses are general eigenvalues (non-zero real and imaginary parts) which give rise to \emph{spiral} relaxation. All eigenvalues with non-zero imaginary parts come in complex conjugate pairs. $\Delta$ is the energy gap between the steady-state and the first non-zero eigenvalue, and it is inversely proportional to the relaxation time.}
\label{fig:spectrum}
\end{figure}

Calculating the steady states and dominant relaxation modes for a given Liouvillian $\cL$ is in general non-trivial, but the possible outcomes can be phenomenologically understood by a closer look at its spectrum \cite{albert14a,albert16a}, see Fig. \ref{fig:spectrum}. Indeed, it is clear from Eq. (\ref{eq:LF-evol}) that for long times only eigenmatrices of $\cL$ corresponding to eigenvalues with zero real part will be present. All eigenoperators with $\text{Re}(\Lambda_k)<0$ will suffer an exponential decay with time, vanishing at the steady state. This decay might be purely exponential for $\text{Im}(\Lambda_k)=0$ or rather an exponentially damped oscillatory mode for $\text{Im}(\Lambda_k)\ne 0$, also known as \emph{spiral} relaxation \cite{albert14a}. There is also the possibility of eigenvalues with $\text{Re}(\Lambda_k)=0$ but $\text{Im}(\Lambda_k)\ne 0$ (oscillating coherences). These eigenvalues correspond to states that are robust under the dissipative character of the Liouvillian, but never reach a stationary state with no time evolution. The variety of possible eigenvalues is represented in Figure \ref{fig:spectrum} (see Ref. \cite{albert14a} for discussion and examples). Note that the slowest relaxation time-scale of the system of interest is proportional to the inverse of the spectral gap $\Delta$ between the steady-states and the first eigenvalue with non-zero real part.

\subsection{Symmetry and degenerate steady states}
\label{sec:sym2}

As described above, for long times all relaxation modes decay and only the steady state remains. Interestingly, the uniqueness of this steady state is not guaranteed a priori \cite{hsiang15a}, and Evan's theorem \cite{evans77a,frigerio78a,davies70a,spohn76a,spohn77a,watanabe79a} specifies the conditions under which a given Lindblad generator exhibits a unique stationary point. Roughly speaking, the steady state will be unique iff the set of operators spanned by the system Hamiltonian $H$ and all Lindblad operators, $L_i$ and $L_i^\dagger$, generates when added and/or multiplied the complete algebra of operators defining the system\footnote[1]{An additional technical requirement for Evan's proof is that the steady-state density matrix is full rank. Note however that there are no known examples when the conditions on the Hamiltonian and Lindblad operators mentioned above hold but the associated steady-state density matrix does not have full rank.
%In all known cases, the stated condition to the Hamiltonian and Lindblad operators implies a full rank density matrix.
}.

Our purpose in this section is to review the effects of symmetry in the steady state properties of open quantum systems described by master equations like Eq. (\ref{eq:red_vn10}), though we will extend our discussion below to treat also symmetries in more general settings. In particular, we say that a system exhibits a symmetry iff there exists a unitary operator $U\in \cBH$ such that
\be
\cor{U,H}=0=\cor{U,S_l}\;\; \forall l \, ,
\label{eq:sym}
\ee
where recall that $H\in\cBH$ is the system Hamiltonian, see Eq. (\ref{eq:hamil}), and $S_l\in\cBH$ are the system operators defining the system-environment interaction $V=\sum_l S_l\otimes E_l$. Note that for Lindblad-type master equations (\ref{eq:red_vn10}), the previous definition extends to the Lindblad operators $L_i\in\cBH$, see Eq. (\ref{eq:lindop}), which also commute with the unitary operator $U$. We stress that this is the case termed \emph{strong symmetry} in the language of Refs. \cite{buca12a,buca15a}, though our definition here is somewhat broader to discuss below the role of symmetry in more general master equations as e.g. the Redfield equation (\ref{eq:red_vn5}).

The commutation relations (\ref{eq:sym}) immediately imply that both the Hamiltonian $H$ and $U$ share a common eigenbasis. Let's denote as  $\ket{\psi_\alpha^{(k)}}$ and $u_\alpha$ the eigenvectors and eigenvalues of $U$, respectively, with $\alpha\in [1,n_U]$ and $k\in[1,d_\alpha]$. Here $n_U$ is the number of distinct eigenvalues of $U$, with $1\le n_U\le D$, and $d_\alpha$ is the dimension of the subspace corresponding to eigenvalue $u_\alpha$, such that $\sum_{\alpha=1}^{n_U} d_\alpha =D$. In this way 
\be
U\ket{\psi_\alpha^{(k)}} =u_\alpha \ket{\psi_\alpha^{(k)}} =  \text{e}^{\ii \Omega_{\alpha}} \ket{\psi_\alpha^{(k)}} \, ,
\label{eq:Ueigen}
\ee
where in the last equality we have used that $U$ is unitary ($U^{-1}=U^\dagger$) so its eigenvalues are pure phases $\text{e}^{\ii \Omega_{\alpha}}$, with $\Omega_\alpha\in \mathbb{R}$. 

The system Hilbert space $\cH$ can be now decomposed in terms of the spectrum of $U$,
\be
\cH=\bigoplus_{\alpha=1}^{n_U} \cH_\alpha \,, \qquad \text{with} \quad \cH_\alpha=\key{ \ket{\psi_\alpha^{(k)}},\; k\in\cor{1,d_\alpha}} \, .
\label{eq:Hdecomp}
\ee
The previous spectral decomposition can be extended to the operator Hilbert space. In order to do so, we first define a superoperator $\cU\in\cBBH$ associated to the adjoint representation of the unitary operator $U$ in $\cBH$, 
\be
\cU \rho\equiv U \rho \, U^\dagger \quad \forall \rho\in \cBH \, . 
\label{eq:defUsup}
\ee
The spectrum of $\cU$ then follows as 
\be
\cU\op{\psi_\alpha^{(n)}}{\psi_\beta^{(m)}} =  \text{e}^{\ii \pare{\Omega_\alpha-\Omega_\beta}} \op{\psi_\alpha^{(n)}}{\psi_\beta^{(m)}} \, ,\qquad (\alpha,\beta=1,\ldots,n_U) \, , 
\label{eq:Usuper1}
\ee
and the adjoint space $\cBH$ can be now decomposed as $\cBH =\bigoplus_{\alpha=1}^{n_U} \bigoplus_{\beta=1}^{n_U} \cBa$, where the symmetry subspaces $\cBa$ are defined as
\be
\cBa=\key{ \op{\psi_\alpha^{(n)}}{\psi_\beta^{(m)}} : n\in\cor{1,d_\alpha}, m\in\cor{1,d_\beta} } \, , 
\label{eq:Usuper2}
\ee
each having a dimension $d_{\alpha\beta}\equiv d_{\alpha}d_{\beta}$.

The existence of a symmetry operator $U$ with the properties (\ref{eq:sym}) then implies the following two simple but important results, namely \cite{buca12a,buca15a,manzano14a}
\begin{itemize}
\item[(i)] The flow induced by the Lindblad-Liouville superoperator $\cL$ leaves invariant the subspaces $\cBa$, i.e. $\cL\cBa \subseteq \cBa$, so $\cL$ can be block-decomposed into $n_U^2$ invariant subspaces.
\item[(ii)] We have at least $n_U$ different (nonequilibrium) steady states or null fixed points of the generator $\cL$, one for each diagonal subspace $\cBaa$, so these steady states can be labelled by the symmetry index $\alpha\in[1,n_U]$.
\end{itemize}
To prove the first result we need to introduce now the right and left adjoint superoperators $\cU_{l,r}\in\cBBH$ associated to the unitary operator $U$. These are defined via
\be
\cU_l \rho = U\rho \qquad ; \qquad \cU_r \rho = \rho \, U^\dagger \qquad \forall \rho\in\cBH \, ,
\label{eq:Ulr}
\ee
and note that $\cor{\cU_l,\cU_r}=0$. Clearly, the subspaces $\cBa$ are the joint eigenspaces of both $\cU_l$ and $\cU_r$, since 
\be
\cU_l \op{\psi_\alpha^{(n)}}{\psi_\beta^{(m)}} = \text{e}^{\ii \Omega_\alpha} \op{\psi_\alpha^{(n)}}{\psi_\beta^{(m)}} \quad , \quad \cU_r \op{\psi_\alpha^{(n)}}{\psi_\beta^{(m)}} = \text{e}^{-\ii \Omega_\beta} \op{\psi_\alpha^{(n)}}{\psi_\beta^{(m)}} \, .
\label{eq:Ulreigen}
\ee
Now, from the commutation relations (\ref{eq:sym}) and the definition of the Lindblad-Liouville superoperator (\ref{eq:red_vn10}), it follows that
\be
\cor{\cU_l,\cL}=0=\cor{\cU_r,\cL} \, , 
\label{eq:symUlr}
\ee
so for any $\rhoa\in\cBa$ we find that $\cL\rhoa$ is still an eigenoperator of both $\cU_{l,r}$, i.e. $\cL\rhoa\in\cBa$, and hence the Lindblad-Liouville evolution superoperator leaves invariant the different symmetry subspaces. This proves result (i) above. 

Next, we note that normalized (physical) density matrices (i.e. with unit trace) can only live in diagonal subspaces $\cBaa$ due to the orthogonality between the different $\cH_\alpha$. This immediately leads to at least $n_U$ distinct NESSs (i.e. $n_U$ different \emph{transport channels}), one for each $\cBaa$ with $\alpha\in[1,n_U]$, which can be labeled according to the symmetry eigenvalues. In particular for any normalized initial density matrix $\rhoaa(0)\in\cBaa$, with $\Tr \left(\rhoaa(0)\right)=1$, we have 
\be
\rhoaa^\text{NESS} \equiv \lim_{t\to\infty} \text{e}^{+t\cL} \rhoaa(0) \in \cBaa \, ,
\label{eq:NESS}
\ee 
and a continuum of possible linear combinations of these NESSs. It is important to notice that the $n_U$ different $\rhoaa^\text{NESS}$ can be further degenerated according to Evans theorem \cite{evans77a,frigerio78a,davies70a,spohn76a,spohn77a,watanabe79a}, as e.g. in the presence of other symmetries which allow to further block-decompose the evolution superoperator, though we will assume here for simplicity that $\rhoaa^\text{NESS}$ are unique for each $\alpha$. As an interesting corollary, note that the dynamical generator $\cL$ will leave invariant one-dimensional symmetry eigenspaces $\op{\psi_\alpha}{\psi_\alpha}$, mapping them onto themselves. This defines decoherence-free, dark states which remain pure even in the presence of enviromental noise, leading to important applications in e.g. quantum computing to protect quantum states from relaxation \cite{buca12a,albert14a,blume-kohout08a,verstraete09a,kraus08a}. We will illustrate below the use of dark states to control quantum transport in arbitrary nonequilibrium settings.

We now turn our attention to the effect of symmetries, defined as in Eq. (\ref{eq:sym}), on the steady state structure of more general Markovian quantum master equations, as e.g. the Redfield equation (\ref{eq:red_vn5}), which in the simpler Dirac (interaction) picture reads
\be
\dot{\trho}(t)= -\eta^2 \int_{0}^\infty ds \Tr_E \cor{\tV(t), \cor{ \tV(t-s),\trho(t)\otimes \trho_{th}}} \equiv \tilde{\cL}_R(t) \trho(t)\, .
\label{eq:red_vn4bis}
\ee
The last equality defines the Redfield superoperator in the interaction picture, $\tilde{\cL}_R(t)$, which we will show next also leaves invariant the symmetry subspaces $\cBa$ and hence exhibits at least $n_U$ different steady states, as in the Lindblad case. As before, the strategy in order to proceed consists in demonstrating that, for any $\trho_{\alpha\beta}(t)=\text{e}^{\ii H t} \rhoa \text{e}^{-\ii H t}\in\cBa$, the operator resulting from the application of the dynamical generator of interest to the original state, $\tilde{\cL}_R(t)\trho_{\alpha\beta}(t)$, remains in the same subspace $\cBa$. Note that $\trho_{\alpha\beta}(t)\in\cBa \Leftrightarrow \rho_{\alpha\beta}\in\cBa$ due to the commutator $\cor{U,H}=0$, see Eq. (\ref{eq:sym}). We hence apply the right and left adjoint symmetry superoperators $\cU_{l,r}$ on $\tilde{\cL}_R(t)\trho_{\alpha\beta}(t)$. Starting with $\cU_{l}$, we find
\ben
\hspace{-2cm}\cU_{l}\tilde{\cL}_R(t)\trho_{\alpha\beta}(t) &=& -\eta^2 \int_{0}^\infty ds \Tr_E \left(U\otimes\ide \cor{\tV(t), \cor{ \tV(t-s),\trho_{\alpha\beta}(t)\otimes \trho_{th}}} \right) \nonumber \\
&=&-\eta^2 \int_{0}^\infty ds \Tr_E  \cor{\tV(t), \cor{ \tV(t-s),U \trho_{\alpha\beta}(t)\otimes \trho_{th}}} = \tilde{\cL}_R(t)\cU_{r} \trho_{\alpha\beta}(t) \, , \nonumber
\een
while for $\cU_{r}$ the calculation is equivalent. Therefore $\cor{\cU_l,\cL_R}=0=\cor{\cU_r,\cL_R}$, and this guarantees that $\forall \trho_{\alpha\beta}(t) \in\cBa \Rightarrow \tilde{\cL}_R(t)\trho_{\alpha\beta}(t) \subseteq \cBa$.

Another interesting issue that we will not treat here in detail concerns the relation between symmetries and conservation laws in open quantum system. This connection, though present, is non-trivial and far less direct than in closed quantum systems subject to coherent dynamics, where it is fully characterized by Noether's theorem \cite{noether1918a,noether71a}. This problem has been recently addressed for Lindblad-type master equations by Albert and Jiang in Ref. \cite{albert14a}. In brief, they show that Noether's theorem does not fully apply for the dissipative (non-unitary) dynamics generated by the Lindblad-Liouville superoperator $\cL$. In particular, and among other peculiarities, they demonstrate that open quantum systems of Lindblad-type may exhibit conservation laws which do not correspond to (strong) symmetries as defined in Eq. (\ref{eq:sym})\footnote[1]{Note however that these conservation laws can be linked in most cases \cite{albert14a} to \emph{weak symmetries} in the language of Refs. \cite{buca12a,buca15a}}. Noether's theorem states that in a closed quantum system under unitary dynamics with Hamiltonian $H$, any unitary symmetry $U=\text{e}^{\ii\Omega}\in \cBH$ with Hermitian generator $\Omega=\Omega^\dagger\in\cBH$ has a corresponding conservation law for the expectation value of this physical observable, $\la \dot{\Omega}\ra=0$, with $\la \Omega\ra=\Tr(\Omega\rho)$, or equivalently: $\cor{U,H}=0 \Leftrightarrow \dot{\la \Omega\ra}=0$. As shown in Ref. \cite{albert14a}, for Lindblad open quantum systems the equivalent proposition holds in general just in one direction, namely: $\cor{U,H}=0=\cor{U,L_i} \, \forall i\Rightarrow \dot{\la\Omega\ra}=0$.

In summary, we have shown in this section that master equations of Lindblad- and Redfield-form can exhibit different invariant subspaces as a consequence of the inner symmetries of the system of interest. Due to Evans theorem \cite{evans77a}, for finite systems each invariant subspace should contain at least one steady-state, which corresponds to zero eigenvalues of the Liouvillian spectrum \cite{albert14a,thingna16a}. Note however that calculating the number of eigenvalues (and their associated eigenmatrices) for arbitrary systems can be a highly nontrivial task. Some difficulties are: 
\begin{itemize}
\item In order to calculate the null eigenvalues of the Liouvillian one has to diagonalise a non-hermitian $D^2\times D^2$ complex matrix, with $D$ the dimension of the pure-states Hilbert space. This is typically a very hard problem for relevant system sizes.

\item Not all eigenfunctions associated to the zero eigenvalue of the Liouvillian correspond to \emph{physical} steady states, as some may have zero trace \cite{buca12a,thingna16a}. Indeed, it is possible to find eigenmatrices of the Liouvillian corresponding to eigenvalue 0 that belong to non-diagonal invariant subspaces $\cBa$ with $\alpha\neq \beta$, see Eq. (\ref{eq:Ulreigen}). Naturally 
%these traceless fixed points still have physical meaning as components in a convex superposition defining a physical NESS density matrix.
non-physical (traceless) fixed points can be still linearly combined with a \emph{real} density matrix, therefore changing the physical properties of the resulting steady state.

\item By diagonalisation one obtains a basis of the zero-eigenvalue subspace of the Liouvillian. The different pairs of biorthogonal left and right eigenmatrices obtained in this way 
%(biorthogonal by pairs) 
do not necessarily correspond to orthogonal steady-states and the physical (unit trace) and non-physical (zero trace) fixed points are mixed. Because of this reason, it is difficult to know how many different physical steady-states exist in a system even after the diagonalisation. To recover the physical states one should apply a Gram-Schmidt orthonormalization procedure within the resulting subspace, a computationally-expensive procedure for high dimension.
\end{itemize}
We will discuss in Sections \ref{sec:LDF} and \ref{sec:LDF2.5} below
%in the next Section 
a complementary approach to the effect of symmetries on the dynamics of open quantum systems, based on full counting statistics, which simplifies this analysis in most cases. Before that, however, we review some particular examples of open quantum systems with symmetries.

\newpage

\section{Examples of dissipative spin systems with symmetries}
%\section{Some dissipative spin systems with symmetries}
%\section{Dissipative spin systems with (strong) symmetries}
\label{sec:symej}

In this section
%, and before moving to describe the effect of symmetries on the fluctuation properties of open quantum systems, 
we describe two different examples of open quantum systems which exhibit strong symmetries in the sense of the previous section. In particular, we will analyze in some detail a driven spin chain based on the Heisenberg XXZ model, and a spin ladder structure.

\subsection{Driven spin chains}
\label{sec:symej1}

%\subsubsection{Symmetry-controlled chain}
Our aim here is to provide a simple example of a driven dissipative open quantum system exhibiting multiple steady states as a result of a symmetry in the sense of \S\ref{sec:sym2}. A first example, already proposed and analyzed in Ref. \cite{buca12a}, is a finite open anisotropic Heisenberg XXZ spin $1/2$ chain. Note that the study of of this model also sheds light on the long-standing problem of normal and anomalous energy transport in one-dimensionas systems \cite{znidaric10a,manzano12a,asadian13a,manzano16b}. 

\begin{figure}
\centerline{\includegraphics[width=8cm]{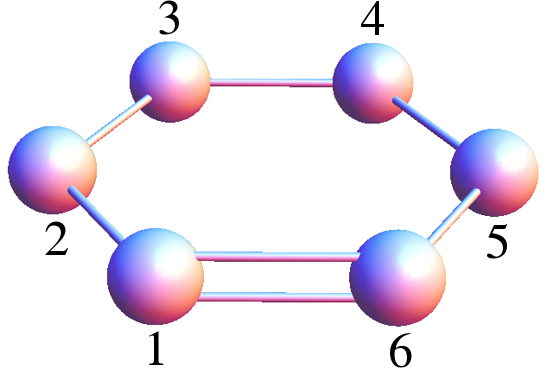}}
\caption{Sketch of the XXZ spin chain described in the text for the particular case of $L=6$ sites. Single bonds represent coherent hopping betwen bulk sites. The double bond represents fully incoherent (and possibly asymmetric) hopping between sites 1 and $L$.}
\label{fig:buca1}
\end{figure}

The XXZ Heisenberg chain is described by the Hamiltonian
\be
H_\text{XXZ}=\sum_{i=1}^{L-1} \sigma_i^x \sigma_{i+1}^x +  \sigma_i^y \sigma_{i+1}^y + \Delta~\sigma_i^z \sigma_{i+1}^z,
\label{eq:chain_xxz}
\ee
where $\sigma_i^{x,y,z}$ are the standard Pauli matrices acting on site $i\in[1,L]$, defined on a Hilbert space $\cH=(\mathbb{C}^2)^{\otimes L}$ of dimension $2^L$, and $\Delta$ is a dimensionless coupling constant. In addition, the system is driven out of equilibrium by two (magnetic) reservoirs acting on the two ends of the chain. We model these reservoirs by a pair of Lindblad \emph{non-local} jump operators \cite{buca12a}
%The same system can present multiple steady states if we choose non-local jump operators  \cite{buca12a}
\ben
L_1^{\text{nl}}= \sqrt{\Gamma \pare{1-\mu}} \sigma_1^+ \sigma_L^- \, ,\nonumber\\
L_2^{\text{nl}}= \sqrt{\Gamma\pare{ 1+\mu }} \sigma_1^- \sigma_L^+ \, .
\label{eq:lind_buca}
\een
Here $\sigma_i^\pm \equiv (\sigma_i^x \pm \text{i}\sigma_i^y)/2$ are the spin flip operators on site $i$, and $\Gamma>0$ and $\mu\in[0,1]$ measure the strength of the reservoir coupling and the nonequilibrium driving. These non-local jump operators incoherently 
%(and asymmetrically for $\mu\neq 0$) 
transfer excitations from the first to the last site of the chain and viceversa.  Note that this model can be interpreted as a spin ring where exciton hopping is fully coherent between bulk bonds, i.e. bonds $(1, 2),(2, 3)\ldots ,(L-1, L)$, while it is fully dissipative (and possibly asymmetric for the particular case $\mu \neq 0$) on one bond $(L, 1)$, see Fig.~\ref{fig:buca1}. This driven dissipative quantum chain hence evolves in time according to a general Lindblad equation $\dot\rho(t)=\cL \rho(t)$, with the coherent part of the Lindblad dynamics defined by the XXZ Hamiltonian (\ref{eq:chain_xxz}) and dissipators defined by the above non-local jump operators (\ref{eq:lind_buca}).
%and is a particular case of quantum exclusion process.

In order to investigate the possible symmetries of this model, let us now define an operator $P$ that exchanges site $i$ with $L-i+1$ for all $i$. In particular, using the computational basis (defined by the eigenvectors of $\sigma_i^z$) $\ket{s_1,\ldots,s_L}$, with $s_i\in\{0,1\}$, we can write
\be
P=\sum_{\pare{s_1,\dots,s_L}\in\{0,1\}^L} \op{s_1,s_2,\dots,s_L}{s_L,s_{L-1},\dots,s_1}
\ee
Combining this operator with spins flips in all the sites we obtain a unitary operator
\be
S\equiv P \prod_{i=1}^L \sigma_i^x,
\ee
which can be interpreted as a parity operator. It is now straightforward to prove that $S$ defines a symmetry of this XXZ chain. In particular, following the definition of a symmetry in \S\ref{sec:sym2}, see Eq. (\ref{eq:sym}), we find that 
%$\cor{S,H^\text{XXZ}}=\cor{S,L^{\text{nl}}_1}=\cor{S,L^{\text{nl}}_2}=0$. 
\be
\cor{S,H^\text{XXZ}}=\cor{S,L^{\text{nl}}_1}=\cor{S,L^{\text{nl}}_2}=0 \, . 
\ee
The unitary operator $S$ has two eigenvalues $s_1=1$ and $s_2=-1$. In this way, as explained in \S\ref{sec:sym2}, the presence of this symmetry gives rise to four different invariant subspaces in the sense of Eq. (\ref{eq:Usuper2}). These subspaces can be labeled by the eigenvalues of $S$ as $\cB_{+1,+1},\;\cB_{+1, -1},\;\cB_{-1, +1},\; \text{and}\;\cB_{-1, -1}$, and recall that only the two diagonal eigenspaces can hold physical steady states as all the elements of $\cB_{+1, -1}$ and $\cB_{-1, +1}$ have zero trace (though they can contribute as components of a real, unit trace density matrix). 

\begin{figure}
\centerline{\includegraphics[width=16cm]{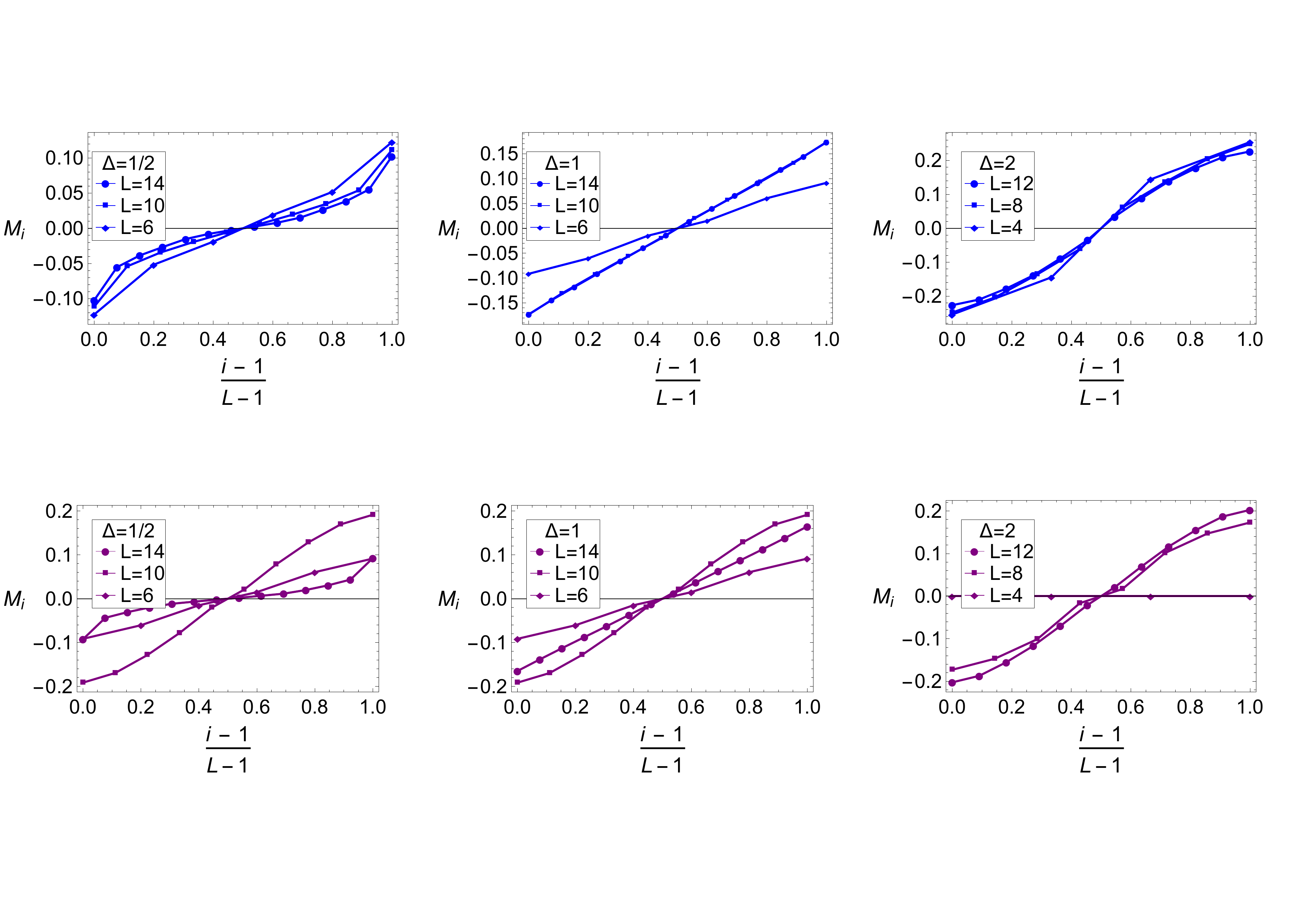}}
\vspace{-1.5cm}
\caption{Fig. 2 from Ref. \cite{buca12a}. Numerical magnetization profiles for different values of $\Delta = 1/2, 1, 2$ in the $\{+1, +1\}$
subspace, top row (blue), respectively, comparing different chain sizes (see legends). Results for the $\{-1, -1\}$ subspace are displayed in the lower row (purple). }
\label{fig:buca_profiles}
\end{figure}

Interestingly, the spin chain here described has another symmetry given by the magnetization operator $S_z=e^{i\phi M}$, with  $M=\sum_{j=1}^L \sigma_j^z$ and $\phi\in \mathbb{R}$, which reflects the conservation of the system total magnetization $M$ \cite{buca12a}. Therefore, for each eigenvalue $s_z\in \pare{-L,-L+2,\dots,L-2,L}$ of the magnetization operator, we have two distinct NESS depending on the eigenvalue of the symmetry $S$. The specific case of zero magnetization ($s_z=0$) and $L$ even is worth analisying with more detail, due to its relevance for transport problems \cite{buca12a}. In this case one can define two orthogonal steady states $\rho^{\scriptscriptstyle\text{NESS}}_{\pm}$, defined by $s_z=0$ and either $s_1=+ 1$ or $s_2=-1$, from which any general steady state of the system can be paramereterized using a real constant $u\in\cor{0,1}$,
\be
\rho^{\scriptscriptstyle\text{NESS}}= u \rho^{\scriptscriptstyle\text{NESS}}_{+} + (1-u) \rho^{\scriptscriptstyle\text{NESS}}_{-}, \qquad \rho^{\scriptscriptstyle\text{NESS}}_{+}\in \cB_{+1,+1},\quad \rho^{\scriptscriptstyle\text{NESS}}_{-} \in \cB_{-1,-1}
\ee

A natural question concerns the net effect of the spin chain symmetries on its transport properties. To further investigate this issue, Bu\v{c}a and Prosen \cite{buca12a} resort to numerical simulations of the XXZ spin chain, as the steady state Lindblad problem does not admit a closed solution in terms of matrix product operators. In particular, they use both exact diagonalization numerical techniques for moderate chain sizes ($L\in[4,10]$) and the method of quantum trajectories for $L\in[12,16]$. Their numerical study focuses on three different anisotropies, $\Delta = 1/2, 1, 2$, for which previous studies have found transport to be ballistic \cite{prosen11a}, anomalous \cite{znidaric11b} and diffusive \cite{heidrich-meisner07a}, respectively. Moreover, the reservoir parameters are fixed to $\Gamma=1$ and $\mu=0.2$ so the transport problem remains close to the linear-response regime. Fig.~\ref{fig:buca_profiles} shows the chain magnetization profiles numerically obtained for different sizes $L$, for the two distinct steady states $\rho^{\scriptscriptstyle\text{NESS}}_{+}$ and $\rho^{\scriptscriptstyle\text{NESS}}_{-}$ in the symmetry subspaces $\{+1, +1\}$ and $\{-1,-1\}$. Though the magnetization profiles exhibit a clear dependence on the symmetry sector for small and moderate chain sizes, a general trend towards convergence to the same average profiles as $L$ increases is observed. A similar convergence is found for the average current traversing the system \cite{buca12a}, suggesting that the spin chain  transport properties might not depend strongly on the symmetry sector in the thermodynamic limit.
%and for density matrices belonging to the two different subspaces. 

\begin{figure}
\centerline{\includegraphics[width=16cm]{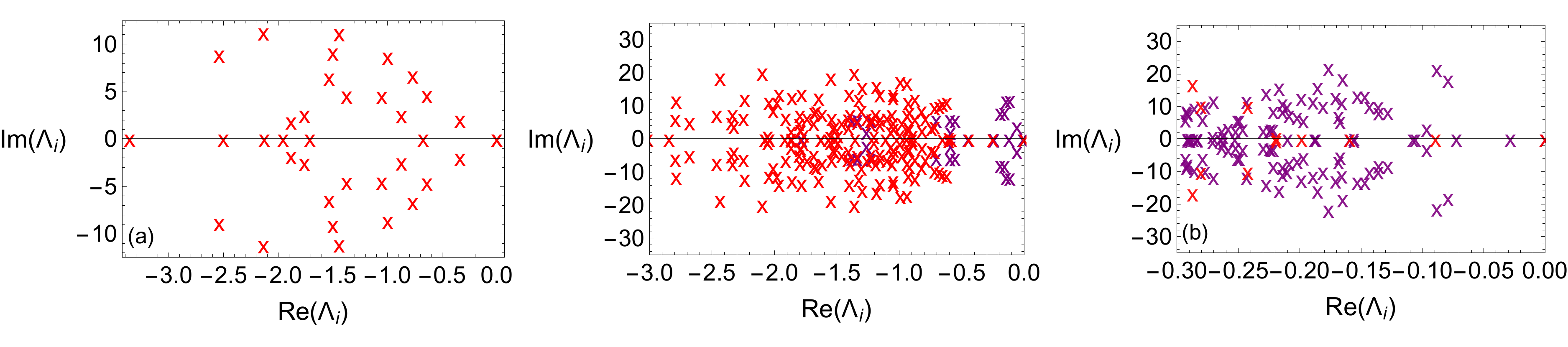}}
\caption{Fig. 4 from Ref. \cite{buca12a}. Complex plane representation of the eigenvalues $\{\Lambda_i\}$ of the Lindbladian $\cL$ of the XXZ dissipative spin chain, for coupling constant $\Delta = 2$ and three different chain sizes, $L = 6$ (left), $L = 8$ (middle), and $L = 10$ (right), see also Fig.~\ref{fig:spectrum} above. Eigenvalues color depend on the symmetry subspace to which they are associated: $\{+1, +1\}$ (red) and $\{-1,-1\}$ (purple).}
\label{fig:buca_spectrum}
\end{figure}

%The transport properties of the systems change depending on the value of the $\Delta$ parameter. The system displays ballistic $\pare{\Delta=1/2}$, diffusive $\pare{\Delta=1}$, and sub-diffusive $\pare{\Delta=2}$ behaviour. Both invariant manifolds behave in a similar way and the results suggest that they should coincide in the thermodynamic limit.  

Bu\v{c}a and Prosen \cite{buca12a} also study the spectral properties of the Lindbladian for the case $\Delta=2$. Figure \ref{fig:buca_spectrum} shows a complex map of the leading eigenvalues, to be compared with the sketch of Fig. \ref{fig:spectrum} above. In particular, they observe as expected the emergence of purely exponential relaxation modes as well as \emph{spiral} relaxation modes, together with the expected steady state null eigenvalue. Interestingly, they find that in the $\{+1, +1\}$ symmetry sector the spectral gap quickly decays to zero as $L$ increases, while a more complex, seemingly non-monotonic behavior is observed for the spectral gap in the $\{-1,-1\}$ sector.
%the full Liouvillian spectra of the XXZ chain with the jump operators given by Eq. (\ref{eq:lind_buca}). 

\subsection{Spin ladders}
\label{sec:symej2}

Spin lattices are a natural generalization of $1d$ chains, and as such they are often used to study quantum transport in multi-dimensional systems \cite{znidaric13b,znidaric13a,asadian13a,manzano16b}. More specifically, spin lattices can be realized in the laboratory \cite{hess07a,hild14a}, and experiments show how lattice dimension deeply affects transport, both for bosons and fermions \cite{hild14a}.
%that they behave in a different way than bosons and fermions lattices \cite{hild14a}. 

The simplest two-dimensional lattice, a ladder, was studied by \v{Z}nidari\v{c} in \cite{znidaric13b}. In particular, this work studies a nonintegrable spin ladder with XX-type interactions along the ladder legs, and XXZ-type coupling along the rungs. Interestingly, this spin ladder system is shown to exhibit a number of symmetries and invariant subspaces \cite{znidaric13b}, some of them capable of supporting ballistic magnetization transport, while diffusive transport is found in complementary subspaces. This coexistence of ballistic and diffusive transport channels can be rationalized in terms of the symmetries of the spin ladder, as described in previous section, constituting an important example of the effect of symmetry on transport properties.

\begin{figure}
\centerline{\includegraphics[width=14cm]{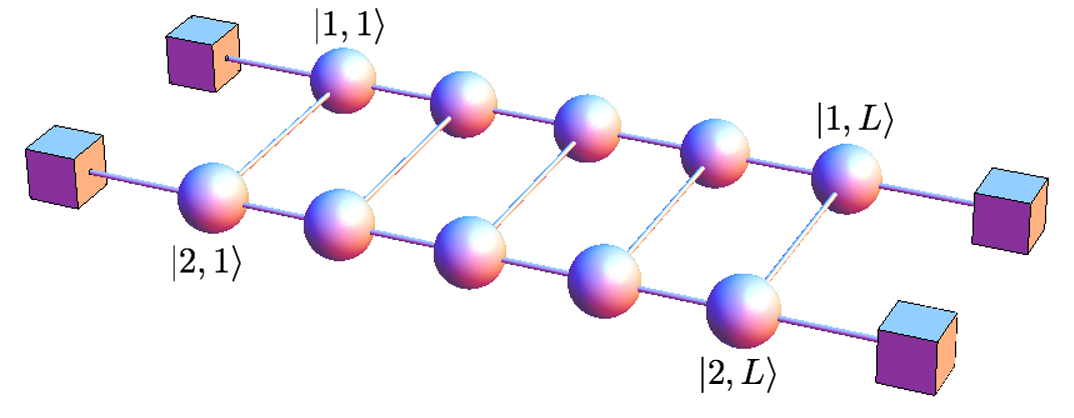}}
\caption{Quantum ladder of length $L$ with driving baths at its boundaries.}
\label{fig:ladder}
\end{figure}

The model studied, represented in Fig.~\ref{fig:ladder}, consists in a two-rungs ladder of length $L$. The total number of spins is $2L$ and we label them as $\pare{i,j}$ with $i=1,2$ being the leg index ($y$-position) and $j=1,\ldots,L$ the rung index ($x$-position). The ladder Hamiltonian corresponds to two spin-$1/2$ chains with XX-type nearest neighbor coupling along the two chains (or legs) and interchain (rung) coupling of the XXZ-type. In particular 
%that of the XYZ model
\be
H^{\text{Ld}}=\sum_{i=1}^2 \sum_{j=1}^{L-1} \pare{ \sigma_{i,j}^x \sigma_{i,j+1}^x + \sigma_{i,j}^y \sigma_{i,j+1}^y} + J~\sum_{j=1}^{L} \pare{ \sigma_{1,j}^x \sigma_{2,j}^x + \sigma_{1,j}^y \sigma_{2,j}^y + \Delta~\sigma_{1,j}^z \sigma_{2,j}^z } \, , 
\ee
%\ben
%H=\sum_{i=1,2} \sum_{j=1}^{L-1}  &\sigma_{i,j}^x \sigma_{i,j+1}^x + \sigma_{i,j}^y \sigma_{i,j+1}^y \nonumber\\
%&+ J\pare{ \sigma_{1,j}^x \sigma_{2,j}^x + \sigma_{1,j}^y \sigma_{2,j}^y + \Delta \sigma_{1,j}^z \sigma_{2,j}^z },
%\een
where $J$ and $\Delta$ are two coupling constants, and $\sigma_{i,j}^{x,y,z}$ represent the different Pauli matrices acting on site $\pare{i,j}$.

Interestingly, this Hamiltonian exhibits several symmetries. As in the spin chain case, the total magnetization along the $z$-axis, defined now as $M=\sum_{i=1,2} \sum_{j=1}^{L} \sigma_{i,j}^z$, is conserved, so the unitary magnetization operator $S_z=e^{i\phi M}$ defines a continuous symmetry ($\phi\in\mathbb{R}$). 
%In the ladder case  $M=\sum_{i=1,2} \sum_{j=1}^{L} \sigma_{i,j}^z$. 
Moreover, due to the ladder topology there are two further symmetries described by an operator $P_1$ exchanging the sites $(i,j)$ and $(i,L+1-j)$ for each chain $i=1,2$, together with an operator $P_2$ that exchanges the two chains. In terms of the computational basis
\ben
P_1=\sum_{i=1}^2 \sum_{\pare{s_1,\dots,s_L}\in\{0,1\}^L} \op{i;s_1,s_2,\dots,s_L}{i;s_L,s_{L-1},\dots,s_1} \nonumber\\
P_2= \sum_{\pare{s_1,\dots,s_L}\in\{0,1\}^L} \op{1;s_1,s_2,\dots,s_L}{2;s_1,s_2,\dots,s_L} 
\een
Furthermore, in the zero magnetisation manifold there is an additional spin-flip symmetry given by the operator $T= \sum_{j=1}^{L} \sigma_{1,j}^x \sigma_{2,j}^x$, and there are also symmetries associated to the $XX$ interaction along the legs. 

\begin{figure}
\centerline{\includegraphics[width=8cm]{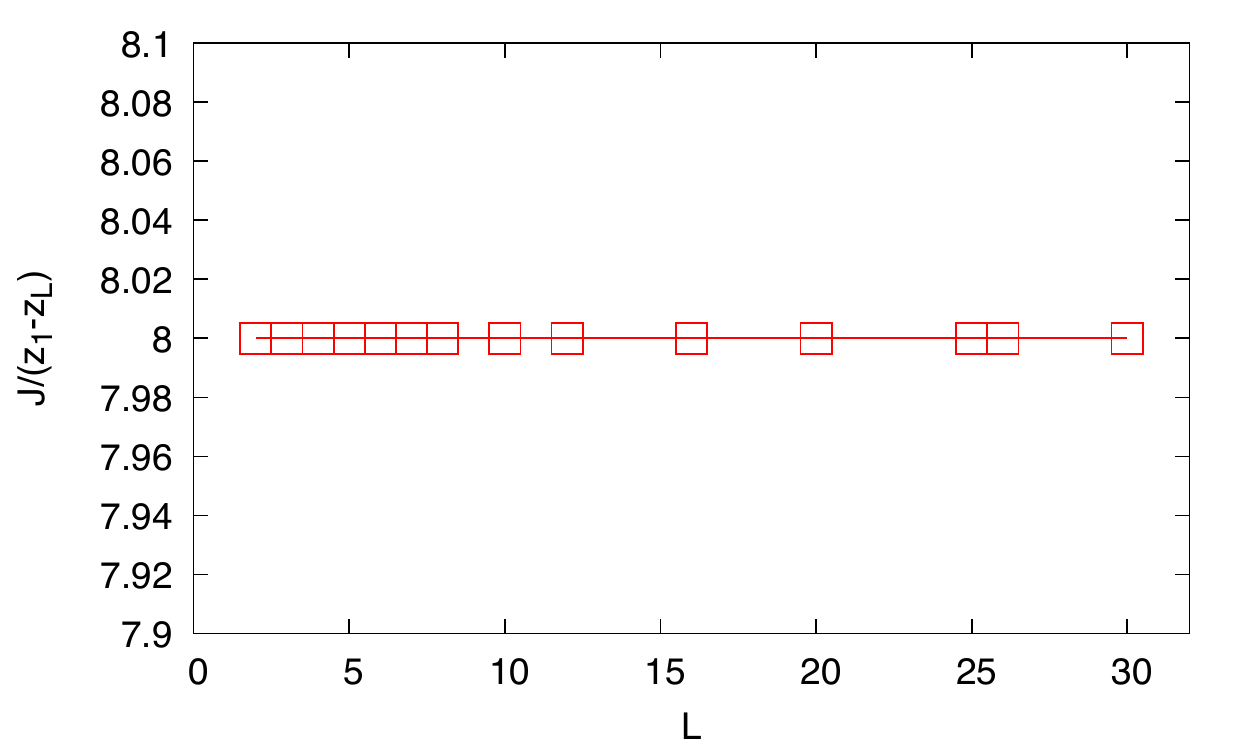}
\includegraphics[width=8cm]{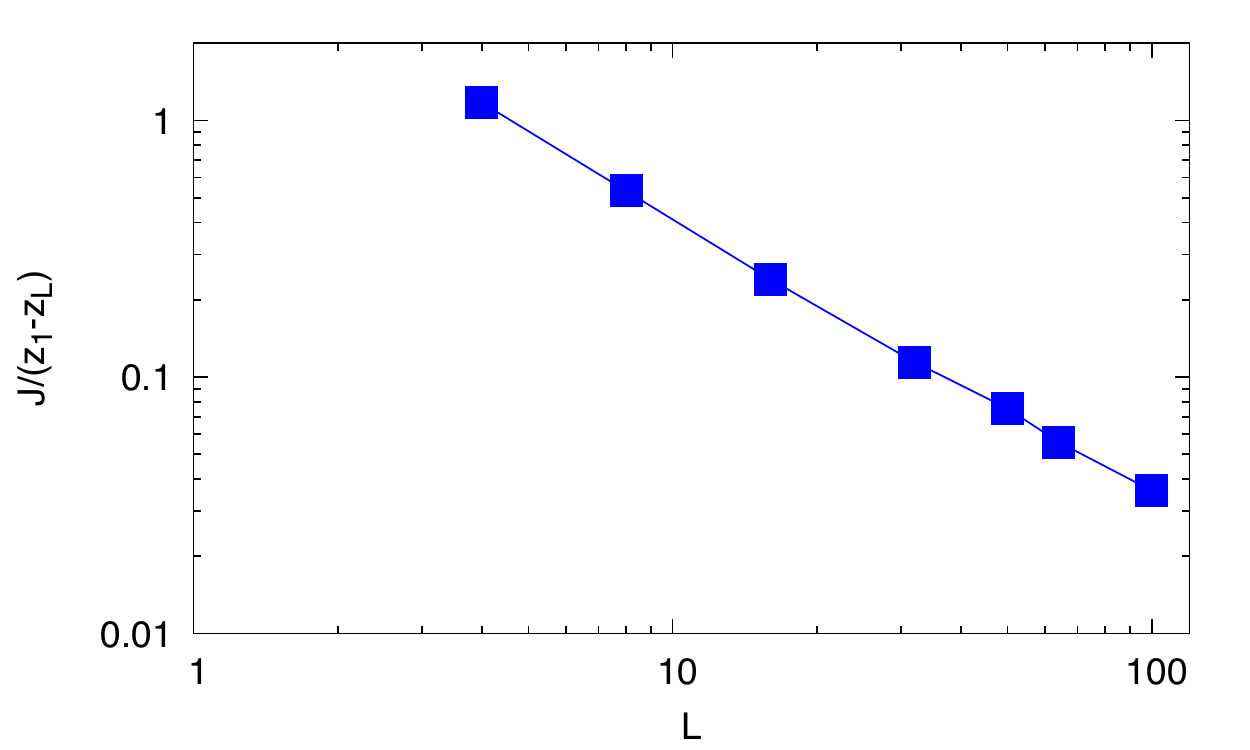}}
\caption{Data from Ref. \cite{znidaric13b}. Scaled current as a function of the spin ladder size for non-local jump operators with $\Delta=0$ (left panel) and local jump operators with $\Delta=0$ and $\gamma=0.2$ (right panel). Flat behaviour indicates ballistic transport. Note that $(z_1-z_L)$ is a measure of the external gradient, with $z_i\equiv\Tr\pare{\rho^{\scriptscriptstyle\text{NESS}}\sigma_i^z}$.}
\label{fig:znidaric_fig1}
\end{figure}

In order to analyse the effect of these symmetries and the resulting invariant subspaces, it is useful to change the notation to the so-called rung eigenbasis \cite{znidaric13b}. On one rung the eigenbasis corresponds to the Bell basis (singlet and triplet states)
\ben
\ket{S} &=& \frac{1}{\sqrt{2}} \pare{ \ket{01} - \ket{10} } \nonumber\\
\ket{T} &=& \frac{1}{\sqrt{2}} \pare{ \ket{01} + \ket{10} } \\
\ket{O} &=&  \ket{00} \nonumber\\
\ket{I} &=& \ket{11}, \nonumber
\een
where the first number (0 or 1) in the ket denotes the state on the upper leg ($i=1$), while the second 0 or 1 corresponds to the lower leg ($i=2$). Using this new basis, it is easy to enumerate some of the simplest invariant subspaces of the spin ladder (see \cite{znidaric13b} for a complete description). Note that, despite possesing a broad set of invariant subspaces, the model of interest is nonintegrable. To asses the transport properties of this model, one may couple the spin ladder with reservoirs so as to study the emerging nonequilibrium steady state, with particular focus on the spin current, a token for transport properties. The coupling to reservoirs is done via Lindblad jump operators that inject/remove excitations to/from the spin ladder. A possibility consists in choosing \emph{non-local} jump operators of the form
%If the baths jump operators are chosen to be non-local in the following form 
\ben
L_1^{\text{nl}}=\op{ITS\ldots ST}{STS\ldots ST} \nonumber\\
L_2^{\text{nl}}=\op{IST\ldots ST}{TST\ldots TS} \\
L_3^{\text{nl}}=\op{STS\ldots ST}{STS\ldots SI} \nonumber\\
L_4^{\text{nl}}=\op{TST\ldots TS}{TS\ldots TI}. \nonumber
\een
Here the idea is to inject/remove one $I$-excitation at the boundaries of the spin ladder, while preserving at the same time the invariant subspace defined by the union of the zero and one $I$-excitation subspaces. Indeed, the jump operators $L_{1,2}^{\text{nl}}$ above inject one $I$-excitation, while $L_{3,4}^{\text{nl}}$ remove one $I$. Not surprisingly, the driven dissipative spin ladder so-defined presents a (strong) symmetry in the sense of \S\ref{sec:sym} under the exchange of the two spin chains that form the ladder. This symmetry hence gives rises to multiple nonequilibrium steady-steates as previously demonstrated, each one with different transport properties. In particular, by choosing the initial state of the spin ladder to be of the form $\ket{STST\ldots}$, one can show that transport exhibits ballistic behaviour \cite{znidaric13b}. Indeed, by numerically diagonalising the ladder Lindbladian up to $L=30$ sites and measuring the spin current $J$, it is found that the current is independent of the system size, as expected for ballistic behaviour, see left panel in Fig.~\ref{fig:znidaric_fig1}.
%given by operator (\ref{eq:spin_chain}). In Figure \ref{fig:znidaric_fig1} it is shown the reescalated spin currents $J/(z_1-z_L)$, with $z_i=\Tr\pare{\rho_{ss}\sigma_i^z}$. The current is independent of the system size, indicating ballistic behaviour. 

%\begin{figure}
%\centerline{\includegraphics[width=12cm]{plots/znidaric_fig1.eps}}
%\caption{ Figure from Ref. \cite{znidaric13b}.  Scaled current as a function of the system size for the non-local jump operators $(\Delta=0)$. The flat behaviour indicates ballistic transport. }
%\label{fig:znidaric_fig1}
%\end{figure}

Alternatively, one may define a \emph{local} coupling to external reservoirs to study the system transport properties. We now choose a local coupling based on 8 jump operators of the form
%If the coupling to the baths is local, in the form 
\ben
L_{1,2}^{\text{lc}} = \sqrt{1\pm\gamma_{1,1} } \; \sigma_{1,1}^\pm \nonumber\\
L_{3,4}^{\text{lc}} = \sqrt{1\pm\gamma_{1,L} } \; \sigma_{1,L}^\pm \\
L_{5,6}^{\text{lc}} = \sqrt{1\pm\gamma_{2,1} } \; \sigma_{2,1}^\pm \nonumber\\
L_{7,8}^{\text{lc}} = \sqrt{1\pm\gamma_{2,L} } \; \sigma_{2,L}^\pm , \nonumber
\een
where the driving parameters $\gamma_{i,j}$ induce a non-zero magnetization at the given ladder boundary site. In  particular, we consider below a symmetric driving around the zero-magnetization manifold, with $-\gamma_{1,1}=-\gamma_{2,1}=\gamma_{1,L}=\gamma_{2,L}\equiv \gamma$ and $\gamma=0.2$ (see Ref. \cite{znidaric13b} for other cases). The boundary driving so-defined breaks the chain-exchange symmetry \cite{znidaric13b}, and hence does not preserve any of the ballistic invariant subspaces, leading to a unique nonequilibrium steady state.
%it can be shown \cite{znidaric13b} that the previous (strong) symmetry is broken, hence leading to a unique nonequilibrium steady state. 
Furthermore, the transport properties of the system in the new steady state change appreciably, and the system becomes diffusive. This can be demonstrated by numerically diagonalising the resulting Lindbladian up to $L=100$. The fitting of the data shows a diffusive $\pare{\sim 1/L}$ scaling of the current with the system size, see right panel in Fig.~\ref{fig:znidaric_fig1}. We note here that more complicated ladders have been studied in Ref. \cite{znidaric13a}, and a general theory explaining the different transport properties of these subspaces and extending these results to general multidimensional lattices can be found at \cite{manzano16b}.

%\begin{figure}
%\centerline{\includegraphics[width=12cm]{plots/znidaric_fig2.eps}}
%\caption{ Figure from Ref. \cite{znidaric13b}.  Scaled current as a function of the system size for the local jump operators $(\Delta=0, \mu=0.2)$. The flat behaviour indicates ballistic transport. }
%\label{fig:znidaric_fig1}
%\end{figure}

We have described in some detail two different examples of open quantum systems exhibiting symmetries. To better understand how symmetries affect the dynamics of open quantum systems and their transport properties, we now present a complementary approach based on full counting statistics (or large deviation theory), which simplifies the analysis in most cases and offers valuable insights on the role of symmetry in transport.

\newpage

\section{Full counting statistics of currents for quantum master equations}
\label{sec:LDF}

\subsection{Current-resolved master equation}
\label{sec:LDF1}

\begin{figure}
\centerline{\includegraphics[scale=0.25]{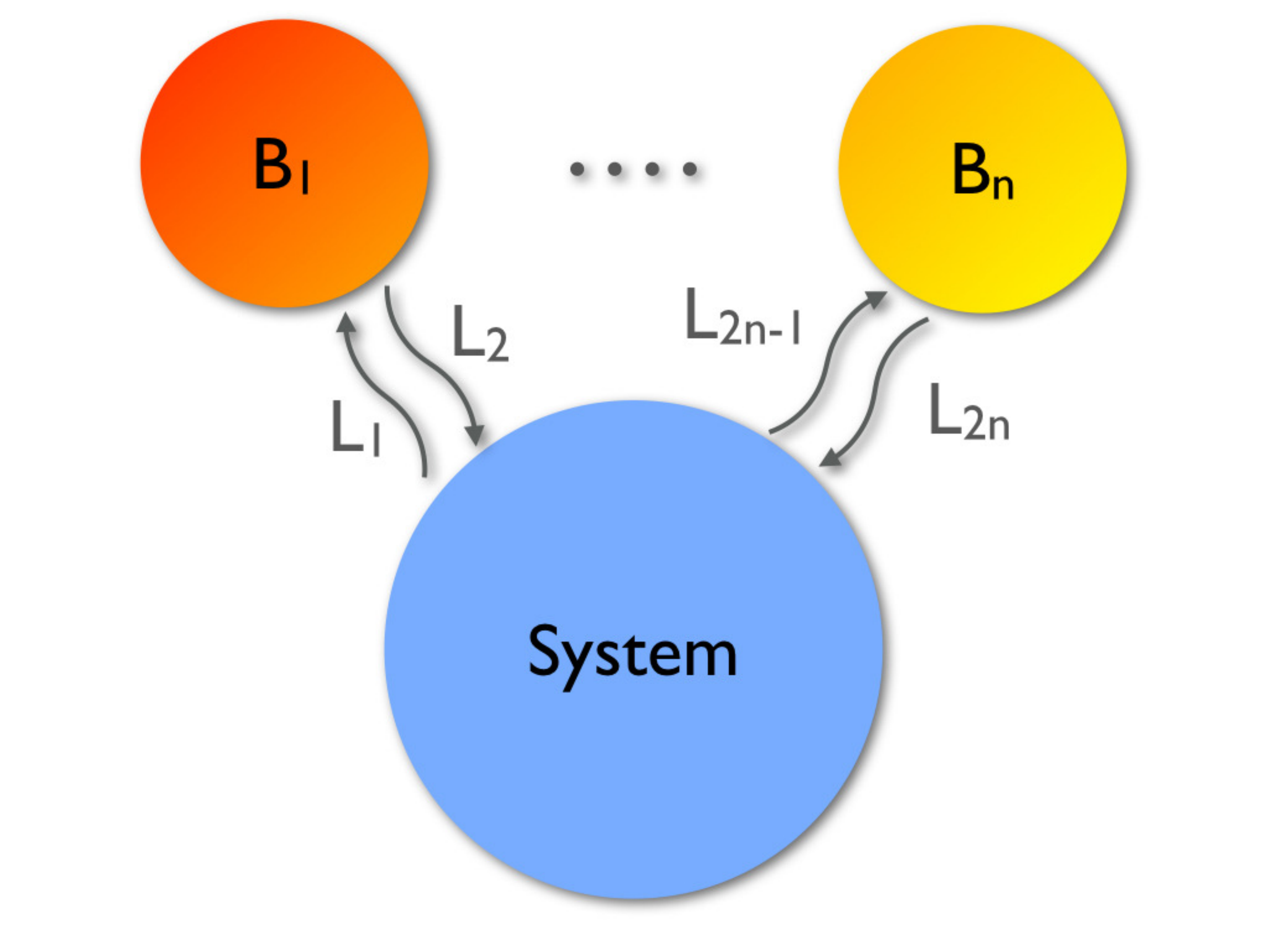}}
\caption{Sketch of the interaction between an open system and $n$ different baths.}
\label{fig:interaction}
\end{figure}

We have already seen how symmetries lead to multiple invariant subspaces and degenerate steady states in general open quantum systems governed by a master equation of the form $\dot{\rho}(t)=\cL \rho(t)$, with $\cL$ a Liouville-like evolution superoperator, see e.g. Eq. (\ref{eq:red_vn10}). Master equations like this can be formally solved to yield
\be
\rho(t)=\textrm{e}^{t\cL} \rho(0) \equiv \cW(t) \rho(0) \, ,
\label{eq:sol}
\ee
which defines the full propagator $\cW(t)\in\cBBH$. Our next aim is to understand how symmetry affects the \emph{thermodynamics of currents} in open systems. Currents generically appear in open quantum systems in response to any driving mechanism pushing the system out of equilibrium, as e.g. an external gradient due to contact with several reservoirs at different temperature and/or chemical potentials. These currents thus play a key role as tokens of nonequilibrium physics, and the distribution of current fluctuations has recently emerged as a central object of investigation, with the associated current \emph{large deviation function} (LDF) \cite{touchette09a} acting as a marginal of the nonequilibrium analog of thermodynamic potential.

To investigate the thermodynamics of currents, we first need a framework capable of dealing with arbitrary current fluctuations. This theory is based on the \emph{current-resolved} quantum master equation obtained from the \emph{unraveling} of the Liouvillian superoperator $\cL$ \cite{derezinski08a} in Eq. (\ref{eq:sol}), an approach related to the input-output formalism \cite{gardiner00a} and connected to matrix product states \cite{osborne10a,lesanovsky13a}.
%. This approach has links with the input-output formalism \cite{gardiner00a} and matrix product states \cite{osborne10a,lesanovsky13a}. 
In particular, we focus now on a $D$-dimensional system connected to $n$ different baths (possibly at varying temperatures and/or chemical potentials, thus leading to net currents), see Fig. \ref{fig:interaction}. 
%Our approach, based on a weak coupling between the system and the baths, is related with the input-output formalism \cite{gardiner00a} and continuous matrix product states of quantum fields \cite{osborne10a,lesanovsky13a}. 
Each bath interacts with the system via different incoherent Lindblad channels which induce quantum jumps associated to the exchange of quanta of different nature (like e.g. photon or exciton emission and absorption) \cite{plenio98a,gardiner00a}. We are interested in analyzing the statistics of the net current of quanta between the system and one of these baths. We can always split the Liouvillian $\cL$ into three well-defined superoperators with respect to their action regarding the selected incoherent channel, namely
\be
\cL=\cL_0+\cL_{+1} + \cL_{-1} \, ,
\label{eq:unrav}
\ee
where the subscripts $0,\pm 1$ refer to the change of quanta in the system induced by the corresponding superoperator through the selected channel. We now define a \emph{trajectory} $\chi$ of duration $t$ as the set of pairs $(t_i,s_i)$, with $i\in [1,m]$ and $0\le t_1\le t_2\le\ldots\le t_m\le t$, which label the times $t_i$ at which a  quantum jump of magnitude $s_i=\pm 1$ happens with the designated reservoir, out of a total of $m$ quantum jumps, i.e.
\be
\chi=\{t_1,s_1; t_2,s_2; \ldots ; t_m,s_m \} \, , \qquad 0\le t_1\le t_2\le\ldots\le t_m\le t \, .
\label{eq:traj}
\ee
Associated to a trajectory, we now introduce a completely positive superoperator $\cW_{\chi}(t)$ defined as
\be
\cW_{\chi}(t) \equiv \Pi_0(t-t_m) \cL_{s_m} \Pi_0(t_m-t_{m-1}) \cdots \cL_{s_2} \Pi_0(t_2-t_{1}) \cL_{s_1} \Pi_0(t_1) \, , 
\label{eq:Wtraj}
\ee
where we have defined the \emph{current-free} propagator $\Pi_0(t)\equiv\exp(t\cL_0)$. The full propagator of the quantum master equation for the system can be now written as
\be
\cW(t)=\textrm{e}^{t\cL} = \int {\cal D} \chi \cW_{\chi}(t) \, ,
\label{eq:prop}
\ee
where the integral over trajectories represents the following sum
\be
\int {\cal D} \chi \equiv \sum_{m=0}^\infty \sum_{s_1\ldots s_m=\pm 1} \int_0^t dt_m \int_0^{t_m}dt_{m-1} \cdots \int_0^{t_2}dt_1 \, .
\label{eq:inttraj}
\ee
Clearly, the superoperator $\cW_\chi(t)$ describes the (unnormalized) evolution of our open quantum system conditioned on a particular trajectory $\chi$. Indeed, using Eqs. (\ref{eq:sol}) and (\ref{eq:prop}),
\be
\rho(t)=\cW(t)\rho(0) = \int {\cal D} \chi \cW_\chi(t) \rho(0) \, ,
\label{eq:rhodesc}
\ee
and this allows us to define $\rho_\chi(t)\equiv \cW_\chi(t) \rho(0)$, the system density matrix at time $t$ conditioned on a particular trajectory $\chi$. The probability of such a quantum trajectory is then given by $\Tr\rho_\chi(t)$, and we can now use this picture to investigate the current statistics through the selected reservoir. For that we first define the current or net flow of quanta associated to a trajectory $\chi$ as 
\be
%Q_\chi\equiv \sum_{k=1}^m (\delta_{s_k,+1} - \delta_{s_k,-1}) \, ,
Q_\chi\equiv \sum_{k=1}^m s_k ,
\label{eq:curr}
\ee
and define $X_Q(t) \equiv \{\chi : Q_\chi=Q \}$ as the set of all trajectories of duration $t$ with a fixed extensive current $Q$. The current-resolved density matrix at time $t$ can be now defined as
\be
\rho_Q(t)\equiv \sum_{\chi\in X_Q} \rho_\chi(t) = \int {\cal D} \chi \, \delta_{Q, Q_\chi}\,  \cW_\chi(t) \rho(0) \, ,
\label{eq:rhocurr}
\ee
with the integral over trajectories defined as in (\ref{eq:inttraj}), and with $\delta_{Q, Q_\chi}$ the Kronecker delta-function. The probability of observing an arbitrary current fluctuation $Q$ during a time $t$ through the selected reservoir is then given as $\PP_t(Q)=\Tr \left(\rho_Q(t)\right)$. From the Dyson-type expansion of the trajectory superoperator $\cW_\chi$, it is then easy to see that $\rho_Q(t)$ obeys a current-resolved master equation of the form
\be
\dot{\rho}_Q(t) = \cL_0 \rho_Q(t) + \cL_{+1}\rho_{Q-1}(t) + \cL_{-1} \rho_{Q+1}(t) \, ,
\label{eq:qmastereq}
\ee
which defines a hierarchy of coupled equations for the current-resolved density matrix. This hierarchy of equations is more easily solved by Laplace-transforming $\rho_Q$, or equivalently by working with the cumulant generating function of the current distribution. For that, we now define
\be
\rho_\lambda(t)\equiv \sum_{Q=-\infty}^{+\infty} \rho_Q(t) \textrm{e}^{-\lambda Q} \, ,
\label{eq:rhol}
\ee
where the parameter $\lambda$ is known as \emph{counting field} conjugated to the current \cite{esposito09a,cao09a,hurtado09c,hurtado10a,hurtado14a}. This unnormalized density matrix evolves according to
\be
\dot{\rho}_\lambda(t) = \cL_0\rho_\lambda(t) + \textrm{e}^{-\lambda} \cL_{+1} \rho_\lambda(t) + \textrm{e}^{+\lambda} \cL_{-1} \rho_\lambda(t) \equiv \cLl \rho_\lambda(t)\, ,
\label{eq:masterrhol}
\ee 
which is a \emph{closed} evolution equation for $\rho_\lambda(t)$ which defines the \emph{deformed} or \emph{tilted} superoperator $\cLl=\cL_0 + \textrm{e}^{-\lambda} \cL_{+1} + \textrm{e}^{+\lambda} \cL_{-1}$ whose spectral properties control the thermodynamics of currents in the system \cite{manzano14a}. In particular, note that if $\PP_t(Q)=\Tr\rho_Q(t)$ is the probability of observing a current fluctuation $Q$ after a time $t$, then the trace 
\be
Z_{\lambda}(t)\equiv \Tr(\rho_{\lambda}(t)) = \sum_{Q=-\infty}^{+\infty} \PP_t(Q) \textrm{e}^{-\lambda Q} \, ,
\label{eq:Zgen}
\ee
is nothing but the moment generating function of the current probability distribution.

\subsection{Large deviations statistics}
\label{sec:LDF2}

For long times, this probability measure obeys a \emph{large deviation principle} (LDP) of the form \cite{touchette09a}
\be
\PP_t(Q)\asymp \exp[+tG(Q/t)],
\label{eq:LDP}
\ee
where the symbol "$\asymp$" means asymptotic logarithmic equality, i.e.
\be
\lim_{t\to\infty}\frac{1}{t}\ln \PP_t(Q) = G(q) \, , \qquad q=\frac{Q}{t} \, ,
\ee
and $G(q)\le 0$ defines the \emph{current large deviation function} (LDF). This key function measures the exponential rate at which the distribution of the time-averaged current peaks around its ensemble average value $\la q \ra$. As a consequence, $G(\la q \ra)=0$. The emergence of a LDP in the long time limit relies in several assumptions, including a non-zero spectral gap and finite correlations times (for a rigorous mathematical derivation see Ref. \cite{touchette09a}, Appendix B). Large deviation functions as the one described here for the current play a fundamental role in nonequilibrium physics, as they generalize the concept of thermodynamic potentials to the realm of nonequilibrium phenomena, where no bottom-up approach exists yet connecting microscopic dynamics with macroscopic properties. Moreover, the LDFs controlling the statistics of macroscopic fluctuations in many classical and quantum systems have been shown to exhibit non-analyticities reminiscent of standard critical behavior, accompanied by emergent order and symmetry-breaking phenomena in the optimal trajectories responsible for a given fluctuation \cite{bodineau05a,bertini05a,bertini06a,garrahan07a,garrahan09a,garrahan10a,hurtado11a,pitard11a,perez-espigares13a,lesanovsky13a,hurtado14a,zannetti14b,lazarescu15a,tizon-escamilla16a,baek17a,abou17a}. In addition, the emergence of coherent structures associated to rare fluctuations implies in turn that these extreme events are far more probable than previously anticipated, a finding of broad implications. These arguments make the investigation of current statistics in open quantum systems a key issue. 

The current moment generating function $Z_\lambda(t)$ also exhibits large deviation scaling for long enough times \cite{touchette09a}, 
\be
Z_{\lambda}(t)\asymp \exp[+t\mu(\lambda)] \, ,
\label{eq:LDPmu}
\ee
where $\mu(\lambda)$ defines a new large deviation function related to the current LDF $G(q)$ via Legendre transform
\be
\mu(\lambda)=\max_q \left[G(q) - \lambda q \right] = G(q_\lambda) - \lambda q_\lambda \, ,
\label{eq:mu}
\ee
with $q_\lambda$ the current associated to a given $\lambda$. The function $\mu(\lambda)$ can be seen as the \emph{conjugate potential} to $G(q)$, a relation equivalent to the free energy being the Legendre transform of the internal energy in thermodynamics. The scaling (\ref{eq:LDPmu}) and the Legendre transform (\ref{eq:mu}) can be easily derived by combining Eqs. (\ref{eq:Zgen})-(\ref{eq:LDP}) above. The new LDF $\mu(\lambda)$ corresponds to the cumulant generating function of the current distribution. Indeed,
\be
\mean{q^k}_c =  \frac{\partial^k \mu(\lambda)}{\partial \lambda^k}\Big\vert_{\lambda=0} = \lim_{t\rightarrow \infty} \frac{1}{t}  \left. \frac{\partial^k \ln Z_\lambda (t)}{\partial \lambda^k}\right|_{\lambda=0} \, ,
\label{eq:cumul}
\ee
with $\mean{q^k}_c$ the $k^\textrm{th}$-order current cumulant \cite{touchette09a}, which correspond to the central moments of the current distribution up to $k=3$.

\newpage

\section{Symmetry and thermodynamics of currents}
\label{sec:LDF2.5}

In the previous section we have derived with some detail the current-resolved master equation for $\rho_Q$ and its (Laplace) dual for $\rho_\lambda$. 
%in terms of the counting field $\lambda$. 
This approach has allowed us to formulate with precision the problem of current statistics in open quantum systems evolving in time according to a general quantum master equation, with the only premise that it can be unraveled in terms of the emission and/or absorption of quanta through a selected incoherent channel.

We now focus on understanding the effect of symmetries on the statistics of the current to a particular reservoir for the case of generic quantum master equations in the Lindblad form 
\be
\dot{\rho}(t) =  - \ii [H,\rho] +  \sum_{i} \left(L_i \rho L^\dagger_i -  \frac{1}{2} \acor{L^\dagger_i L_i,\rho}\right) \, , 
\label{eq:lindblad}
\ee
though our results can be easily extended to more general settings. For this family of systems, the $\lambda$-resolved master equation for the unnormalized density matrix $\rho_\lambda(t)$ reads
\be
\dot{\rho}_{\lambda}(t) =  - \ii [H,\rho_{\lambda}] +  \text{e}^{-\lambda} L_1 \rho_{\lambda} {L_1}^\dagger  +  \text{e}^{+\lambda} L_2 \rho_{\lambda} {L_2}^\dagger + \sum_{j\neq 1,2} L_j \rho_{\lambda} L^\dagger_j -  \frac{1}{2}\sum_{j} \{L^\dagger_j L_j,\rho_{\lambda}\} \equiv \cLl \rho_\lambda(t)\, .
\label{eq:rhol}
\ee
%where we are interested in the injection and extraction of quanta by two channels. 
Without loss of generality we assume that $L_1$ and $L_2$ are respectively the Lindblad jump operators responsible of the injection and extraction of quanta through the reservoir of interest\footnote[1]{This formalism can be easily generalized to study the statistics of currents to $K$ different reservoirs. In the general case the counting field would be a vector $\vlamb=(\lambda_1,\ldots,\lambda_K)$ of dimension $K$, and the deformed superoperator $\cL_{\vlamb}$ should include a term $e^{\lambda_k}$ for the injecting $k$-channel and a term $e^{-\lambda_k}$ for the extracting $k$-channel, with $k\in[1,K]$ \cite{manzano14a}.}. As described above, this evolution defines a \emph{deformed} (or \emph{tilted}) superoperator $\cLl$ which no longer preserves the trace, $\dot{\rho}_\lambda(t)=\cLl \rho_\lambda(t)$. For completeness, the identification with the general unraveling  (\ref{eq:masterrhol}) of the master equation corresponds to 
\ben
\cL_{+1} \bullet &=& L_1\bullet L_1^\dagger \, ,  \nonumber \\
\cL_{-1} \bullet &=& L_2\bullet L_2^\dagger \, ,  \\
\cL_{0} \bullet &=& - \ii [H,\bullet] +  \sum_{i\neq 1,2} L_i \bullet L^\dagger_i -  \frac{1}{2}\sum_{i} \acor{L^\dagger_i L_i,\bullet} \, . \nonumber 
\een
In close analogy with our discussion in \S\ref{sec:sym2}, the existence of a symmetry $U$ obeying the commutation relations (\ref{eq:sym}) implies that the adjoint right and left symmetry superoperators $\cU_{l,r}$, defined in Eq. (\ref{eq:Ulr}), and the tilted superoperator $\cLl$ all commute
\be
\cor{\cU_l,\cLl} = 0 = \cor{\cU_r,\cLl} \, ,
\label{eq:symlambda}
\ee
so there exists a complete biorthogonal basis of common right ($\oa$) and left ($\loa$) eigenfunctions in $\cBH$ for these three superoperators, linking eigenvalues of $\cLl$ to particular symmetry eigenspaces. In particular, 
\ben 
\cU_l\oa&=&\text{e}^{\ii\Omega_{\alpha}}\oa \, , \qquad \;\;\, \loa\cU_l = \text{e}^{\ii\Omega_{\alpha}}\loa \, , \label{eq:oa1}\nonumber\\
\cU_r\oa&=&\text{e}^{-i\Omega_{\beta}}\oa\, , \qquad \, \loa\cU_r = \text{e}^{-\ii\Omega_{\beta}}\loa \, , \label{eq:oa2}\\
\cLl\oa&=&\mu_{\nu}(\lambda)\oa \, , \qquad \loa\cLl = \mu_{\nu}(\lambda)\loa \, .\nonumber \label{eq:oa3}
\een
Note that, due to orthogonality of symmetry eigenspaces, $\Tr(\oa)\propto \delta_{\alpha\beta}$, and we introduce the normalization $\Tr (\omega_{\alpha\alpha\nu}(\lambda)) =1$ for simplicity. The solution to Eq. (\ref{eq:rhol}) can be formally written as $\rho_{\lambda}(t)=\exp(+t \cLl)\rho(0)$, so a spectral decomposition of the initial density matrix $\rho(0)$ in terms of the common biorthogonal basis, $\rho(0)=\sum_{\alpha,\beta,\nu} \bbraket{\loa}{\rho0} \, \oa$,
allows us to write 
\be
Z_{\lambda}(t)=\sum_{\alpha\nu}\text{e}^{+t\mu_\nu(\lambda)} \bbraket{\hat{\omega}_{\alpha\alpha\nu}(\lambda)}{\rho(0)}\, . 
\label{eq:Zlambsp}
\ee
For long times
\be
Z_{\lambda}(t) \xrightarrow{t\to\infty} \text{e}^{+t\mu_0^{(\alpha_0)}(\lambda)} \bbraket{\hat{\omega}_{\alpha_0\alpha_0 0}(\lambda)}{\rho(0)} \, ,
\label{eq:zsum1}
\ee
where $\mu_0^{(\alpha_0)}(\lambda)$ is the eigenvalue of $\cLl$ with largest real part and symmetry {index} $\alpha_0$ among all symmetry diagonal eigenspaces $\cBaa$ with nonzero projection on the initial $\rho(0)$. In this way, comparing this expression with Eq. (\ref{eq:LDPmu}), we realize that this eigenvalue is nothing but the Legendre transform of the current LDF, 
\be
\mu(\lambda)\equiv\mu_0^{(\alpha_0)}(\lambda) \, . 
\label{eq:mul}
\ee
Note that the projection $\bbraket{\hat{\omega}_{\alpha_0\alpha_0 0}(\lambda)}{\rho(0)}$ in Eq. (\ref{eq:zsum1}) above just amounts to a subleading ${\cal O}(t^{-1})$ correction to the LDF $\mu(\lambda)$ which disappears in  the $t\to\infty$ limit.

Some comments are now in order. Interestingly, the long time limit in Eq. (\ref{eq:zsum1}) \emph{selects} a particular symmetry sector $\alpha_0$ among all symmetry subspaces present in the initial state $\rho(0)$, effectively \emph{breaking at the fluctuating level} the original symmetry of our open quantum system. Note that we assume here the symmetry subspace $\alpha_0$ to be unique in order not to clutter our notation; this is however unimportant for our conclusions below. The resulting picture is that, if starting from a state $\rho(0)$ we happen to observe a current fluctuation of magnitude $q$, the transport channel (or symmetry sector) overwhelmingly responsible of this current fluctuation will be $\alpha_0(\lambda)$, with $\lambda= \lambda(q)$ the conjugate counting field to the observed current.
%$q=q_\lambda$ (i.e. the current conjugated to the counting field $\lambda$), the transport channel (or symmetry sector) overwhelmingly responsible of this current fluctuation will be $\alpha_0$. 
As we show next, distinct symmetry eigenspaces may dominate different fluctuation regimes, separated by first-order-type dynamic phase transitions. Note that different types of spontaneous symmetry breaking scenarios at the fluctuating level have been recently reported in classical diffusive systems \cite{bertini01a,bertini02a,bertini05a,bertini06a,hurtado14a,hurtado11b,bodineau05a,perez-espigares13a,tizon-escamilla16a,baek17a}.

\subsection{Effects of symmetry on the average current}
\label{sec:LDF3}

The previous discussion already hints at how to control both the statistics of the current and the average transport properties of an arbitrary open quantum system. Indeed, this can be accomplished by playing with the symmetry decomposition of the initial state $\rho(0)$, which in turn controls the amplitude of the scaling in Eq. (\ref{eq:zsum1}) (see Ref. \cite{kundu11a} for a discussion of this amplitude in a classical context). The previous idea is most evident by studying the average current, defined as 
\be
\la q \ra = \lim_{t\to\infty} \frac{1}{t} \partial_{\lambda} \ln Z_{\lambda}(t)\vert_{\lambda=0} \, ,
\label{eq:curr}
\ee
see also the generic cumulant expressions (\ref{eq:cumul}) above. The $\lambda$-derivative in the previous equation can be made explicit now by recalling that $Z_\lambda(t)=\Tr \rho_{\lambda}(t)$ and noting that $\rho_\lambda(t)=\exp(+t\cLl)\rho(0)$, leading to
\be
\la q \ra = \lim_{t \to \infty} \left. \frac{\Tr\big((\partial_\lambda \cLl) \rho_\lambda (t)\big)}{\Tr \rho_\lambda (t)} \right\vert_{\lambda=0} \, , %\nonumber
\label{eq:qq1}
\ee
where the new superoperator $\partial_\lambda \cLl$ is defined via
\be
(\partial_\lambda \cLl) \sigma = \text{e}^{+\lambda} L_2 \sigma {L_2}^\dagger - \text{e}^{-\lambda} L_1 \sigma {L_1}^\dagger  \, , \quad \forall \sigma \in \cBH \, , \nonumber
\ee
as derived from the definition of $\cLl$ in Eq. (\ref{eq:rhol}) above. If we now restrict the initial density matrix to a particular (diagonal) symmetry subspace, $\rho(0)\in \cBaa$, we have that $\lim_{t\to \infty} \rho_\lambda(t)\vert_{\lambda=0}=\rho_{\alpha}^{\scriptscriptstyle\text{NESS}}$, which is normalized, $\Tr(\rho_{\alpha}^{\scriptscriptstyle\text{NESS}})=1$, and therefore
\be
\la q_{\alpha}\ra = -\partial_{\lambda} \mu_0^{(\alpha)}(\lambda)\vert_{\lambda=0} = \Tr\big(L_2\rho_{\alpha}^{\scriptscriptstyle\text{NESS}}L_2^\dagger\big)-\Tr\big(L_1\rho_{\alpha}^{\scriptscriptstyle\text{NESS}}L_1^\dagger\big) \, . 
\label{eq:qq2}
\ee
On the other hand, for a general $\rho(0)\in \cBH$ we may use in Eq. (\ref{eq:qq1}) the following spectral decomposition 
\be
\rho_\lambda(t)=\sum_{\alpha\beta\nu} \text{e}^{+t\mu_\nu(\lambda)} \bbraket{\hat{\omega}_{\alpha\beta\nu}(\lambda)}{\rho(0)} \, \omega_{\alpha\beta\nu}(\lambda) \, . 
\ee
Moreover, similarly to $\cLl$, the new superoperator $\partial_\lambda \cLl$ also leaves invariant the symmetry subspaces because it commutes with both $\cU_{l,r}$, i.e. $(\partial_\lambda \cLl) \cBa \subset \cBa$, so $\Tr((\partial_\lambda \cLl)\sigma) =0$ for any $\sigma\in\cBa$ with $\alpha\ne\beta$, and hence, using the previous spectral decomposition,
\be
\Tr\big((\partial_\lambda \cLl) \rho_\lambda (t)\big)=\sum_{\alpha\,\nu} \text{e}^{+t\mu_\nu(\lambda)} \bbraket{\hat{\omega}_{\alpha\alpha\nu}(\lambda)}{\rho(0)} \Tr\big((\partial_\lambda \cLl)\omega_{\alpha\alpha\nu}(\lambda)\big) \, . 
\ee
Using this expression in Eq. (\ref{eq:qq1}) above and noting that for $\lambda=0$ the largest eigenvalue of $\cLl$ within each symmetry eigenspace $\cBaa$ is necessarily 0, with associated normalized right eigenfunction $\omega_{\alpha\alpha 0}(\lambda=0)=\rho_{\alpha}^{\scriptscriptstyle\text{NESS}}$ and dual $\hat{\rho}_{\alpha}^{\scriptscriptstyle\text{NESS}}$, we hence obtain 
\be
\la q \ra = \frac{\sum_{\alpha} \la q_{\alpha}\ra \bbraket{\hat{\rho}_{\alpha}^{\scriptscriptstyle\text{NESS}}}{\rho(0)}}{\sum_{\alpha} \bbraket{\hat{\rho}_{\alpha}^{\scriptscriptstyle\text{NESS}}}{\rho(0)}} \, ,
\label{eq:qmed}
\ee
with $\la q_{\alpha}\ra$ the average current of the NESS $\rho_{\alpha}^{\scriptscriptstyle\te{NESS}}\in\cBaa$, see Eq. (\ref{eq:qq2}). This is nothing but a weighted average of the currents of the different NESSs or transport channels, with weights proportional to the projection of the initial density matrix on each symmetry sector. In this way,  Eq. (\ref{eq:qmed}) opens the door to the symmetry-based controllability of quantum currents in general open quantum systems. Indeed, nonequilibrium steady states $\rho_{\alpha}^{\scriptscriptstyle\text{NESS}}$ with different $\alpha$ will typically have different average currents $\la q_{\alpha}\ra$, so the manipulation of the projections $\bbraket{\hat{\rho}_{\alpha}^{\scriptscriptstyle\text{NESS}}}{\rho(0)}$ by adequately preparing the symmetry of the initial state will lead to symmetry-controlled transport properties. We will show below several examples of this control mechanism.

\subsection{Symmetry-induced dynamic phase transitions}
\label{sec:LDF4}

We next demonstrate that, remarkably, the existence of a \emph{symmetry under nonequilibrium conditions} also implies non-analyticities in the LDF $\mu(\lambda)$ which can be interpreted as dynamical phase transitions, or phase transitions at the trajectory level, separating regimes where the original symmetry is spontaneously broken in different ways. Interestingly, this dynamical symmetry-breaking scenario is exclusive of nonequilibrium physics, disappearing in equilibrium. 

To show this explicitly, we first note that for $|\lambda|\ll 1$ the leading eigenvalue of $\cLl$ with symmetry index $\alpha$ can be expanded as 
\be
\mu_0^{(\alpha)}(\lambda)\approx \mu_0^{(\alpha)}(0) + \lambda \partial_{\lambda} \mu_0^{(\alpha)}(\lambda)\vert_{\lambda=0}=-\lambda \la q_{\alpha}\ra \, , 
\label{eq:muexpand}
\ee
where we have used in the second equality the definition (\ref{eq:qq2}) of the average current for NESS $\rho_{\alpha}^{\scriptscriptstyle\te{NESS}}$. This shows that, as expected, $\mu_0^{(\alpha)}(\lambda)$ hits the origin for $\lambda=0$, with a local slope corresponding to $\la q_{\alpha}\ra$. Now, the Legendre transform of the current LDF is given by
\be
\mu(\lambda)=\max_{\alpha}[\mu_0^{(\alpha)}(\lambda)] \, , 
\label{eq:mumax}
\ee
the maximum taken over the symmetry eigenspaces with nonzero overlap with $\rho(0)$. Therefore, depending on the sign of $\lambda$, see Eq. (\ref{eq:muexpand}), $\mu(\lambda)$ will correspond to different symmetry sectors $\alpha$ as dictated by their average current $\la q_{\alpha}\ra$. In particular, we find that
\be
\mu(\lambda) \underset{|\lambda|\ll 1}{=} \left\{
\begin{array}{l l}
  +|\lambda| \la q_{\alpha_{\text{max}}} \ra  & \quad \text{for} \, \, \lambda \lesssim 0\\
  \phantom{aaa} \\
  -|\lambda| \la q_{\alpha_{\text{min}}} \ra & \quad \text{for} \, \, \lambda \gtrsim 0
  \end{array} \right. \, ,
  \label{eq:kink}
\ee
where $\alpha_{\text{max}}$ ($\alpha_{\text{min}}$) denotes the symmetry eigenspace with maximal (minimal) average current $\la q_{\alpha_{\text{max}}} \ra$ ($\la q_{\alpha_{\text{min}}} \ra$) among those with nonzero overlap with $\rho(0)$. The previous argument proves that the LDF $\mu(\lambda)$ must exhibit a kink (i.e. a discontinuity in its first derivative) at $\lambda=0$ whenever a symmetry $U$ exists. This kink will be characterized by a finite jump in the \emph{dynamic order parameter} $q(\lambda)\equiv -\mu'(\lambda)$ at $\lambda=0$ of magnitude $\Delta q=\la q_{\alpha_{\text{max}}} \ra - \la q_{\alpha_{\text{min}}} \ra$, a behavior reminiscent of first order phase transitions \cite{garrahan10a}. 

\begin{figure}
\centerline{\includegraphics[width=14.5cm]{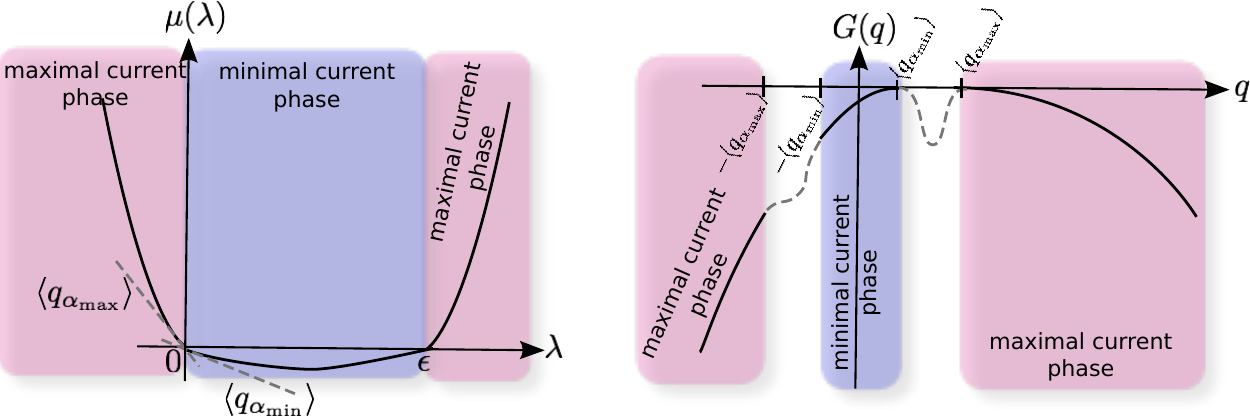}}
\caption{\small Sketch of the twin dynamic phase transitions in the current statistics of an open quantum system with a symmetry $U$, as appears for the current cumulant generating function $\mu(\lambda)$ (left), and the associated current large deviation function $G(q)$ (right). Notice the twin kinks at $\lambda=0,\epsilon$ in $\mu(\lambda)$ and the corresponding non-convex regimes for $q\in[\la q_{\alpha_{\text{min}}} \ra,\la q_{\alpha_{\text{max}}} \ra]$ and $q\in[-\la q_{\alpha_{\text{max}}} \ra,-\la q_{\alpha_{\text{min}}} \ra]$ in $G(q)$.}
\label{fig:ldf1}
\end{figure}

Next we explore the consequences of another (inherent) symmetry, \emph{microscopic time reversibility}, on the thermodynamics of currents in the presence of a symmetry $U$. Indeed, the reversible character of the microscopic coherent dynamics \cite{lu16a} leaves a footprint in the dissipative dynamics of open quantum systems in the form of a local detailed balance condition for the Lindblad-Liouville dynamical superoperator $\cL$. In brief, this condition states that $\cL$ is equal to its time-reversal dynamical map, suitably defined in terms of a time-reversal (anti-linear and anti-unitary) superoperator  \cite{chetrite12a,manzano15a}. If this detailed balance condition for $\cL$ holds \cite{chetrite12a,agarwal73a,andrieux09a}, it is now well-understood that the system of interest will obey the \emph{Gallavotti-Cohen fluctuation theorem} for currents, which links the probability of an arbitrary current fluctuation with its time-reversal event \cite{evans93a, gallavotti95a, gallavotti95b, kurchan98a, lebowitz99a,hurtado11b}, namely 
\be
\lim_{t\to\infty} \frac{1}{t} \ln \frac{\displaystyle \PP_t(qt)}{\displaystyle \PP_t(-qt)} = G(q)-G(-q) =  \epsilon q \, ,
\label{eq:GC}
\ee
where $\epsilon$ is a constant related to the rate of entropy production in the system \cite{solano-carrillo16a}. Equivalently, this fluctuation theorem can be stated as 
\be
\mu(\lambda)=\mu(\epsilon-\lambda) 
\label{eq:GCmu}
\ee
for the Legendre transform of the current LDF.  This time-reversal symmetry relation for the current LDF and its dual $\mu(\lambda)$ has a direct impact of their analyticity properties. In particular, the Gallavotti-Cohen relation (\ref{eq:GCmu}) implies that the symmetry-induced kink in $\mu(\lambda)$ observed at $\lambda=0$, see Eq. (\ref{eq:kink}), is reproduced at $\lambda=\epsilon$, where a \emph{twin dynamic phase transition} emerges, see left panel in Fig. \ref{fig:ldf1}. By inverse Legendre transforming $\mu(\lambda)$ we obtain the convex envelope of the current LDF \cite{touchette09a}, $G_{ce}(q)$, i.e.
\be
G_{ce}(q)=\max_{\lambda}[\mu(\lambda) + q\lambda] \, , 
\ee
%it is straightforward to show \cite{touchette09a} that 
The twin kinks in $\mu(\lambda)$ correspond to two different current intervals, 
\be
|q|\in[|\la q_{\alpha_{\text{min}}} \ra|,|\la q_{\alpha_{\text{max}}} \ra|] \, , 
\ee
related by time-reversibility (or $q\leftrightarrow -q$), where $G_{ce}(q)$ is affine, meaning that the real $G_{ce}(q)$ in this current regime can be either affine as $G_{ce}(q)$ or \emph{non-convex}\footnote[1]{Note that this cannot be directly inferred from $\mu(\lambda)$ \cite{touchette09a}.},
%(or at least affine\footnote[1]{A technical comment is needed here. The current LDF $G(q)$ can be either non-convex or affine in the intervals $|q|\in[|\la q_{\alpha_{\text{min}}} \ra|,|\la q_{\alpha_{\text{max}}} \ra|]$, and this cannot be directly inferred from $\mu(\lambda)$ \cite{touchette09a}. However, a current distribution with a completely flat regime between $\la q_{\alpha_{\text{min}}} \ra$ and $\la q_{\alpha_{\text{max}}} \ra$ is an unlikely physical situation, so we expect generically a non-convex $G(q)$.}), 
see right panel in Fig. \ref{fig:ldf1}. This typically corresponds to a multimodal current distribution $\PP_t(Q=qt)$, 
%with several peaks  
reflecting the \emph{coexistence} of multiple transport channels, each one associated with a different NESS in our open quantum system with a symmetry $U$ \cite{buca12a}. This coexistence is again reminiscent of the phenomenology of first-order phase transitions, but now at the dynamical level. 

Remarkably, the original symmetry of the system is broken at the fluctuating level, where the quantum system selects a symmetry sector that maximally facilitates a given current fluctuation (other symmetry sectors are still present in the dynamics, but only one dominates the given current fluctuation, see Eqs. (\ref{eq:zsum1})-(\ref{eq:mul}) and related discussion). In particular, the statistics during a current fluctuation with $|q|>|\la q_{\alpha_{\text{max}}} \ra|$ is dominated by the symmetry eigenspace with maximal current ($\alpha_{\text{max}}$), whereas for $|q|<|\la q_{\alpha_{\text{min}}} \ra|$ the minimal current eigenspace ($\alpha_{\text{min}}$) prevails. This regimes are termed maximal/minimal current phases in Fig. \ref{fig:ldf1}. The previous symmetry-breaking scenario is best captured by the effective density matrix 
\be
\rho_{\lambda}^{\text{eff}}\equiv \lim_{t\to\infty} \frac{\rho_{\lambda}(t)}{\Tr \left(\rho_{\lambda}(t)\right)}=\omega_{\alpha_0\alpha_0 0}(\lambda) \, ,
\label{eq:rhoeff}
\ee
with $\alpha_0=\alpha_{\text{max}}$ ($\alpha_{\text{min}}$) for $|\lambda-\frac{\epsilon}{2}|>\frac{\epsilon}{2}$ ($|\lambda-\frac{\epsilon}{2}|<\frac{\epsilon}{2}$). This normalized density matrix represents the typical state of the system during a current fluctuation with conjugated parameter $\lambda$, and its structure and properties typically change acutely from one current phase to the other, see \S\ref{sec:networks}  and Fig.~\ref{fig:symmetry_network} below for a detailed example in quantum spin networks.  

To end this section we want to discuss several features associated to the twin dynamic phase transitions (tDPTs) discussed above. First, these tDPTs are fragile against  environmental decoherence, as this noisy interaction typically destroys existing internal symmetries in the system of interest. For instance, the local noise operator in some cases does not commute with the symmetry operator and the multiplicity of steady-states (and the associated tDPTs) is lost. However, the existence of invariant subspaces in the noise-free case remains important even under environmental decoherence, as e.g. it crucially affects the short-time behavior of the system (see \S\ref{sec:transient} below for a detailed discussion). In this way symmetry signatures can be found even if there is decoherence. Furthermore, certain systems can be engineered to preserve the symmetries even in a noisy environment \cite{lidar98a,beige00a,metz06a,diehl08a}. Note also that similar dynamical phase transitions have been recently reported in literature \cite{pigeon15a,hossein-nejad13a}, whose origin can be traced back to the presence of an underlying symmetry.

In addition, and interestingly, the previous twin dynamic phase transitions in current statistics only happen out of equilibrium, disappearing in equilibrium (i.e. in the absence of boundary driving). In the latter case, the average currents for the multiple steady states are zero in all cases as expected, $\la q_{\alpha}\ra =0 \,\, \forall \alpha$, so no symmetry-induced kink appears in $\mu(\lambda)$ at $\lambda=0$ in equilibrium\footnote[1]{The LDF $\mu(\lambda)$ in equilibrium might still exhibit a (symmetric) kink at $\lambda=0$ due to some other singular behavior of current fluctuations in the dominant symmetry subspace in equilibrium, as e.g. a symmetric double-hump $G(q)$. This potential kink would be however unrelated to the underlying symmetry of the open quantum system.}. Moreover, an expansion for $|\lambda|\ll 1$ of the leading eigenvalues yields to first order 
\be
\mu_0^{(\alpha)}(\lambda)\approx  \frac{\lambda^2}{2} (\partial_{\lambda}^2 \mu_0^{(\alpha)}(\lambda))\vert_{\lambda=0}=\frac{\lambda^2}{2} \sigma_{\alpha}^2 \, , \nonumber
\ee
where $\sigma_{\alpha}^2$ is the variance of the current distribution in each steady state, so for equilibrium systems the overall current statistics is dominated by the symmetry eigenspace with maximal variance among those present in the initial $\rho(0)$. Therefore it is still possible to control the statistics of current fluctuations in equilibrium by an adequate preparation of $\rho(0)$, though $G(q)$ is convex around $\la q\ra=0$ and no dynamic phase transitions are expected (provided there is no other singular mechanism at play, see our previous footnote). 

To end this section, note also that this approach to symmetry based on full counting statistics simplifies considerably the study of multiple steady states in comparison to diagonalizing the full Liouvillian. Some advantages are: 
\begin{itemize}
\item It is no longer necessary to compute the full spectrum of the Liouvillian to conclude that there are different steady-states with varying currents. We only need to calculate the eigenvalue with the largest real part of the modified Liouvillian $\cLl$ (\ref{eq:masterrhol}) around $\lambda=0$.  
 
\item The eigenfunctions are not necessary to evaluate the maximum and minimum currents and their moments, as this information can be inferred from the large deviation function $\mu(\lambda)$, see Eq. (\ref{eq:cumul}). The orthonormalization process described at the end of Section \S\ref{sec:sym} to obtain the physical steady-states is no longer necessary.  
\end{itemize}

\newpage

\section{Transport and fluctuations in qubit networks}
\label{sec:networks}

%\subsection{Spin networks}
%\label{sec:networks}
In this section we apply what we have learned previously to study energy transport in a particular example of broad interest, open quantum networks; see Refs. \cite{manzano14a, scholak11a}. 

In recent years, several experiments have shown strong indications of coherent transport at room temperature in a number of photosynthetic network complexes \cite{adolphs06a,lee07a,engel07a,collini10a}, as e.g. the Fenna-Matthews-Olson complex of green sulfur bacteria (see Ref. \cite{scholes11} for a comprehensive review). In these complexes energy is transported with a very high efficiency from the antenna, where photons are absorbed, through a heterogeneous chromophore network to the reaction center, where the photosynthetic reaction takes place. These networks can be considered as open systems due to their interaction with incoming light and the vibrational degrees of freedom of the surrounding medium, and recent experiments have reported strong coherences and oscillations in this transport process whose quantum interpretation and potential role in photosynthesis are still under intense debate \cite{mancal13a,brumer12a,manzano13a,witt13a}.

In any case, these experimental results have motivated an intense study of transport in quantum networks, both in the transient regime \cite{mohseni08a,caruso09a,caruso10a,scholak11a,walschaers13a} and at the steady state \cite{manzano13a,witt13a,jesenko13a,pelzer14a,manzano14a}. The focus now is not only to understand transport across natural chromophore networks, but also to design specific network architectures for optimal transport \cite{scholak11b}, engineer noise sources to enhance transport efficiency \cite{chin10a}, or even to create artificial light-harvesting quantum antennae using genetic engineering techniques for enhanced exciton transport \cite{park16a}. Further recent advances also include the study of the interplay between complex network structure and quantum dynamics \cite{biamonte17a}, as well as the development of quantum photonic networks using tools from chiral quantum optics \cite{lodahl17a}. Motivated by these transport problems, we now proceed to study the thermodynamics of currents in homogeneous open quantum networks. These are simplified models of quantum transport which have proven extremely useful in the past to understand e.g. the functional role of noise and dephasing in enhancing coherent energy transfer \cite{plenio08a,caruso11a,biggerstaff16a}.

\begin{figure}
\centerline{\includegraphics[width=15cm]{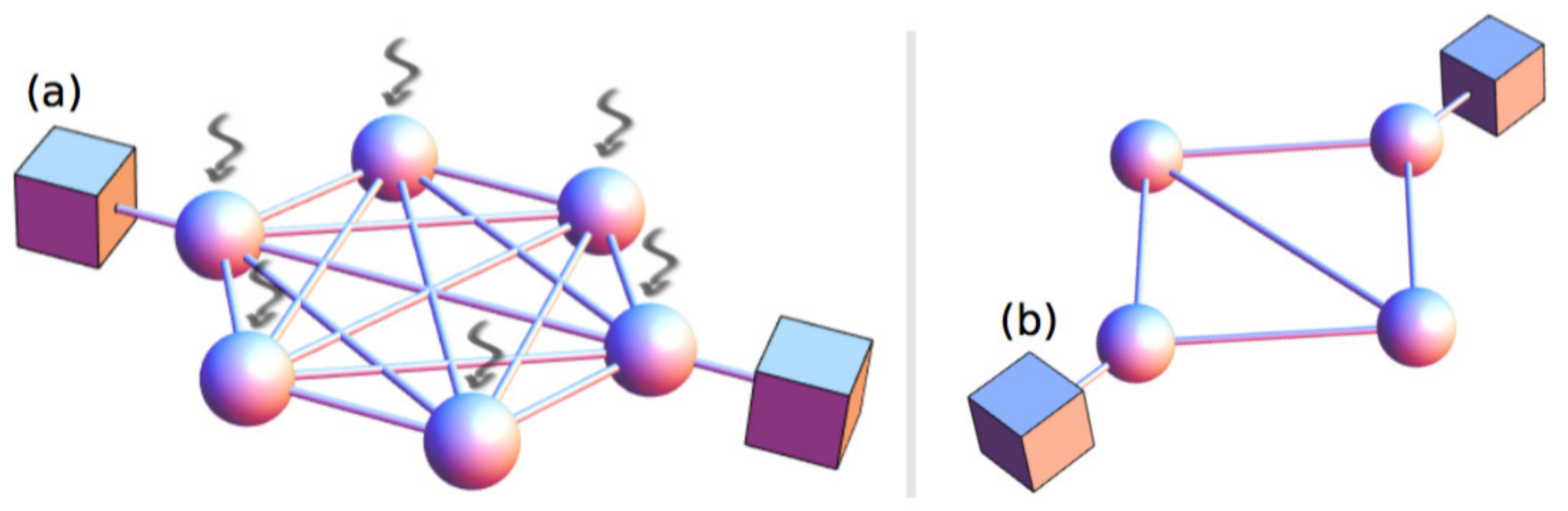}}
\caption{\small (a) Fully-connected network of 6 qubits (spheres) in contact with two thermal baths (boxes) and possibly subject to dephasing noise (wavy arrows). Symmetries in this case correspond to permutations of pairs of bulk qubits (i.e. qubits not connected to the external baths). (b) Sketch of a symmetry-controlled quantum thermal switch. Manipulating the symmetry of the pair of bulk qubits in this case enables full control of the energy current between the baths. See also Ref. \cite{manzano14a}.
}
\label{fig:network}
\end{figure}

\subsection{Model and current statistics}
\label{sec:QNmod}

Fig. \ref{fig:network}.a depicts an example of the model of interest. It consists in a fully-connected network of $N$ spins or two-level systems (\emph{qubits}) with dipole-dipole interaction of equal strengths and homogeneous on-site energies. The system Hamiltonian is
\be 
H^{\text{net}}= h \sum_{i=1}^N \sigma_i^+ \sigma_i^- + J \sum_{\substack{i,j=1 \\  j< i}}^N  \left( \sigma_i^+ \sigma_j^- +  \sigma_i^- \sigma_j^+  \right) \, , 
\label{eq:fcnHamil}
\ee 
where $\sigma_i^\pm=\sigma^x_i\pm\ii\sigma^y_i$ are the raising ($+$) and lowering ($-$) operators acting on spin $i$, with $\sigma^{x,y}$ the corresponding spin-$1/2$ Pauli matrices,
\be
\sigma^x =\begin{pmatrix} 0 & 1 \\ 1 & 0 \end{pmatrix} \, , \quad \quad 
\sigma^y =  \begin{pmatrix} 0 & -\ii \\ \ii & 0 \end{pmatrix}   \, , 
\ee
$h$ is the on-site energy, and $J$ represents the coupling strength between the different spins. We will focus on $N$ even for simplicity, though similar results hold for odd $N$. 

To model the interaction with an energy source and a reaction center, we couple the quantum network at (arbitrary) qubits $1$ and $N$ to two \emph{bosonic} heat baths working at different temperatures. We refer to the sites connected to the baths as \emph{terminal spins}, while the remaining sites constitute the \emph{bulk}, see Fig. \ref{fig:network}.a. The reservoirs locally pump and extract excitations in the system in an incoherent way, triggering the nonequilibrium dissipative dynamics of the quantum network complex. The dynamics of the system is then given by a Lindblad master equation (\ref{eq:lindblad}) with Lindblad operators 
\ben
L_1 &= \sqrt{a_1}\sigma_1^+ \, , \qquad L_2 &= \sqrt{b_1}\sigma_1^- \, , \nonumber \\
L_3 &= \sqrt{a_N}\sigma_N^+ \, , \qquad L_4 &= \sqrt{b_N}\sigma_N^- \, , 
\een
where the coefficients $a_{i}$ represent the pumping rate of quanta to the system due to the action of the corresponding bath at qubit $i=1,N$, while the coefficients $b_{i}$ represent the corresponding rate of quanta absorption. Whenever $a_1 b_N\neq a_N b_1$, a temperature gradient sets in that drives the system out of equilibrium, with an associated net exciton current in the steady state. This external nonequilibrium drive can be quantified by $\epsilon=\ln[a_1 b_N/(a_N b_1)]$.

Note that the Hamiltonian (\ref{eq:fcnHamil}) is related with that of the Lipkin-Meshkov-Glick model introduced in the 60's to describe phase transitions in nuclei \cite{lipkin65a,meshkov65a,glick65a}. This model can be solved exactly in the purely coherent, closed case using Bethe equations \cite{pan99a,ribeiro07a,ribeiro08a}, though an analytical solution in the presence of an environment is still lacking. Moreover, similar open spin models with dipole-dipole interactions have been recently studied to analyze quantum Fourier's law and energy transfer in quantum networks \cite{manzano12a,znidaric11a}. 

Interestingly, this model exhibits not just one but many symmetries in the sense of Section \S\ref{sec:sym2}, as there are a number of independent unitary operators that fulfill the condition defined in Eq. (\ref{eq:sym}). In particular, any \emph{permutation operator} $\pi_{ij}\in\cBH$ corresponding to the exchange of two \emph{bulk} spins $i,j\in[2,N-1]$ leaves invariant the system and hence commutes with all the elements of the Liouvillian for this model, $\cor{\pi_{ij},H}=0=\cor{\pi_{ij},L_m}\; (\forall m)$.  Therefore, following our general discussion above, this leads to a myriad of possible nonequilibrium steady states that depend on the symmetry sectors populated by the initial state of bulk spins. 

\begin{figure}
\centerline{\includegraphics[width=17cm]{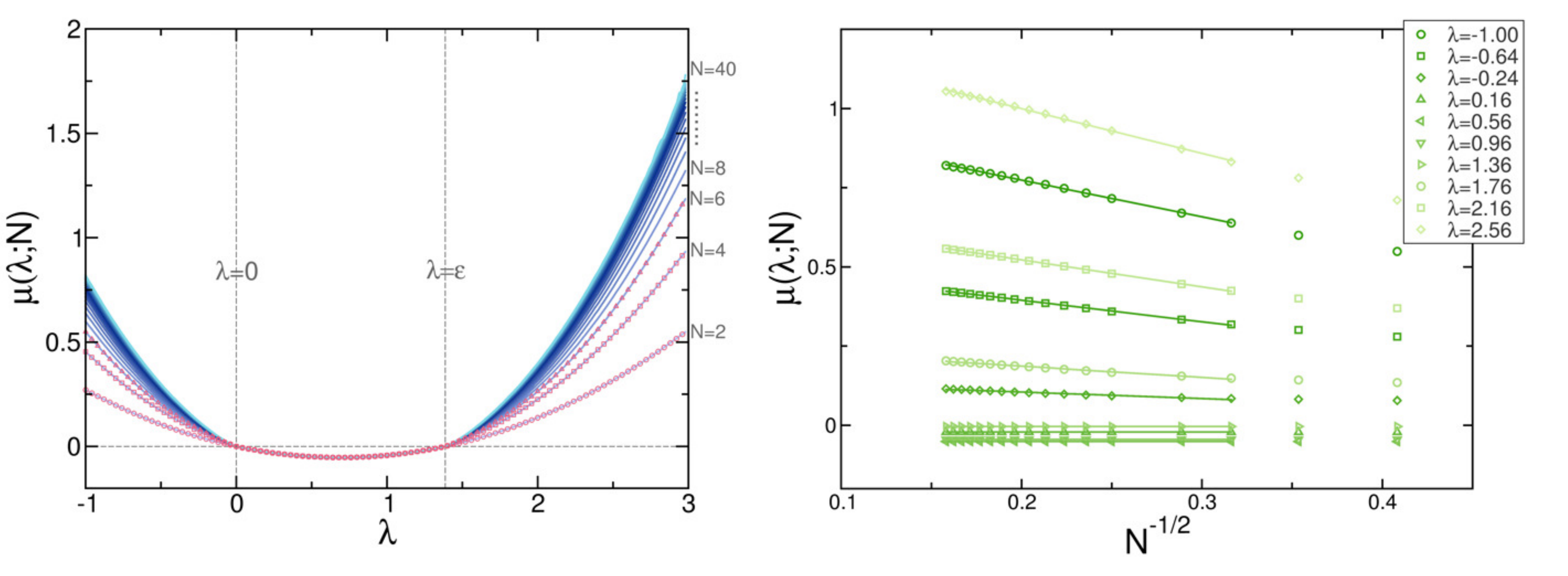}}
\caption{Left: Large deviation function $\mu(\lambda;N)$ of a fully-connected open quantum network of $N$ spins with Hamiltonian (\ref{eq:fcnHamil}). The curves for $N=2,4,6$ (open symbols) are calculated by direct diagonalization of the full deformed superoperator $\cLl^{\text{net}}$. The remaining lines are obtained after a symmetry-induced dimensional reduction of the system Hilbert space, see Section \S\ref{sec:QNsym} below.
The parameters of the simulation are: $a_1=b_2=1,\; b_1=a_2=1,\; \text{and}\; J=h=1$. The vertical dashed lines signal the critical points $\lambda=0,\epsilon$. While no $N$-dependence is observed for $0<\lambda<\epsilon$, a rapid increase with size appears outside this interval, suggesting the emergence of two kinks in $\mu(\lambda;N)$ at $\lambda=0,\epsilon$. Right: Dependence of $\mu(\lambda;N)$ with the system size $N$ for different fixed $\lambda$. A clear $N^{-1/2}$ scaling for large enough $N$ is evident in all cases. This allows us to infer the thermodynamic limit of the finite-size LDF $\mu(\lambda;N)$, see Fig. \ref{fig:LDF_dephasing} below.
}
\label{fig:LDF_Networks}
\end{figure}

As described in Section \S\ref{sec:LDF2.5}, these multiple symmetries have also an effect on the current flowing through the system and its fluctuations. In particular, we expect twin dynamic phase transitions in the current statistics, which should appear as a pair of kinks in the cumulant generating function $\mu(\lambda)$ (associated to the current large-deviation function $G(q)$ via Legendre transform) at $\lambda=0$ and $\lambda=\epsilon=\ln[a_1 b_N/(a_N b_1)]$. For completeness, we recall that the LDF $\mu(\lambda)$ is nothing but the eigenvalue with largest real part of the deformed Liouvillian $\cLl^{\text{net}}$, with
\be
\cLl^{\text{net}} \rho_{\lambda} \equiv -\ii \cor{H^{\text{net}},\rho_\lambda} + e^{-\lambda} L_1^{\phantom{\dagger}} \rho_\lambda L_1^{\dagger} + e ^{\lambda} L_2^{\phantom{\dagger}} \rho_\lambda L_2^{\dagger} + \sum_{m=3,4} L^\dagger_m \rho_\lambda L_m^{\phantom{\dagger}}
- \frac{1}{2} \sum_{m=1}^4 \key{L_m^{\phantom{\dagger}} L_m^{\dagger}, \rho_\lambda} \, .
\label{eq:mod_liou}
\ee

To test our predictions in this particular model, we diagonalized numerically the superoperator $\cLl^{\text{net}}$ in Fock-Liouville space for $N=2,~4,~6$ and a particular set of parameters ($J=h=1$ and $a_1=b_2=1,\; b_1=a_2=1$ so $\epsilon\approx 1.39$), focusing on its leading eigenvalue $\mu(\lambda;N)$ and the associated right eigenmatrix. Note that the Hilbert space of interest grows exponentially with the system size ($\cLl^{\text{net}}$ is a $4^N\times 4^N$ matrix in this representation), so direct numerical evaluation of its leading spectral properties is only possible for these relatively small system sizes. However we will explain below how symmetry can be used to simplify the problem and reach much larger system sizes. In the meantime, Fig. \ref{fig:LDF_Networks} displays the measured LDF $\mu(\lambda;N)$ for different values of the network size. Remarkably this function does not depend on $N$ for $0<\lambda<\epsilon$, while it grows steeply with $N$ outside this interval. This behavior strongly suggests the presence of two kinks in $\mu(\lambda;N)$ at $\lambda=0$ and $\lambda=\epsilon$, as expected. 
 
\begin{figure}
\centerline{\includegraphics[width=16cm]{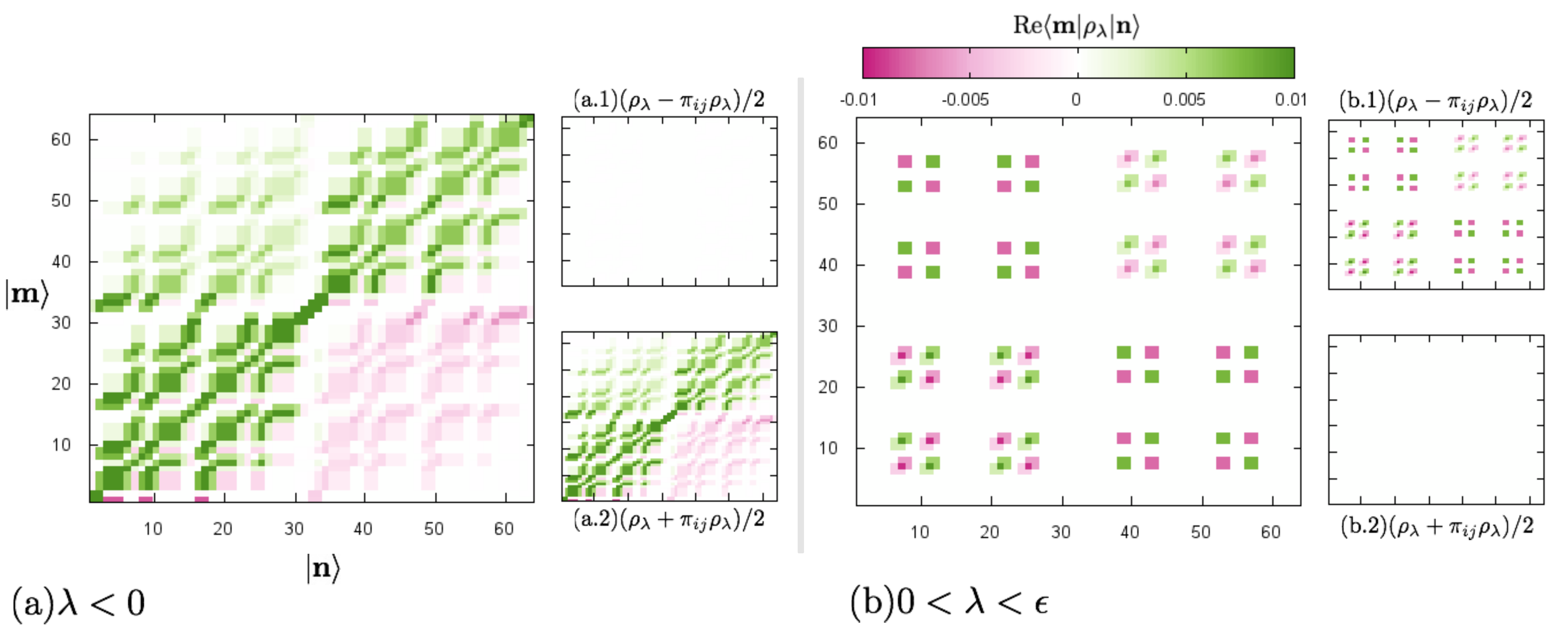}}
\vspace{-0.25cm}
\caption{Real part of the $N=6$ normalized right eigenmatrix $\omega_{\alpha_0\alpha_0 0}(\lambda)$ associated with the eigenvalue of $\cLl$ with largest real part, for (a) $\lambda=-0.4$ and (b) $\lambda=0.2$. Panels (a.1)-(b.1) and (a.2)-(b.2) show respectively the $(i,j)$-antisymmetrized and -symmetrized eigenmatrices, with $(i,j)$ an arbitrary pair of bulk qubits. For $\lambda<0$ (and $\lambda>\epsilon$) the leading eigenmatrix is completely symmetric, while for $0<\lambda<\epsilon$ it is pair-antisymmetric. System parameters as in Fig. \ref{fig:LDF_Networks}.
}
\label{fig:symmetry_network}
\end{figure}

The qualitative change of behavior at $\lambda=0$ and $\lambda=\epsilon$ is better understood at the configurational level, i.e.~by studying the eigenmatrix associated to the leading eigenvalue. This is shown in Fig. \ref{fig:symmetry_network}, which displays the real part of the leading right eigenmatrix $\rho_\lambda=\omega_{\alpha_0\alpha_0 0}(\lambda)$ --the one associated to the leading eigenvalue $\mu(\lambda;N)$-- in the computational basis ($\ket{\mathbf{n}}\equiv \otimes_{i=1}^N \ket{n_i}\in \cH$, with $\ket{n_i}=\ket{0}$ or $\ket{1}$) for $N=6$ and two different values of $\lambda$, one for $\lambda<0$ (Fig. \ref{fig:symmetry_network}.a) and another for $0<\lambda<\epsilon$  (Fig. \ref{fig:symmetry_network}.b). Clearly, the leading eigenmatrices are structurally very different across the kink at $\lambda=0$. This qualitative difference is confirmed by studying their symmetry properties under permutations $\pi_{ij}$ of bulk spins, $i,j\in[2,N-1]$. In particular, for $\lambda<0$ (as well as for $\lambda>\epsilon$) the measured eigenmatrix is \emph{completely symmetric} under any permutation of bulk qubits, 
\be
\rho_\lambda = \pi_{ij} \, \rho_\lambda \, , 
\ee
see Figs. \ref{fig:symmetry_network}.a.1-2. On the other hand, in the regime $0<\lambda<\epsilon$ the resulting eigenmatrix is instead \emph{antisymmetric by pairs}. This means that non-overlapping pairs of bulk qubits are in antisymmetric, singlet state, so
\be
\rho_\lambda = - \pi_{ij} \, \rho_\lambda \, , 
\ee
see Figs. \ref{fig:symmetry_network}.b.1-2. Interestingly this pair-antisymmetric regime is degenerate for $N>4$, as the $N_\text{b}\equiv N-2$ bulk qubits can be partitioned by pairs in different ways, though this degeneracy does not affect the results. 

The previous observations confirm the presence of a pair of twin dynamic phase transitions happening at $\lambda=(0,\epsilon)$, where the symmetry of the original system is broken at the fluctuating level\footnote[1]{Note that equivalent results have been numerically checked also for $N=4$, and are expected to hold for arbitrary even $N$.}. For large enough current fluctuations such that $\lambda<0$ or $\lambda>\epsilon$ (equivalently $|\lambda-\frac{\epsilon}{2}|>\frac{\epsilon}{2}$), the open quantum system of interest is expected to select the symmetry subspace with \emph{maximal current} among all symmetry subspaces present in the system initial state, see the discussion in Section \S\ref{sec:LDF4}. For the open quantum network here studied, this maximum current manifold corresponds to the \emph{totally symmetric} subspace, a sort of \emph{bosonic transport regime} which can be easily understood at a heuristic level. Indeed, it is sufficient to note that a totally symmetric bulk can absorb a maximal number of excitations from the terminal qubit, hence leaving it free to receive further excitations from the reservoir, maximizing in this way the exciton current across the system. On the other hand, the minimal current symmetry subspace dominating current statistics for $0<\lambda<\epsilon$ is antisymmetric by pairs. This \emph{pair-fermionic transport regime} can be again easily rationalized by noting that pairs of bulk qubits in singlet state are \emph{dark states} of the dynamics (decoherence-free subspaces): such states remain frozen in time, decoupling from the rest of the system, and hence cannot accept excitations from the terminal qubits effectively reducing the size of bulk and thus leading to a minimal current. In fact, this observation explains why $\mu(\lambda;N)$ does not depend on $N$ for $0<\lambda<\epsilon$: all bulk spins are paired in singlet states, resulting in an system with an \emph{effective size} of 2 spins, the number of terminal qubits, independently from $N$. 

The previous argument is an example of the severe dimensional reduction induced by symmetry in a particular fluctuation regime ($0<\lambda<\epsilon$). We now extend this idea to the bosonic transport regime $|\lambda-\frac{\epsilon}{2}|>\frac{\epsilon}{2}$, a procedure that allows us to investigate the thermodynamics of currents in networks of size much larger than previously anticipated.

\subsection{Dimensional reduction and finite-size scaling of current fluctuations}
\label{sec:QNsym}

In this section we exploit the totally symmetric nature of the maximal current phase, $|\lambda-\frac{\epsilon}{2}|>\frac{\epsilon}{2}$, to drastically reduce the dimension of the Hilbert space in this region. Moreover, we explicitly demonstrate the dimensional reduction associated to the emergence of dark states in the pair-antisymmetric phase. In this way, for a network of $N$ qubits, the dimension of the spectral problem drops from an exponential $O(2^N)$ to a linear scaling $O(N)$, and we use this effect to explore current statistics in networks of size much larger than anticipated. We mention that a similar dimensional reduction was already described in previous studies of the related Lipkin-Meshkov-Glick model \cite{lipkin65a,meshkov65a,glick65a,ribeiro07a,ribeiro08a}. 

We start by noting that a completely symmetric state of bulk qubits is univocally described by the total number of excitations in the bulk. In this way, an arbitrary state of the quantum network with a totally symmetric bulk can be thus written as
\be
\ket{K;n_1,n_N} = \frac{1}{\sqrt{N_{\text{b}} \choose K}}\sum_{n_2\ldots n_{N-1}=0,1} \ket{\mathbf{n}} \, \delta \Big(K-\sum_{i=2}^{N-1} n_i \Big) \, 
\label{nsim}
\ee
where $N_{\text{b}}\equiv N-2$ is the number of bulk qubits, $\ket{\mathbf{n}}\equiv \otimes_{i=1}^N \ket{n_i}\in \cH$, with $\ket{n_i}=\ket{0}$ or $\ket{1}$, and $K\in[0,N_{\text{b}}]$ is the total number of excitations in the bulk in this symmetric state. The combinatorial number ${N_{\text{b}} \choose K}$ in the normalization constant counts the number of ways of distributing $K$ excitations among $N_{\text{b}}$ bulk qubits. The strategy here will consist in proving that the Hamiltonian of the open quantum network with a completely symmetric bulk can be fully written in terms of the low-dimensional basis formed by vectors (\ref{nsim}), thus radically reducing the complexity of the problem.

The dichotomy between bulk and terminal qubits allows to decompose now the qubit network Hamiltonian (\ref{eq:fcnHamil}) as $H^{\text{net}} = H_0^{\text{net}} + H_{\text{b}}^{\text{net}} + H_{\text{I}}^{\text{net}}$, where
\be
H_0^{\text{net}} \equiv h \sum_{i=1}^N \sigma_i^+ \sigma_i^- \, , \qquad \qquad H_{\text{b}}^{\text{net}} \equiv J \sum_{i=2}^{N-2} \sum_{j=i+1}^{N-1} \Delta_{ij} \, , \label{H0b}
\ee
with the definition $\Delta_{ij} \equiv (\sigma_i^+\sigma_j^- + \sigma_i^-\sigma_j^+)$, and
\be
H_{\text{I}}^{\text{net}} = J  \Big[ (\sigma_1^+ + \sigma_N^+)\Delta_- + (\sigma_1^- + \sigma_N^-)\Delta_+ + \Delta_{1N}\Big] \, ,
\label{Hint}
\ee
where we further define $\Delta_{\pm} \equiv \sum_{i=2}^{N-1} \sigma_i^{\pm}$. Trivially, the on-site contribution to the Hamiltonian ($H_0^{\text{net}}$) is diagonal in the basis defined by the states (\ref{nsim}), i.e. $H_0^{\text{net}} \ket{K;n_1,n_N} = h (K+n_1+n_N) \ket{K;n_1,n_N}$, so we can write
\be
H_0^{\text{net}} = h \sum_{\substack{K=0 \\ n_1,n_N=0,1}}^{N_{\text{b}}} (K+n_1+n_N)  \ket{K;n_1,n_N} \bra{K;n_1,n_N}
\label{aH0}
\ee
For the bulk self-interaction part $H_{\text{b}}^{\text{net}}$, first note that the operators $\Delta_{ij}$ simply exchange the states of qubits $i$ and $j$ whenever they are different, yielding zero otherwise, i.e. $\Delta_{ij} \ket{\mathbf{n}} = \delta_{n_i,1-n_j}\ket{\mathbf{n}}_{ij}$, where $\ket{\mathbf{n}}_{ij}$ is the state resulting from exchanging $n_i \leftrightarrow n_j$ in $\ket{\mathbf{n}}$. Using this, one can easily show that $H_{\text{b}}^{\text{net}}$ is also diagonal in the basis defined by $\ket{K;n_1,n_N}$, namely
\be
H_{\text{b}}^{\text{net}} = J \sum_{\substack{K=0 \\ n_1,n_N=0,1}}^{N_{\text{b}}} K (N_{\text{b}}-K) \ket{K;n_1,n_N} \bra{K;n_1,n_N} \, ,
\label{aHb}
\ee
where the prefactor counts the number of distinct computational basis bulk states that we can form with $K$ $\ket{1}$'s and $(N_{\text{b}}-K)$ $\ket{0}$'s. It is now straightforward to show that the operators $\Delta_{\pm}$ defined above move the state $\ket{K;n_1,n_N}$ to $\ket{K\pm 1;n_1,n_N}$, with a prefactor that counts the number of ways of distributing the pertinent excitations among $(N_{\text{b}}-1)$ bulk sites and takes into account the different normalizations. In particular, $\Delta_{\pm}\ket{K;n_1,n_N} = D_K^{\pm} \ket{K\pm 1;n_1,n_N}$, with 
\be
D_K^\pm = \sqrt{(K+1-k_\pm)(N_{\text{b}}-K+k_\pm)} \, , \label{dpm}
\ee
with $k_{\pm}\equiv (1\mp 1)/2$, so we may write
\be
\Delta_{\pm} = \sum_{\substack{K=k_{\pm} \\ n_1,n_N=0,1}}^{N_{\text{b}}-(1-k_{\pm})} D_K^{\pm} \ket{K\pm 1;n_1,n_N} \bra{K;n_1,n_N} \, .
\label{aDpm}
\ee
In this way the Hamiltonian (\ref{eq:fcnHamil}) of the open quantum network with a completely symmetric bulk can be fully written in terms of the low-dimensional basis formed by vectors (\ref{nsim}). Moreover, as the Lindblad operators in the network master equation only act on the terminal spins, the dimension of the problem in this totally symmetric regime drops from the original $2^N$ to a much lower dimension $4(N-1)$, which scales linearly with the number of qubits.

For completeness, we show now explicitly that the spectral problem associated to the deformed Lindblad superoperator for a network of size $N$ with a pair of bulk qubits in antisymmetric state is equivalent to that of a network with $N-2$ qubits. This results from the frozen dynamics of the antisymmetric qubit pair, which forms a dark state which effectively decouples them from the rest of the system. We hence consider our network with $N$ qubits, such that the pair formed by the (otherwise arbitrary) bulk qubits $a$ and $b$ is in an antisymmetric state. The initial density matrix can thus be written in a direct product form, $\rho_-\equiv \ket{-}\bra{-}_{ab} \otimes \rho_{N-2}$, where $\ket{-}=\frac{1}{\sqrt{2}}(\ket{10}-\ket{01})$ is the singlet state, and $\rho_{N-2}$ is an arbitrary reduced density matrix for the remaining $N-2$ qubits. We next decompose the Hamiltonian (\ref{eq:fcnHamil}) in three natural parts, $H=H_{ab} + H_{N-2} + H_{\text{int}}$ with
\ben
H_{ab} &=& h(\sigma_a^+ \sigma_a^- + \sigma_b^+ \sigma_b^-) + J \Delta_{ab} \, ,  \\
 H_{\text{int}} &=& J \Big[ (\sigma_a^+ + \sigma_b^+) \sum_{\substack{k=1 \\ k\neq a,b}}^N \sigma_k^- +  (\sigma_a^- + \sigma_b^-) \sum_{\substack{k=1 \\ k\neq a,b}}^N \sigma_k^+ \Big]  \, , \nonumber
\een
and $H_{N-2}$ the standard Hamiltonian (\ref{eq:fcnHamil}) for $N-2$ qubits (excluding qubits $a$ and $b$). It is now easy to show that the terms $H_{ab}$ and $H_{\text{int}}$ commute with any pair-antisymmetric density matrix of the form $\rho_-$, i.e. $[H_{ab},\rho_-]=0=[H_{\text{int}},\rho_-]$, so
\ben
\dot{\rho}_- &=& -\ii [H,\rho_-] + {\cal L}_1^{(\lambda)} \rho_- + {\cal L}_N \rho_- \equiv \cLl^{(N)} \rho_- \, \nonumber \\
&=& \ket{-}\bra{-}_{ab} \otimes \Big( -\ii [H_{N-2},\rho_{N-2}] + {\cal L}_1^{(\lambda)} \rho_{N-2} + {\cal L}_N \rho_{N-2} \Big)  \\
&=& \ket{-}\bra{-}_{ab} \otimes \Big(\cLl^{(N-2)} \rho_{N-2}\Big) = \ket{-}\bra{-}_{ab} \otimes \dot{\rho}_{N-2} \, , \nonumber
\een
where ${\cal L}_1^{(\lambda)}$ and ${\cal L}_N$ can be defined from Eq. (\ref{eq:lindblad}) above. Interestingly, using this method in a recursive manner it can be proved that the eigenvalue problem for any open quantum network of arbitrary size with a pair-antisymmetric bulk (i.e. with non-overlapping pairs of bulk qubits in singlet state) can be reduced to the case $N=2,3$, depending on $N$ being even or odd. 

\begin{figure}
\centerline{\includegraphics[width=16cm]{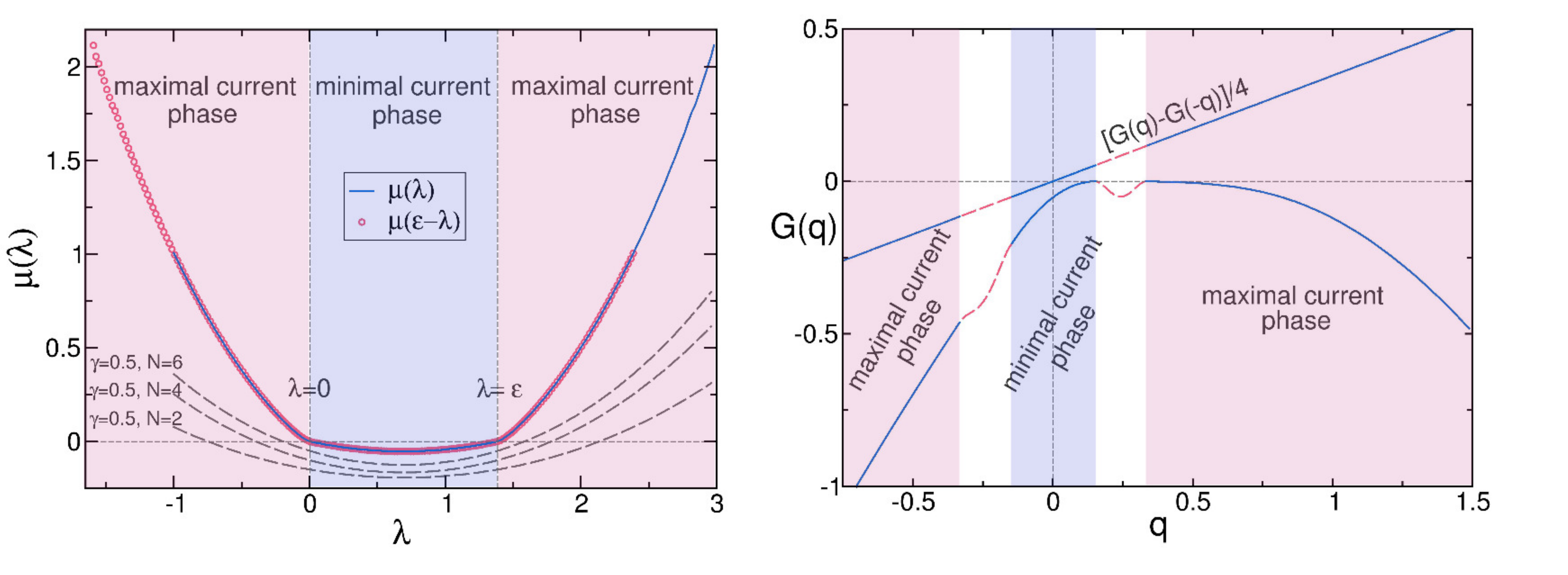}}
\caption{\small Left: Current statistics for the open quantum network in the thermodynamic limit ($N\to\infty$) as captured by $\mu(\lambda)$, see Eq. (\ref{muN}) and Fig. \ref{fig:LDF_Networks}. A first-order-type twin dynamic phase transition happening at $\lambda=(0,\epsilon)$ is apparent, signaled by twin kinks in $\mu(\lambda)$ (thin vertical dashed lines) separating the maximal and minimal current phases of the fluctuation phase diagram (shaded areas). Note that the current distribution obeys the Gallavotti-Cohen fluctuation theorem, $\mu(\lambda)=\mu(\epsilon-\lambda)$. Dashed thick lines show $\mu(\lambda)$ measured for small networks ($N=2, 4, 6$) with dephasing noise ($\gamma=0.5$), which breaks the permutation symmetry and hence destroys the twin dynamic phase transitions. Curves have been shifted downward for clarity (recall that $\mu(0)=0$ in all cases). Right: Asymptotic current LDF $G(q)$ obtained from the numerical inverse Legendre transform of $\mu(\lambda)$ in the left panel. Dashed curves sketch the non-convex regimes of $G(q)$ for which $\mu(\lambda)$ offers no information. Gallavotti-Cohen symmetry, $G(q)-G(-q)=\epsilon q$, is again clearly satisfied.
}
\label{fig:LDF_dephasing}
\end{figure}

\begin{figure}
\centerline{\includegraphics[width=9cm]{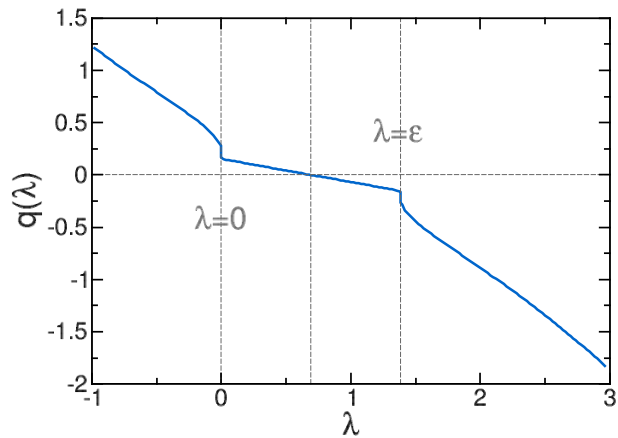}}
\caption{\small Current as a function of the counting field $\lambda$ for the open quantum network. Note the discontinuous jump in $q(\lambda)$ at the twin DPTs, $\lambda=0,\epsilon$.
}
\label{fig:LDF_qlamb}
\end{figure}

The drastic dimensional reduction just demonstrated can be now used to compute from the resulting simplified spectral problem the cumulant generating function of the current LDF, $\mu(\lambda;N)$, for quantum networks of size $N\leq 40$, see lines in left panel of Fig. \ref{fig:LDF_Networks}. These curves clearly exhibit convergence toward some well-defined, limiting behavior as $N\to \infty$, allowing us to characterize in detail the finite-size corrections in current statistics. Indeed, for each fixed value of the conjugate parameter $\lambda$, the data strongly suggest the following finite-size scaling behavior
\be
\mu(\lambda;N)=\mu(\lambda) + \frac{a(\lambda)}{\sqrt{N}} \, , \label{muN}
\ee
with $a(\lambda)$ some amplitude, see right panel in Fig. \ref{fig:LDF_Networks}. Note that $a(\lambda)=0$ for $0<\lambda<\epsilon$, as the current statistics in the minimal current phase does not depend on $N$. The previous scaling law yields an estimate of the current LDF $\mu(\lambda)$ for the open quantum network in the thermodynamic ($N\to\infty$) limit. This limiting behavior is shown is Fig. \ref{fig:LDF_dephasing}, left panel, where the presence two clear kinks at $\lambda=(0, \epsilon)$ is apparent. Notice that the estimated $\mu(\lambda)$, as well as all finite-size LDFs in left panel of Fig. \ref{fig:LDF_Networks}, obey the Gallavotti-Cohen fluctuation theorem $\mu(\lambda)=\mu(\epsilon-\lambda)$, which results from the time-reversibility of microscopic dynamics \cite{evans93a, gallavotti95a, gallavotti95b, kurchan98a, lebowitz99a,hurtado11b,andrieux09a,agarwal73a,chetrite12a}. In order to obtain a direct estimate of the current LDF $G(q)$ in the $N\to\infty$ limit, we also performed numerically the inverse Legendre transform of $\mu(\lambda)$, see right panel in Fig. \ref{fig:LDF_dephasing}. As expected two different current regimes emerge, $|q|\in[|\la q_{\alpha_{\text{min}}} \ra|,|\la q_{\alpha_{\text{max}}} \ra|]$, related to the kinks in $\mu(\lambda)$, where $G(q)$ is non-convex (or at least affine, see footnote in \S\ref{sec:LDF4}). This corresponds to a multimodal current distribution which results from the coexistence of different transport channels classified by symmetry. For completeness, Fig.~\ref{fig:LDF_qlamb} shows the relation between the current $q$ and the counting field $\lambda$ as obtained in the numerical inverse Legendre transform.

\subsection{Symmetry-controlled transport}
\label{sec:QNcontrol}

\begin{figure}
\centerline{\includegraphics[width=16cm]{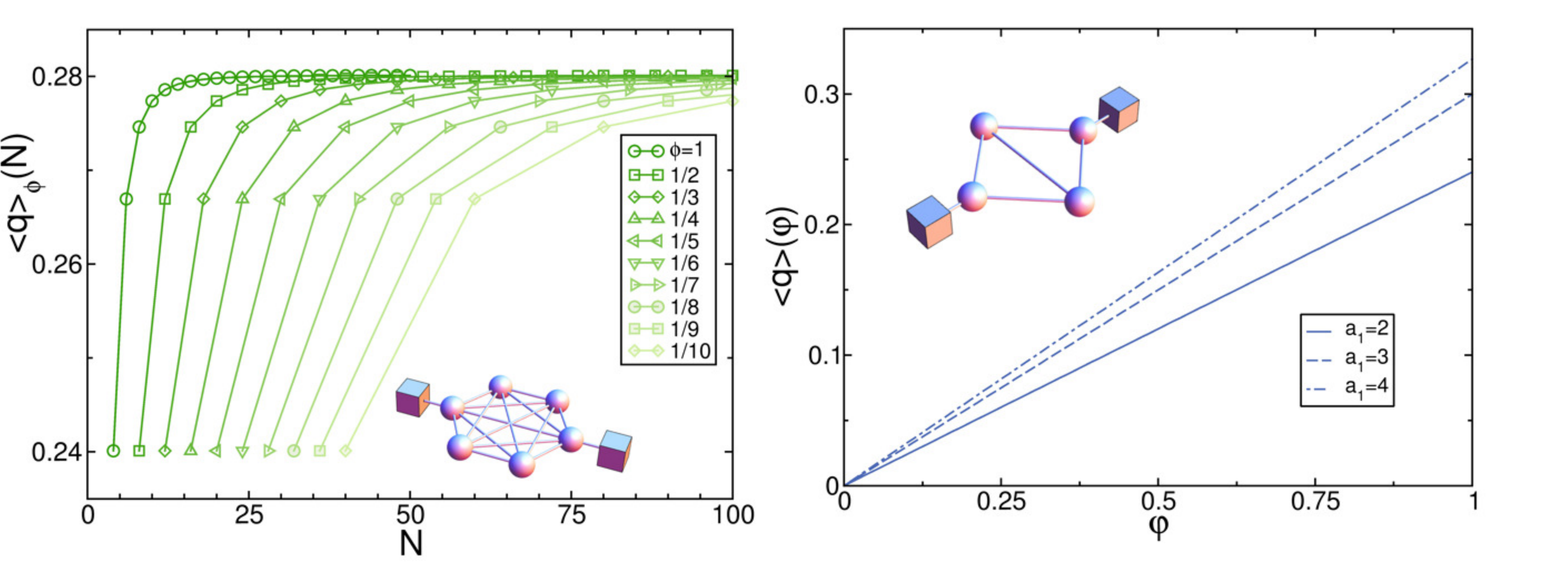}}
\caption{Left: Average current $\la q \ra_\phi(N)$ as a function of the network size for a system with an even number $(1-\phi)N$ of bulk qubits initialized in pair-antisymmetric states. The current is an increasing function of both $N$ and $\phi$, demonstrating symmetry-controlled transport. System parameters are $a_1=b_2=1,\; b_1=a_2=1,\; \text{and}\; J=h=1$, as in Fig.~\ref{fig:LDF_Networks}. The inset shows an example of the topology of the fully-connected quantum network. Right: For the four-qubit quantum thermal switch sketched in the inset, we plot the average current $\la q \ra(\varphi)$ as a function of $\varphi$, the projection of the initial density matrix on the subspace of $\cBH$ corresponding to a totally symmetric bulk, for different excitation pumping rates $a_1$ (the other parameters as above). This shows how the heat current between hot and cold reservoirs can be completely blocked, modulated, or turned on by preparing the symmetry of the initial state.}
\label{fig:current_control}
\end{figure}

The presence of different invariant subspaces in quantum networks can be used to control the excitonic current through the network via \emph{initial state preparation}, see Eq. (\ref{eq:qmed}) in \S\ref{sec:LDF3} and related discussion. For this particular case, we note that a network with a totally-symmetric bulk has maximal current for a given network size and hence, by preparing pairs of bulk spins in antisymmetric states (which effectively decouple playing no role in the system dynamics), it is possible to reduce this current. This effect is illustrated in the left panel of Fig. \ref{fig:current_control}, where we plot the average current flowing through a fully-connected qubit network as a function of the system size for different values of the ratio $\phi$ of spins that are initially in a totally symmetric state. In particular, the initial state $\rho(0)$ is prepared in a direct product configuration such that an \emph{even} number $(1-\phi)N$ of bulk qubits are initialized in antisymmetric, singlet states by pairs, while the complementary set of bulk qubits are initially in a totally symmetric state. As expected from our previous argument, for a fixed ratio $\phi$ the current increases with the network size and saturates for large values of $N$. In this way, by tuning the initialization parameter $\phi$ we can control the average current for each $N$, though the range of currents that we can access in this case is limited.

However, a simple modification of the network Hamiltonian allows us now to gain \emph{full control} of the heat current traversing the quantum system. In particular, by removing the interaction between the terminal qubits, it is possible to block completely the exciton current which flows from the hot to the cold reservoir by initializing the bulk qubits in a pair-antisymmetric state. The simplest case illustrating this idea is a network of $N=4$ qubits connected as in Fig. \ref{fig:network}.b, see also the inset of the right panel of Fig. \ref{fig:current_control}. Indeed, for this particular network topology, if we start from a \emph{pure} initial state such that the two bulk qubits form a singlet (totally-antisymmetric) state, this singlet forms a dark state of the dynamics, see \S\ref{sec:QNsym}. Due to the lack of connection of the terminal qubits in this topology, this dark state disconnects the hot reservoir from the cold one, hence effectively blocking any exciton flow in the system. In order to use this effect to control the current, we now initialize the system in a mixed state $\rho(0)=\varphi \rho_+ + (1-\varphi)\rho_-$, with $\varphi \in [0,1]$ a fixed parameter and $\rho_\alpha \in \cBaa$ arbitrary density matrices, with $\alpha=+$ ($\alpha=-$) for the diagonal subspace of $\cBH$ totally-symmetric (totally-antisymmetric) with respect to permutations of the two bulk spins\footnote[1]{Note that this particular choice for the initial state assumes for simplicity that the projections of the initial density matrix on the non-diagonal subspaces $\cBa$ (with $\alpha\neq \beta$) is zero. Allowing for non-zero projections on these non-diagonal subspaces would allow for further transport control.}.
%such that the projection of the initial density matrix $\rho(0)$ on the subspace of $\cBH$ corresponding to a totally symmetric bulk is fixed and equal to $\varphi \in [0,1]$. The formal expression of the initial density matrix can be expressed as $\rho(0)=\varphi \rho^{\text{sym}}+ (1-\varphi)\rho^{\text{ant}}$, being $\rho^\text{sym}$ and $\rho^\text{ant}$ the totally symmetric and antisymmetric density matrices. 
In this case, one can easily show that the flowing current is just $\la q \ra= \varphi \la q_+\ra$, with $\la q_+\ra$ the average current of the totally symmetric NESS $\rho^\text{NESS}_{\alpha_\text{max}}$, see Eq. (\ref{eq:qmed}). Of course this is so because $\la q_-\ra = 0$ due to the dynamical decoupling between terminal qubits produced by the frozen, dark state of the antisymmetric bulk. The right panel in Fig. \ref{fig:current_control} shows this current as a function of $\varphi$ for different values of the exciton pumping rate $a_1$ at the first reservoir, demonstrating that the combination of a simple network topology (see inset) with our symmetry results leads to a \emph{symmetry-controlled quantum thermal switch}, where the exciton current flowing between hot and cold reservoirs can be completely blocked, modulated or turned on by just tuning the symmetry of the initial state. Moreover, a nonlinear control of the heat current can be obtained by introducing a weighted interaction between terminal qubits. We will discuss in Section \S\ref{sec:switch} below an experimental realization of this symmetry-controlled quantum switch using a pair of cold $\Lambda$-atoms in an optical cavity, see also Ref. \cite{manzano16a}.

\subsection{Role of dephasing noise}
\label{sec:QNdeph}

To continue our discussion of open quantum networks we note that, as explained in Section \S\ref{sec:LDF4}, the presence of the symmetry-induced twin dynamic phase transitions in the current statistics is fragile against environmental decoherence. In particular, any finite amount of dephasing noise in the system can violate the network permutation symmetry, thus leading to a unique steady-state according to our discussion in \S\ref{sec:LDF4}. This results in a smooth current LDF with no kinks and hence no dynamic phase transition. The interaction with a dephasing environment, that reduces the quantum coherent character of the system at hand, has been probed very important for the energy transfer in different nonequilibrium quantum networks, where noise-enhanced transport has been recently reported \cite{manzano13a,witt13a}.

We briefly analyze this idea now in quantum networks by including a dephasing channel that reduces the coherent character of the transport in the system \cite{gardiner00a,manzano14a}. This dephasing channel is modelled by adding a new set of Lindblad operators to the deformed Liouville-Lindblad superoperator (\ref{eq:mod_liou}) of the form
\be
L_j^{\text{deph}}=\sqrt{\gamma} \sigma_j^+\sigma_j^-,
\ee
with $j$ running over all the spins in the network, and $\gamma$ being the dephasing parameter. Since the dephasing operators act locally on each qubit, it is straightforward to prove that  the symmetry operator $\pi_{ij}$ of bulk spins does not commute with all Lindblad operators as $\cor{\pi_{ij},L_{i}^{\text{deph}}}\neq0$ as well as $\cor{\pi_{ij},L_{j}^{\text{deph}}}\neq0$. This breaks the symmetry-induced multiplicity of steady states, see \S\ref{sec:sym2} above, with the new evolution equation now mixing the original symmetry eigenspaces. In the absence of any other symmetries, and provided that Evans theorem holds \cite{evans77a},  one can then show that the system eventually forgets the information on the initial state, converging to a unique NESS.
%The new evolution equation hence mixes the original symmetry eigenspaces, thus leading to a unique NESS, independent of the initial steady state. 
%Because of that, there are no multiple steady-states anymore and the system eventually forgets the information on the initial state, converging to a unique NESS. 
This violation of the permutation symmetry also implies immediately the disappearance of the twin dynamic phase transitions. This is shown in Fig. \ref{fig:LDF_dephasing}, where the LDF $\mu(\lambda)$ is plotted  for $\gamma=0.5$ and different sizes ($N=2,~4$). It is clear that $\mu(\lambda)$ is smooth and analytic for all values of $\lambda$ and that the dynamic phase transition vanishes whenever the system is subject to dynamical noise.

%\begin{figure}
%\centerline{\includegraphics[width=10cm]{plots/3-qubit-pigeon-v2.png}}
%\caption{Sketch of the system studied in Ref \cite{pigeon15a}. The box represents the harmonic oscillator that acts as a bath.  }
%\label{fig:3-qubits}
%\end{figure}

\begin{figure}
\centerline{\includegraphics[width=16cm]{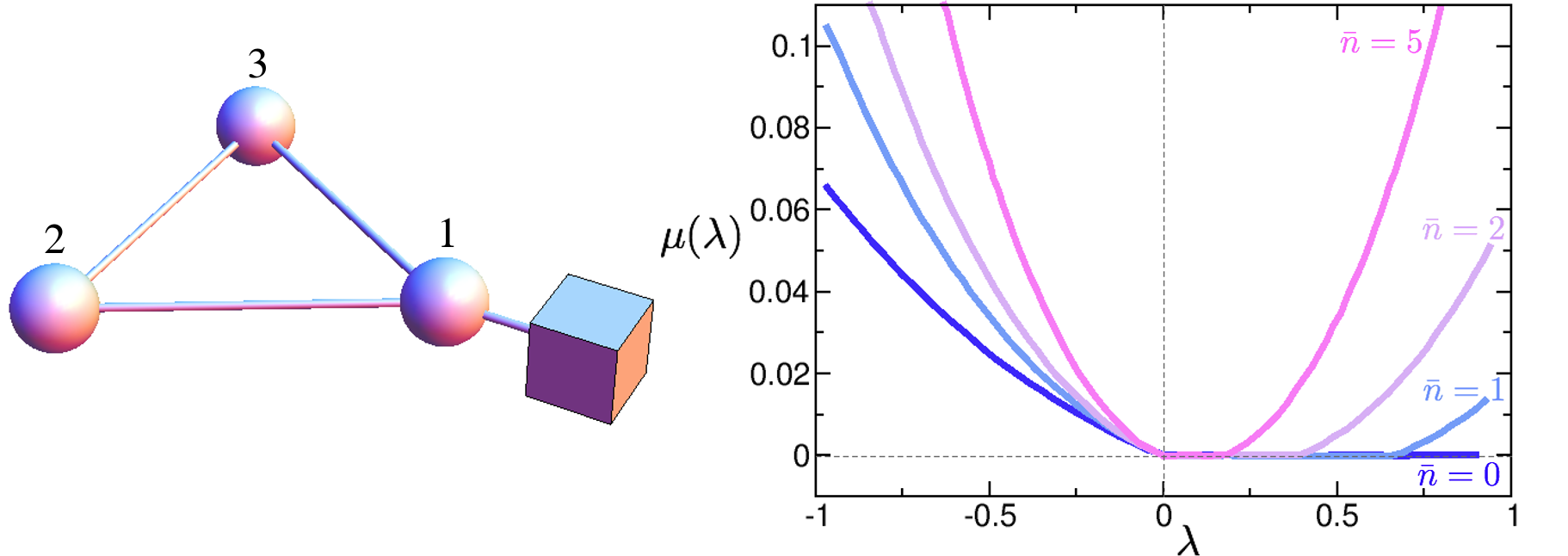}}
\caption{Left panel: Sketch of the system studied in Ref \cite{pigeon15a}. The box represents the harmonic oscillator that acts as a bath on the first spin of the network. Right panel: Figure 2 from Ref. \cite{pigeon15a}, LDF $\mu(\lambda)$, corresponding to the Legendre transform of the current LDF, as a function of the counting field $\lambda$ for the three-spin network of left panel, and for different average number of excitations in the bath $\bar{n}$. Note the two kinks that appear in $\mu(\lambda)$ which reflect the twin dynamic phase transitions expected for current statistics in this model due to the presence of a symmetry under the exchange of spins 2 and 3.}
\label{fig:3-qubits}
\end{figure}

\subsection{A three-spin network with collective dissipation}

To end this section, we note that in Ref. \cite{pigeon15a} a specific three-spin network as the one displayed in the left panel of Fig. \ref{fig:3-qubits} has been studied. The proposed model, which exhibits a symmetry opening the door to transport control, can in principle be engineered in hybrid electro/opto-mechanical settings, which makes this model particularly interesting for practical purposes. The specific spin Hamiltonian for this system corresponds to a spin-$1/2$ ring under the effect of two magnetic fields $\alpha$ and $\beta$ \cite{pigeon15a}
\be
H^{\text{3-spins}}=\alpha \sum_{i=1}^3\sigma_x^i - \beta \sum_{i=1}^3 \sigma^z_i + \pare {\sigma^x_1 \sigma^x_2 + \sigma^x_2 \sigma^x_3 +\sigma^x_1 \sigma^x_3 + H.c. },
\ee
with $\beta\ll\alpha$. Additionally, spin 1 interacts with a harmonic oscillator that acts as a reservoir, see left panel in Fig. \ref{fig:3-qubits}. In fact, the motion of spin 1 due to this harmonic coupling can be adiabatically eliminated \cite{pigeon15a}, leading to an effective dissipative dynamics for the three-spin network described by a Lindblad master equation (\ref{eq:lindblad}) with jump operators 
\ben
L_1=\sqrt{\Gamma \pare{\bar{n}+1} }~\sigma^-_1 \, , \nonumber \\
L_2=\sqrt{\Gamma \bar{n}}~\sigma^+_1 \, ,\nonumber
\een
being $\Gamma$ the coupling strength between spin 1 and the oscillator bath, and $\bar{n}$ the average number of excitations in the reservoir. In Ref. \cite{pigeon15a} this system was reported to exhibit coexistence of two dynamical phases with different activity levels, and the symmetry perspective adopted in this paper easily explains this remarkable behavior, as this spin network clearly exhibits a symmetry under the exchange of spins 2 and 3. Indeed, one can define an unitary operator $\pi=\exp{[\text{i} (\op{2}{3}+\op{3}{2}})]$ associated to this exchange symmetry, and this operator fulfils $\cor{\pi,H}=\cor{\pi,L_1}=\cor{\pi,L_2}=0$, hence defining a symmetry of the dynamics. As described in previous sections, the presence of this symmetry leads to a pair of twin dynamic phase transitions which appear as a non-analytic (kink) behavior of its LDF $\mu(\lambda)$, see Eq. (\ref{eq:mul}). This is exactly what was measured in Ref. \cite{pigeon15a}, see right panel in Fig. \ref{fig:3-qubits}, were the LDF $\mu(\lambda)$ is displayed for different average number of excitations in the bath $\bar{n}$, and the emergence of twin kinks is apparent in all cases. The experimental verification of these results seems feasible in hybrid electro/opto-mechanical settings where this system can be prepared, and it would further support the use of symmetry as a resource to control quantum transport.

%\begin{figure}
%\centerline{\includegraphics[width=10cm]{plots/pigeon_fig1.pdf}}
%\caption{CHANGE CAPTION!! Sketch of the system studied in Ref \cite{pigeon15a}. The box represents the harmonic oscillator that acts as a bath.  }
%\label{fig:pigeon1}
%\end{figure}

\newpage

\section{An atomic switch controlled by symmetry}
\label{sec:switch}

We have seen in previous sections how symmetry can be harnessed to control energy transport in generic nonequilibrium open quantum systems. In particular, we have proposed in Section \S\ref{sec:QNcontrol} a schematic model of symmetry-controlled quantum thermal switch, i.e. a system capable of modulating at will the exciton current flowing from a hot to a cold reservoir via initial state preparation. In this section we explore further this idea and review an experimental setup proposed in Ref. \cite{manzano16a} where such symmetry-enabled control of transport can be realized in detail.

\begin{figure}[b]
\bc
\includegraphics[width=13cm]{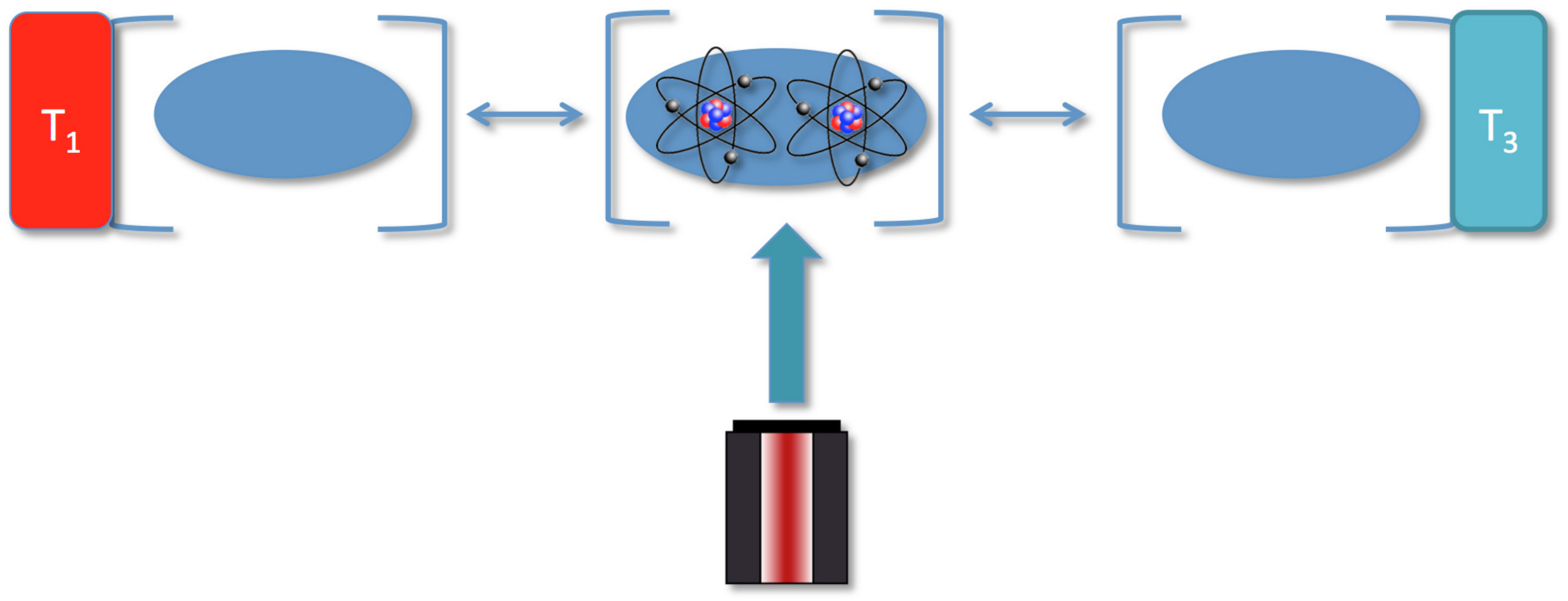}
\caption{Sketch of an atomic optical switch controlled by symmetry. It is composed by three different optical cavities coupled in a linear topology. The terminal cavities interact locally with thermal baths at temperatures $T_1\ne T_3$ which drive the linear cavity array out of equilibrium, while the middle cavity is doped with two identical atoms that act as a symmetry-controlled switch for transport.}
\label{fig:switch}
\ec
\end{figure}

The idea is simple and based on a recent proposal to use macroscopic jumps for robust entangled-state preparation \cite{metz06a,metz07a}. In particular the setup consists in three optical cavities coupled in linear topology, as sketched in Fig. \ref{fig:switch}. Due to cavity leakage, there is a coherent hopping of photons between neighboring cavities. The two cavities at the edges are locally coupled to thermal baths at different temperatures, $T_1\ne T_3$. This coupling drives the system out of equilibrium, so in the long time limit the system is expected to reach a nonequilibrium steady state characterized by a net photon current flowing through the cavity array, related to the external gradient imposed by the thermal reserviors \cite{dhar08a,garrido01a,garrido02b,manzano12a,manzano13a}. In addition, the central cavity is doped with two identical $\Lambda$-atoms that are externaly driven by a laser. We assume that the laser and the cavity act equally on both atoms, so all coupling constants (described below) are the same for both atoms. 

As shown in \cite {metz06a,metz07a}, the internal state of the two atoms can switch between symmetric and antisymmetric manifolds due to the laser interaction. When the atoms are in the antisymmetric (maximally-entangled) state they form a dark state and the cavity photons do not interact with them. As we will show below, this effect allows to control deterministically the photon current up to four orders of magnitude \cite{manzano16a}.

\subsection{Master equation for the laser-controlled cavity-atom system}
\label{sec:switchQM}

We start by describing the dynamics of the cavity-atom array. The system evolution is described by a master equation of Lindblad-type 
\be
\dot{\rho}(t) = -\ii \cor{H^{\text{SW}},\rho} + \sum_{k=1}^4 \cL_k^\text{at}\rho + \sum_{b=1,3} \left(\cL_{b,+}^\text{th} + \cL_{b,-}^\text{th}\right) \rho \, 
\label{lindbladAT0}
\ee
where $H^{\text{SW}}$ is the system Hamiltonian, while $\cL_k^\text{at}$ and $\cL_{b,\pm}^\text{th}$
%$\cL \rho \equiv L\rho L^\dagger - \frac{1}{2}\acor{L^\dagger L,\rho}$ 
are dissipators\footnote[1]{Note that in this case the Hamiltonian $H$ and the total Liouvillian $\cL$ are not bounded as they involve bosonic operators.}  which describe the interactions of the system with the different incoherent channels. In particular, there are four incoherent channels $\cL_k^\text{at}$, $k\in[1,4]$, related to the spontaneous decay of atomic states, and two incoherent channels $\cL_{b,\pm}^\text{th}$ for each thermal bath ($b=1,3$) which describe photon pumping ($+$) and extraction ($-$) at the corresponding terminal cavity.
 
The Hamiltonian of the full system, once in the interaction picture and using the rotating wave approximation described in Section \S\ref{sec:subspaces}, can be decomposed as
\be
H^{\text{SW}} =  H_\text{hop} + H_\text{ctrl}\, ,
\ee
where $H_\text{hop}$ describes the coherent hopping of photons between coupled cavities due to leakage effects, while $H_\text{ctrl}$ describes the interaction of the atoms in the central cavity with the control laser fields. The hopping Hamiltonian reads
\be
H_\text{hop} = J~\pare{ a_1^{\dagger} a_2^{\phantom{\dagger}} + a_2^{\dagger} a_3^{\phantom{\dagger}} + H.c.}  \, ,
\ee
where $J$ is the hopping coupling constant between neighboring cavities, $a_k$ ($a_k^{\dagger}$) are the destruction (creation) bosonic operators acting on cavity $k=1,~2,~3$ (with $k=2$ indicating the central, atom-doped cavity), and $H.c.$ stands for \emph{Hermitian conjugate}.
\begin{figure}[t]
\bc
\includegraphics[width=7cm]{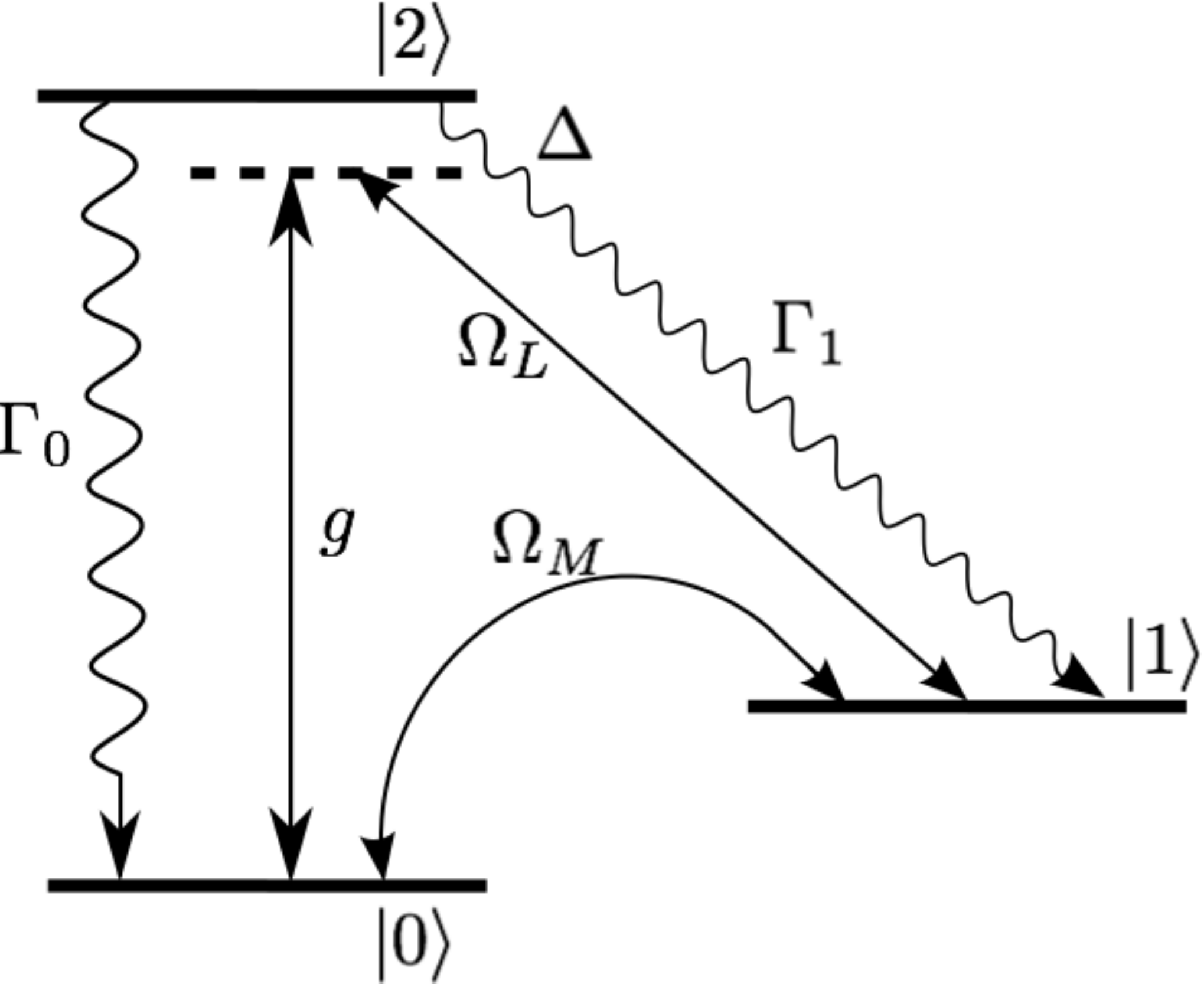} \hspace{1cm}
\includegraphics[width=5.4cm]{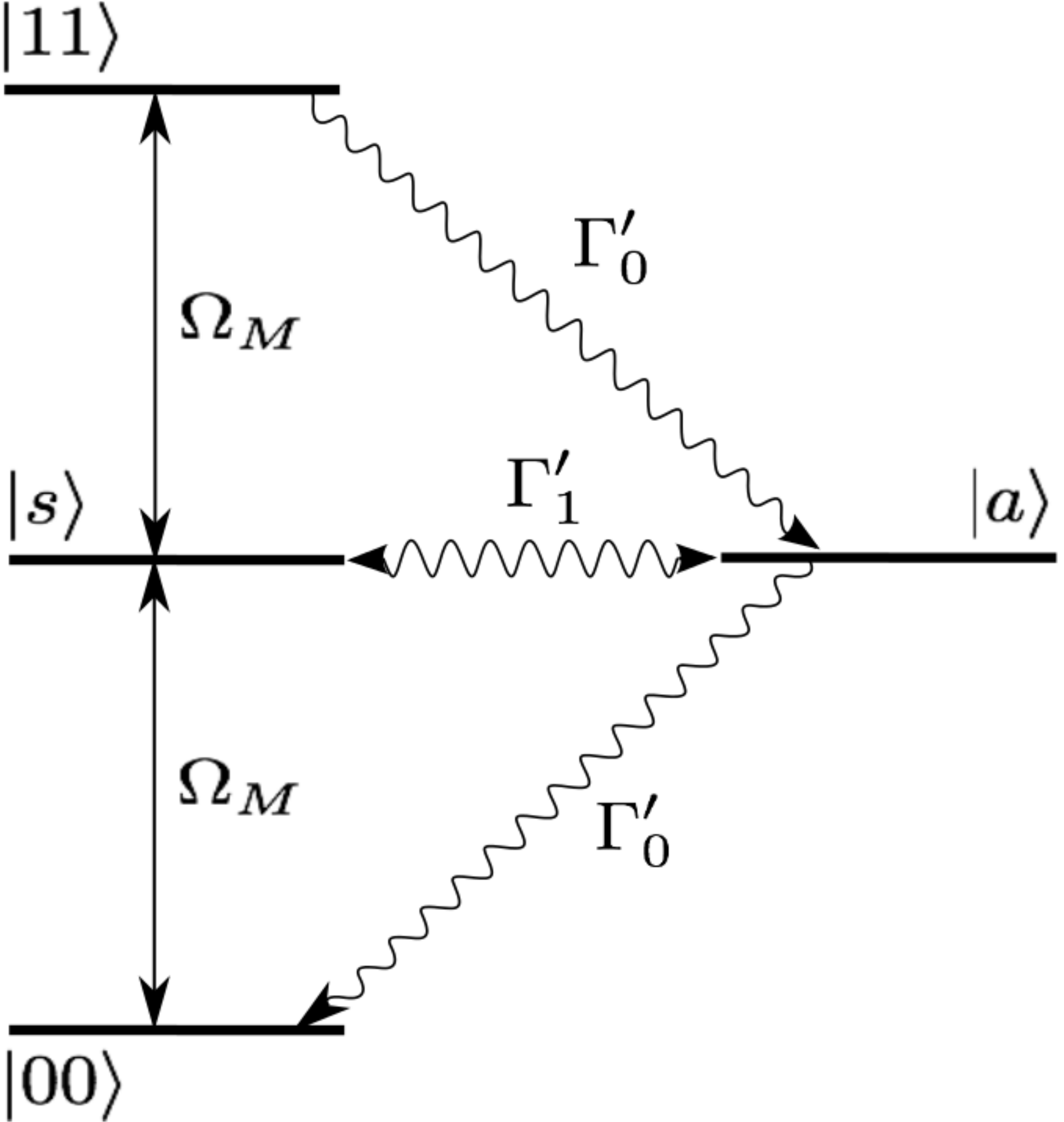}
\caption{Left: Energy level diagram of each of the three-level $\Lambda$-atoms. Right: Simplified energy level diagram of the resulting four-level system after the adiabatic elimination of the fast excited states. Straight (wavy) arrows represent coherent (incoherent) transitions. For simplicity, this panel shows only incoherent transitions mixing the symmetric manifold $\left( \ket{11}, \ket{s}, \ket{00} \right)$ and the antisymmetric one $(\ket{a})$. Other incoherent transitions are present as it is clear from direct inspection of Eqs. (\ref{ref:lindbladAT1}) and (\ref{ref:lindbladAT2}). See Refs. \cite{metz06a,metz07a,manzano16a} for a more detailed explanation.}
\label{fig:levels}
\ec
\end{figure}

To better understand the atom-laser interaction in the central cavity, $H_\text{ctrl}$, we show in the left panel of Fig. \ref{fig:levels} the typical energy level diagram of a $\Lambda$-atom in the presence of laser (light) fields. The transition between states $\ket{0} \leftrightarrow \ket{2}$ is coupled to the cavity photon field with a coupling strength $g$. On the other hand, the transitions $\ket{2} \leftrightarrow \ket{1}$ and $\ket{0} \leftrightarrow \ket{1}$ are driven by laser fields with Rabi frequency $\Omega_L$ and $\Omega_M$, respectively. Finally, the spontaneous (incoherent) decay from level $\ket{2}$ has a total rate $\Gamma=\Gamma_0 + \Gamma_1$ (with $\Gamma_0$ and $\Gamma_1$ the rates for the spontaneous decays $\ket{2} \leadsto \ket{0}$ and $\ket{2} \leadsto \ket{1}$), and there is also a laser detuning $\Delta$. The resulting control Hamiltonian reads
\be
H_\text{ctrl} = \sum_{i=1}^2 \pare{\frac{\Omega_L}{2}\ket{1}_i \bra{2}_{i} + \frac{\Omega_M}{2}\ket{0}_i \bra{1}_i + H.c.  }
+ \sum_{i=1}^2 g\pare{ \ket{0}_i \bra{2}_i  a_2^{\dagger} + H.c. } + \Delta\ket{2}_i\bra{2}_i \, ,
\ee
where the sum over $i=1,2$ refers to the two atoms in the central cavity. 

In the particular limit where the excited atomic states are far off-resonant
\be
\Omega_M<g,\Gamma,\Omega_L<<\Delta \, ,
\ee
it can be shown that these excited states with population in level $\ket{2}$ evolve much faster than any other state in the atom \cite{metz06a,metz07a}, meaning that the state $\ket{2}$ can be \emph{adiabatically eliminated} and the two-atoms system can be described by a vector in a 4-dimensional spece (i.e. as a four-level system), as shown in the right panel of Fig \ref{fig:levels}. In particular, since the interaction of both atoms with the laser fields is the same, it is most convenient to work in the Bell basis of the joint Hilbert space. This basis naturally splits into two orthogonal subspaces, one \emph{fully symmetric} including the state $\ket{s}=\frac{1}{\sqrt{2}} \pare{\ket{01}+\ket{10}}$ as well as $\ket{00}$ and $\ket{11}$, and another \emph{antisymmetric} subspace consisting in a singlet state $\ket{a}=\frac{1}{\sqrt{2}} \pare{\ket{01}-\ket{10}}$. The atoms-laser (control) Hamiltonian hence becomes\footnote[1]{For a detailed presentation of the adiabatic elimination technique, we refer the interested reader to Ref. \cite{metz07a} and references therein.} 
%(see \cite{metz07a} for a detailed derivation)
\ben
\hspace{-1cm} H_\text{ctrl}&=&\frac{\Omega_M}{\sqrt{2}} \Big( \op{00}{s} +\op{s}{11} + H.c. \Big) + g' \Big( \op{00}{s} a_2^{\dagger}  + \op{s}{11} a_2^{\dagger} + H.c. \Big)  \nonumber\\
&-&\Delta'  \Big( \op{00}{00} -\op{11}{11} \Big) \, , 
\een
with $\Delta'=-\frac{g^2}{\Delta} a_2^\dagger a_2 - \frac{\Omega_L^2}{4\Delta}$ and $g'=-\frac{\Omega_L g}{\sqrt{2}\Delta}$. 

Interestingly, the transition between the symmetric and antisymmetric manifolds of the two-atoms system occur as a result of two different incoherent decay channels with jump operators\footnote[2]{Recall that a dissipator $\cL$ acts of an arbitrary density matrix $\rho$ as $\cL \rho \equiv L\rho L^\dagger - \frac{1}{2}\acor{L^\dagger L,\rho}$, where $L$ are the associated jump operators.} \cite{metz06a,metz07a,manzano16a}
\be
L_1^\text{at}=\sqrt{ \Gamma'_0}  \Big( \op{00}{a} -\op{a}{11}\Big) \, ,  \qquad L_2^\text{at}=\sqrt{ \frac{\Gamma'_1}{2}  } \Big( \op{s}{a} + \op{a}{s}\Big) \, ,
\label{ref:lindbladAT1}
\ee
while there exists another two decay channels which do not mix the symmetric and antisymmetric manifolds, namely \cite{metz06a,metz07a,manzano16a}
\be
L_3^\text{at}= \sqrt{ \Gamma'_0  }  \Big( \op{00}{s} +\op{s}{11}\Big) \, , \qquad L_4^\text{at}=\sqrt{ \frac{\Gamma'_1}{2}  } \Big( \op{a}{a} + \op{s}{s} + 2 \op{11}{11}\Big) \, .
\label{ref:lindbladAT2}
\ee
The effective rates appearing in the previous equations for the spontaneous decay between the symmetric and antisymmetric manifolds are 
\be
\Gamma'_{\ell}=\frac{\Omega_L^2 \Gamma_\ell}{4 \Delta^2} \, ,  \label{Geff}
\ee
see also Fig.  \ref{fig:levels} (right). Remarkably, these effective rates are in all cases proportional to $\Omega_L^2$, with $\Omega_L$ the Rabi frequency of one of the lasers, so mixing between the symmetric and antisymmetric atomic subspaces is only possible whenever $\Omega_L \ne 0$. Hence, switching on and off the laser with Rabi frequency $\Omega_L$ enables us to externally manipulate the mixing between the symmetric and antisymmetric manifolds and therefore control the transport properties of the atom-cavity array, as explained below.

Finally, the coupling of the terminal optical cavities ($b=1,3$) to the different thermal baths is captured by the following Lindblad operators
\ben
L_{1,-}^{\text{th}} &=&\sqrt{\Gamma_{\text{th}} (\la n_1\ra+1)} a_1, \qquad L_{1,+}^{\text{th}}=\sqrt{\Gamma_{\text{th}}\la n_1\ra} a^{\dagger}_1 \nonumber \\
L_{3,-}^{\text{th}} &=&\sqrt{\Gamma_{\text{th}} (\la n_3\ra+1)} a_3, \qquad L_{3,+}^{\text{th}}=\sqrt{\Gamma_{\text{th}}\la n_3\ra} a^{\dagger}_3,
\een
with $\Gamma_{\text{th}}$ the coupling constant between the terminal cavities and the corresponding thermal bath, and $\la n_b\ra$ is the average excitation number at the cavity resonance frequency $\omega$ in each bosonic bath, $\la n_b\ra = [\exp(\frac{\omega}{k_B T_b})-1]^{-1}$ with $b=1,3$.

\subsection{Photon current statistics and laser control}
\label{sec:switchC}

Interestingly, the laser-controlled atom-cavity array described in the previous section exhibits a \emph{symmetry} in the language of Section \S\ref{sec:sym2} above. Indeed, it can be easily proved that the following operator $\pi$, acting on the atoms Hilbert space and defined in the Bell basis as 
\be
\pi \equiv \op{00}{00}+\op{s}{s}+\op{11}{11}-\op{a}{a} \, ,
\ee
commutes with all the elements entering the system dynamics, Eq. (\ref{lindbladAT0}), \emph{whenever} $\Omega_L=0$, see also Eq. (\ref{Geff}). Note that the physical effect of the operator $\pi$ is to exchange the state of the two atoms, i.e. $\pi \ket{\alpha} = \ket{\alpha}$ for $\ket{\alpha} = \ket{00},\ket{11},\ket{s}$ while $\pi \ket{a} = -\ket{a}$. 

In particular one can easily check that, for arbitrary values of the lasers Rabi frequencies, the commutators obey
\be
\cor{\pi,H} = \cor{\pi,L_k^\text{at}}= \cor{\pi,L_{b,\pm}^\text{th}} = 0 \, \qquad  \forall k=3,4, \quad \forall b=1,3 \, , \label{commu1}
\ee
while
\be
\cor{\pi,L_1^\text{at}} = \sqrt{\frac{2\Gamma_0}{\Gamma_1}} L_2^\text{at} \ne 0 \, , \qquad \cor{\pi,L_2^\text{at}} = \sqrt{\frac{\Gamma_1}{2 \Gamma_0}} L_1^\text{at}  \ne 0 \, . \label{commu2}
\ee
Therefore the operator $\pi$ is not in general a symmetry of the laser-controlled atom-cavity array. However, by switching off laser $L$ we in fact fix $\Omega_L= 0$, meaning that $\Gamma'_{0,1} = 0$, see Eq. (\ref{Geff}), and hence $L_k^\text{at}=0$ $\forall k\in[1,4]$. In this case all commutators (\ref{commu1})--(\ref{commu2}) vanish, turning $\pi$ into a symmetry.

\begin{figure}[t]
\bc
\includegraphics[width=7cm]{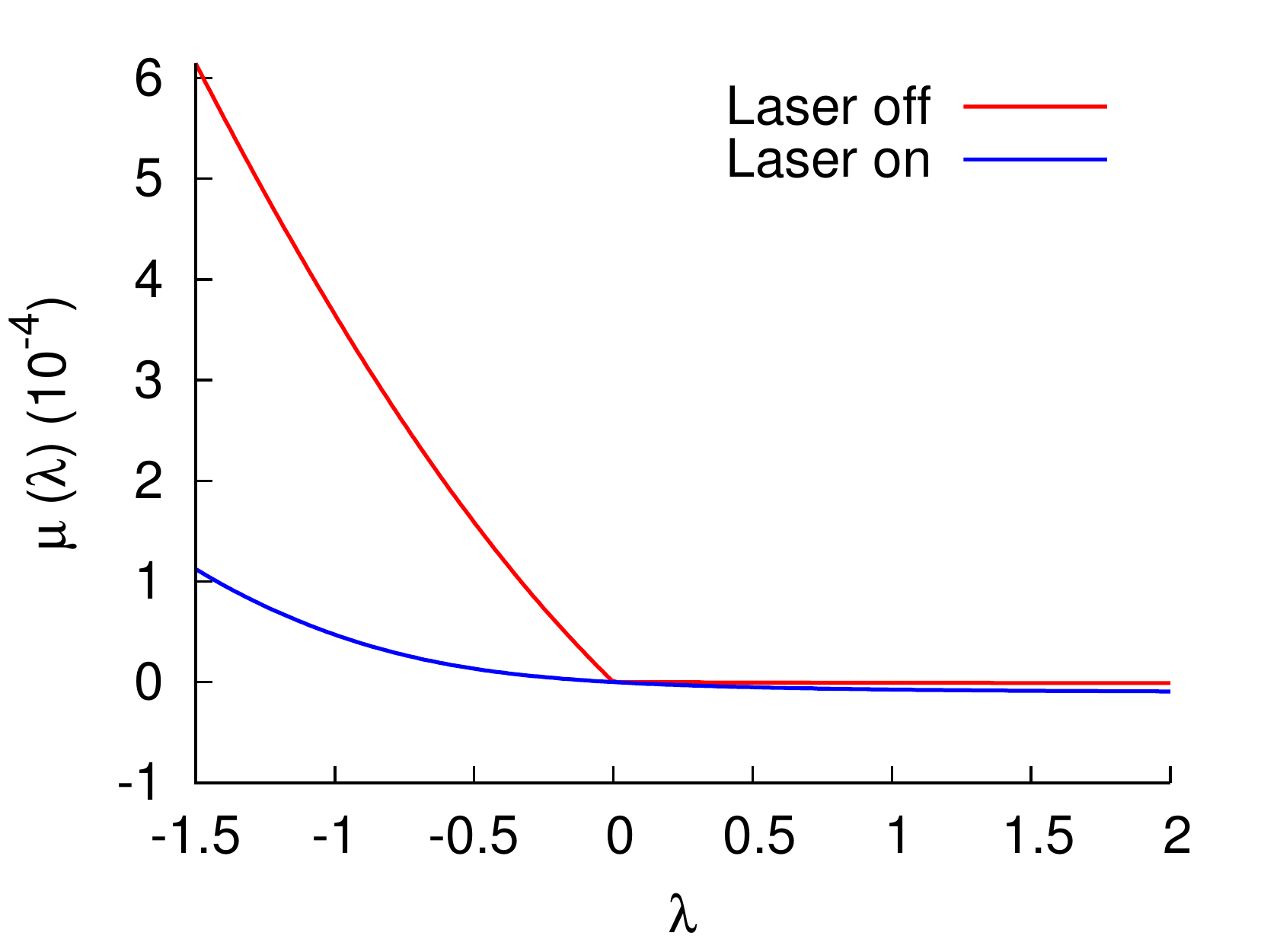}
\includegraphics[width=7cm]{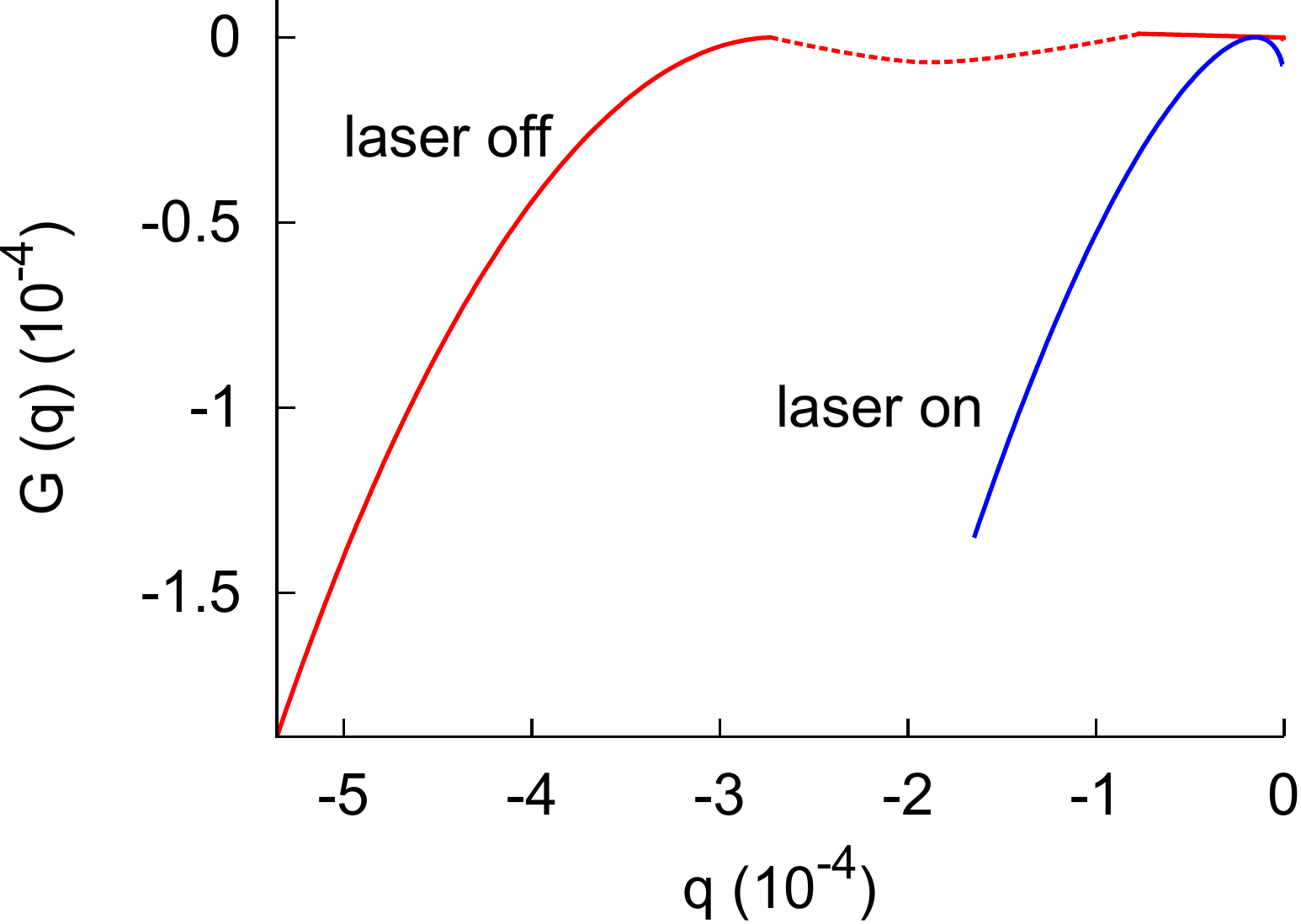}
\caption{Cumulant generating function of the photon current distribution, $\mu(\lambda)$ (left), and current large deviation function, $G(q)$ (right) for the laser-controlled atom-cavity array.  In both panels the red line represents the corresponding function when the control lasers are off ($\Omega_R= \Omega_L=0$), while blue lines correspond to control lasers on ($\Omega_R=0.005 g,\; \Omega_L=g$). The remaining parameters are fixed to: $\Delta=75g, \; \Gamma_0=\Gamma_1,\; \Gamma=\Gamma_0+\Gamma_1=g,\; J=10^{-3}g, \; \la n_1\ra=0.005,\; \la n_2\ra=10^{-6}$. The dashed red line in the right panel signals the non-convex (or at least affine) region in the current LDF associated to the kink in $\mu(\lambda)$, see also Section \S\ref{sec:LDF4}. No information on $G(q)$ can be derived from $\mu(\lambda)$ in this current interval. See also Fig. 2 in Ref. \cite{manzano16a}.}
\label{fig:LDF_switch}
\ec
\end{figure}

As explained in previous sections, the presence of a symmetry leads to the appearance of \emph{multiple steady state}, corresponding to different transport channels associated to each one of the different symmetry sectors present in the initial state. Furthermore, the symmetry induces a pair of twin dynamic phase transitions in the current statistics, which show up as two kinks in the cumulant generating function of the current distribution, $\mu(\lambda)$, or equivalently as a pair of non-convex (or at least affine) regions in the current large deviation function $G(q)$, see Section \S\ref{sec:LDF4}. These effects can be readily tested in the laser-controlled atom-cavity array. Fig. \ref{fig:LDF_switch} shows both $\mu(\lambda)$ (left) and $G(q)$ (right) obtained for the atom-cavity array both under the effect of laser control (blue lines) and when the lasers are switched off (red line). The LDF  $\mu(\lambda)$ has been numerically calculated as the eigenvalue with highest real part of the deformed or tilted Lindbladian $\cLl$ corresponding to the system dynamics (\ref{lindbladAT0}), which in this case reads
\be
\cL_\lambda \rho= -\ii \cor{H,\rho} + \sum_{k=1}^4 \cL_k^\text{at}\rho +  \left(\cL_{1,+}^\text{th} + \cL_{1,-}^\text{th}\right) \rho 
+ \left( \tilde{\cL}_{3,+}^\text{th}(\lambda) + \tilde{\cL}_{3,-}^\text{th}(\lambda)\right) \rho,
\ee
where we have defined $\tilde{\cL}^\text{th}_{3,\pm}(\lambda)\equiv e^{\mp \lambda} L_{3,\pm}^{\text{th}} \rho L_{3,\pm}^{\text{th} \dagger}-\frac{1}{2}\key{L_{3,\pm}^{\text{th} \dagger} L_{3,\pm}^{\text{th}}, \rho}$, see Eq. (\ref{eq:rhol}) and \S\ref{sec:LDF2.5}. 
%see Eq. (\ref{eq:masterrhol}) and \S\ref{sec:LDF}. 
On the other hand, the LDF $G(q)$ is calculated by numerically Legendre-transforming $\mu(\lambda)$ following the inverse of Eq. (\ref{eq:mu}). As clearly shown in Fig. \ref{fig:LDF_switch}, the photon current statistics is strongly affected by the presence of the lasers, and their absence leads to a distinct non-analyticity in $\mu(\lambda)$ and a non-convex $G(q)$, as expected. Note also that, interestingly, switching on the control lasers strongly suppresses current fluctuations, as reflected in as much narrower $G(q)$, see right panel in Fig. \ref{fig:LDF_switch}.

The presence of a symmetry and the associated coexistence of multiple transport channels also enable the direct control of the photon current through the atom-cavity array, see the general results of Section \S\ref{sec:LDF3}. The control range for the average current is determined by the typical currents in the maximal and minimal current phases, i.e. the maximum and minimum currents in the system, see Section \S\ref{sec:LDF3} and our discussion on the symmetry-controled thermal switch in Section \S\ref{sec:QNcontrol}. These currents can be obtained from the cumulant generating function as
\be
\mean{q}_{\text{max}} = - \lim_{\lambda \to 0^-} \partial_\lambda \mu(\lambda) \, , \qquad \mean{q}_{\text{min}} = - \lim_{\lambda \to 0^+} \partial_\lambda \mu(\lambda) \, ,
\ee 
see Eq. (\ref{eq:qq2}) in \S\ref{sec:LDF3} and Fig. \ref{fig:ldf1} (left) above. 

To control effectively photon transport in the system one would start an experiment with the control lasers on, a situation in which the atom-cavity array exhibits a unique steady state. As demonstrated in \cite{metz06a,metz07a}, in this state the central atom-doped cavity exhibits a random telegraph, blinking fluorescence signal between bright and dark periods (i.e. periods of intense fluorescence interrupted by intervals with no emitted photons at all). Such blinking is associated to the quantum incoherent jumps between the symmetric (bright) and antisymmetric (dark) manifolds of the two-atoms system, a mixing enabled by the presence of the laser fields. Now, switching off the laser control field with Rabi frequency $\Omega_L$ during a dark (bright) period leaves the two-atoms system entrained into the antisymmetric (symmetric) subspace, since these manifolds do not mix in the absence of the control laser field, and this allows us to control at will the photon current flowing through the atom-cavity array by turning off the control laser at the right moment. In this way the current can be modulated between $\mean{q}_{\text{max}}$ in bright periods and $\mean{q}_{\text{min}}$ during dark intervals.

\begin{figure}[t]
\bc
\includegraphics[width=10cm]{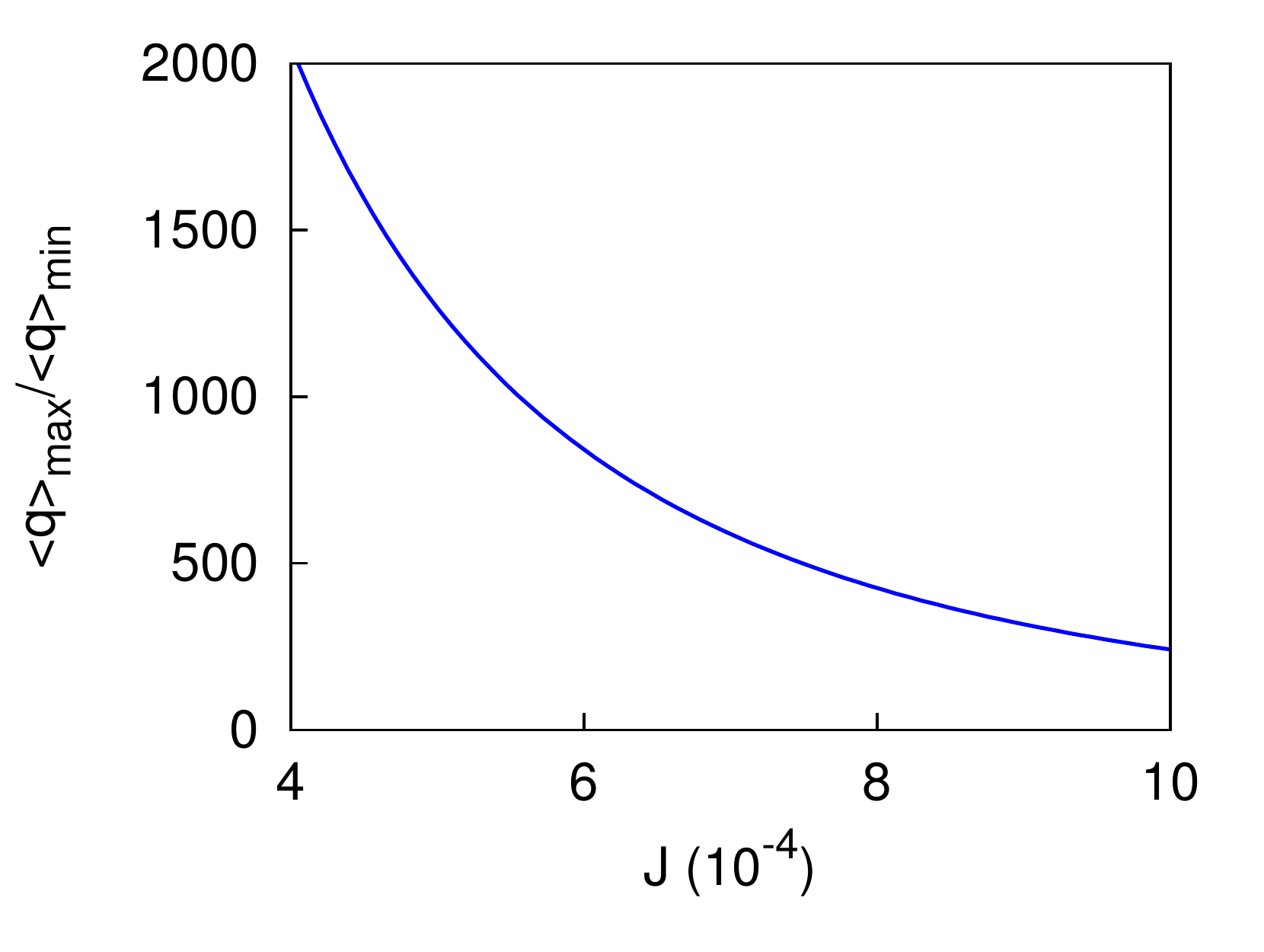}
\vspace{-0.5cm}
\caption{Ratio between the maximum and minimum currents, associated respectively to the symmetric and antisymmetric subspaces of the two-atoms system when the control laser is off ($\Omega_L=0$), as a function of the hopping coupling constant $J$ between neighboring optical cavities. All other parameters are as in Fig. \ref{fig:LDF_switch}. See also Fig. 3 (left) in Ref. \cite{manzano16a}.}
\label{fig:ratio}
\ec
\end{figure}

The control capacity depends on the hopping coupling constant $J$ between neighboring optical cavities. In the large-$J$ limit, photons can hop between coupled cavities easily and the effect of the interaction with the atoms doping the central cavity is weak. On the other hand, for the small values of $J$ typical of realistic (i.e. experimentally relevant) conditions, the control capacity increases steeply and the ratio between the maximum and minimum photon current, $\mean{q}_{\text{max}}/\mean{q}_{\text{min}}$, can grow up to four orders of magnitude. This is shown in Fig. \ref{fig:ratio} for the same parameters as in Fig. \ref{fig:LDF_switch}.

In summary, we have shown in this section how a simple diatomic system trapped inside an optical cavity and subject to two laser fields can be harnessed to control photon transport (and hence energy flow) across a linear array of optical cavities under a temperature gradient. This control mechanism relies on the symmetry of atomic states and how it is affected by the driving laser fields, and is a experimentally-feasible example of the symmetry-controlled quantum thermal switch proposed in Section \S\ref{sec:QNcontrol}.

\newpage

\section{Signatures of molecular symmetries at the dynamical level}
\label{sec:transient}

We have seen in previous sections how the presence of a symmetry-breaking element (as e.g. decohering noise) in an otherwise symmetric open quantum system strongly affects its transport properties. In particular, while a symmetric open quantum system exhibits multiple, coexisting steady states and a twin dynamic phase transition in its transport properties, whenever the original symmetry is externally broken these exotic effects disappear and \emph{standard} behavior is recovered, i.e. a unique steady state with a well-defined current. This simple observation leads to the possibility of detecting inherent molecular symmetries by studying the change in transport properties under the action of a localized external probe (a \emph{B\"uttiker probe} \cite{buttiker86a}) which acts as a symmetry-breaking perturbation \cite{li14a} on the molecular dynamics, as recently proposed by Thingna \emph{et al} in Ref. \cite{thingna16a}.

\begin{figure}
\centerline{\includegraphics[width=16cm]{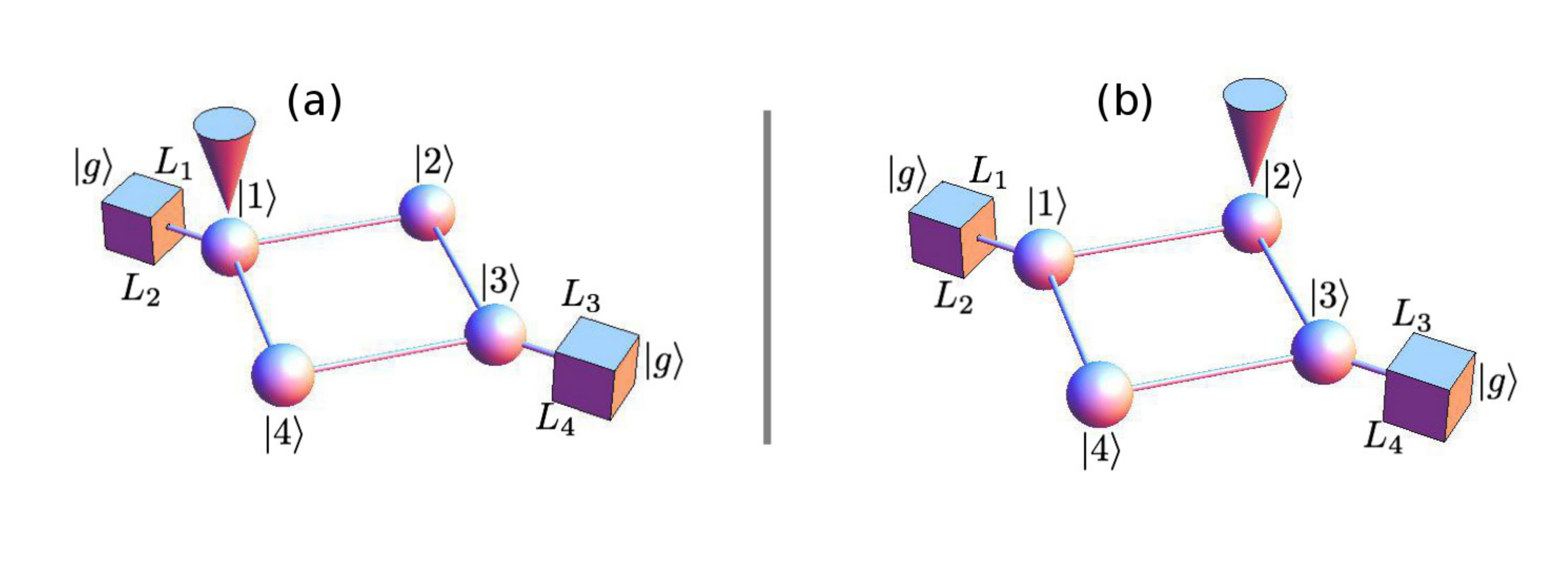}}
\caption{Sketch of the 4-site toy model used to study the detection of molecular symmetries with the help of a B\"uttiker probe. The different atomic sites (spheres) interact with their nearest neighbors (links), and possibly with an external probe (depicted here as a cone). In addition the system is connected to two excitonic reservoirs (cubes) working at different temperatures. Two probe configurations, one acting on site 1 (left) and another on site 2 (right), are depicted. Note that the second configuration breaks the internal molecular symmetry under the exchange of sites 2 and 4, hence affecting molecular transport properties. Sketch inspired by Fig. 1 of Ref. \cite{thingna16a}.
}
\label{fig:4-level}
\end{figure}

We now discuss this possibility in detail using a simple toy model, a four-site molecular system as the one depicted in Fig.~\ref{fig:4-level}. Later on we will also discuss briefly the application of this molecular-symmetry detection method to the more realistic case of a benzene molecule, see Fig.~\ref{fig:benzene} below.

\subsection{Detecting symmetries in a four-site toy molecule}
\label{sec8:four}

The first system is represented in Fig \ref{fig:4-level}, and it consists in 4 two-level atomic sites interacting among them according to the sketched geometry and connected to two excitonic reservoirs as depicted. We will assume hereafter that the temperature at the reservoirs is such that at most one excitation can populate the system at any time. In this \emph{single-excitation limit}, the system of interest can exhibit up to five different states, denoted here as $\key{\ket{1},\ket{2},\ket{3},\ket{4},\ket{g}}$, where $\ket{n}$ represents the state with the excitation at site $n\in[1,4]$, while $\ket{g}$ is the ground state corresponding to the absence of excitations in the molecular complex. Excitations can coherently hop from one site to a neighboring site, so the associated 4-site Hamiltonian reads
\be
H^{\text{4s}} = \epsilon \sum_{n=1}^4 \op{n}{n} + J\big( \op{1}{2}+\op{2}{3}+\op{3}{4}+\op{4}{1}+ H.c.\big) \, , \label{H4s}
\ee
with $\epsilon$ the on-site energy and $J$ the hopping coupling constant. Hamiltonians like this one are typically used within the H\"uckel theory of molecular orbitals to model atomic sites with nearest neighbour interactions \cite{streitwieser61a}, and have already been used with success to model realistic systems as e.g. the benzene molecule \cite{chen14c}. 

In addition to the coherent dynamics generated by the previous Hamiltonian, the system is driven out of equilibrium by the action of two incoherent baths at different temperatures. The dynamics of the system in the absence of external probe is then given by a  Lindblad-type master equation (\ref{eq:lindblad}) with jump operators of the form 
\ben
L_{1}=\op{g}{1} \, ,&\qquad L_{2}=\op{1}{g} \, ,\nonumber\\
L_{3}=\op{g}{3}\, , &\qquad L_{4}=\op{3}{g} \, .
\een
Here $L_1$ and $L_2$ describe the extraction and pumping of a single excitation at site 1, respectively, while $L_3$ and $L_4$ act similarly on site 3\footnote[1]{The validity  and range of applicability of Lindblad-type equations with \emph{local} jump operators has been analytically and numerically studied in detail, see e.g. Refs \cite{rivas10a,asadian13a}. We note in particular that a generic assumption of weak internal system interactions is needed in order for this coupling to remain local.}. Each of these jump operators act with a fixed rate $\Gamma_k$, $k\in[1,4]$, see below, and these rates control the overall temperature gradient. 

For the probe we choose a standard B\"uttiker description \cite{buttiker86a}. In this approach, the action of the probe is modelled by a Redfield-like term in the master equation \cite{thingna16a,redfield57a,breuer02a}. In particular, the system-probe interaction Hamiltonian from which the Redfield term derives (see Section \S\ref{sec:subspaces} above) is assumed to have a direct product form $H_{SP}=S\otimes Y$, with $S$ an operator acting on the Hilbert space of the system and $Y$ acting on the probe space. In what follows we will particularize to a probe acting on the first site ($S=S_1\equiv\op{1}{1}$) or the second one ($S=S_2\equiv \op{2}{2}$), see Fig.~\ref{fig:4-level}, though the results can be trivially extended to any other site\footnote[2]{Note that the local coupling of a probe with a molecular structure is experimentally feasible, see e.g. Ref. \cite{rai11a}.}. The dynamics of the probe itself is determined by the probe Hamiltonian, which we will choose to be that of a set of harmonic oscillators, $H_P = \sum_k \frac{p_k^2}{2m_k} + m_k (\omega_k x_k)^2$. Moreover, the interaction with the system will be given by the collective position operator $Y=-\sum_k c_k x_k$, being $c_k$ the strength of the coupling of oscillator $k$ with the system.

The reduced dynamics of the 4-site system is then given by a master equation of the form
\ben
\hspace{-2.3cm} \dot{\rho} &=& -\ii \cor{H^{\text{4s}},\rho} + \sum_{k=1}^4 \Gamma_k  \Big( L_k^{\phantom{\dagger}} \rho L_k^{\dagger} - \frac{1}{2} \big\{L_k^{\dagger} L_k^{\phantom{\dagger}},\rho \big\} \Big) + \int_0^{\infty} dt \Big(\cor{S(t),\rho\, S(t)} C(t) + H.c. \Big) \nonumber \\
\hspace{-2.3cm} &\equiv& \mathcal{L}_{LR} \rho \, ,
\label{eq:L_RL}
\een
which defines the Lindblad-Redfield Liouvillian $\mathcal{L}_{LR}$ for this monitored molecular system. Here $S(t)=e^{\ii H^\text{4s} t} S e^{-\ii H^\text{4s} t}$, and the probe information is encapsulated in the time-correlator $C(t)=\Tr_P \cor{Y(t) \, Y e^{-\beta H_P} }$, with $Y(t)=e^{\ii H_P t} Y e^{-\ii H_P t}$ and $\beta$ the inverse probe temperature. For a harmonic probe as the one described above, all parameters are determined by its spectral density. In Ref. \cite{thingna16a} an ohmic density with a Lorentz-Drude cutoff frequency $\omega_D$ and dissipation strength $\gamma$ is used  
\be
\zeta(\omega)= \pi\sum_{k=1}^{\infty} \frac{c_k^2}{2m_k \omega_k} \delta \pare{\omega - \omega_k}= \frac{\gamma \omega}{1+\pare{\omega/\omega_D}^2} \, ,  \label{eq:gamma}
\ee
allowing to write down explicitly the correlator $C(t)$, namely
\be
C(t)=\frac{1}{\pi}\int_0^{\infty} dw\, \zeta(\omega) \cor{ \coth\pare{\frac{\beta \omega}{2}} \cos \pare{\omega t} - \ii \sin\pare{\omega t}}.
\ee
Importantly, this B\"uttiker probe does not pump or extract excitations (or energy) into the system \cite{buttiker86a}, but is still capable of breaking the symmetry of the molecular complex by interferring with its coherent dynamics. Moreover, B\"uttiker probes are experimentally feasible and hence realistic from a practical point of view. The derivation of the evolution equation (\ref{eq:L_RL}) assumes a weak coupling between the molecular system and both the probe and the baths, as well as Markovian or memory-less dynamics in the baths and the probe. In addition, we have assumed that baths dynamics is fast so it does not affect their incoherent interactions with the molecular system. We stress that we do not apply this assumption to the probe dynamics, so we can study the interplay between the probe dynamics and the molecule transport properties, an interaction relevant in laboratory conditions.

This minimal molecular model may exhibit a symmetry in the language of Section \S\ref{sec:sym} due to the existence of a unitary operator $\pi=\exp [\text{i}\pare{\op{2}{4} + \op{4}{2}}] $ obeying
\be
\cor{\pi,H^{\text{4s}}}=\cor{\pi,L_i}=0 \quad \pare{\forall i} \, . 
\label{eq:sec3_commutation}
\ee
The argument in this exponential operator acts by exchanging the states of atomic sites 2 and 4, see Fig. \ref{fig:4-level}. Now, when the B\"uttiker probe acts on site 1 (Fig. \ref{fig:4-level}.a), we have that $S=S_1\equiv \op{1}{1}$ so the symmetry operator does commute with the probe operator, $\cor{\pi,S}=0$, and hence the monitored molecular complex is expected to exhibit multiple nonequilibrium steady states and different invariant subspaces (as well as a pair of twin dynamic phase transitions in its current statistics, as described previously). On the other hand, if the probe is acting on site $2$  (Fig. \ref{fig:4-level}.b) we have that $\cor{\pi,S_2}\neq 0$ so the probe-molecule interaction breaks the molecular internal symmetry, thus leading to a unique steady state. This fact can be then engineered to study the symmetry of the molecular system by analyzing its steady state and/or transient transport properties as a function of the initial state of the molecule and the probe location.

A first possibility consists in detecting underlying molecular symmetries by studying the molecule's stationary excitonic current as a function of the initial state $\rho(0)$. We have demonstrated that, whenever a symmetry is present, the average current can be modulated by controlling the projection of the initial density matrix on the maximal/minimal current subspaces. For our particular 4-site example, we may choose for the initial state
\be
\rho(0)=\rho_+ \cos^2 \theta + \rho_- \sin^2 \theta \, 
\label{rhoini4}
\ee
where $\rho_\pm = \frac{1}{2}\left[\pare{\op{2}{2}+\op{4}{4}} \pm \pare{\op{2}{4} + \op{4}{2}}\right]$ is symmetric ($+$) or antisymmetric ($-$) under the exchange $2\leftrightarrow 4$, and $\theta$ is a mixing angle that allows to control at will the initial overlap with the different symmetry sectors. One can then easily show that, when the probe acts on site 1 (symmetric case), the current varies continuously as a function of $\theta$, being maximal for $\theta=0^{\circ},180^{\circ}$ (i.e. for $\rho(0)=\rho_+$) and reaching 0 for $\theta=90^{\circ}$ (where $\rho(0)=\rho_-$). On the other hand, when the probe acts on site 2 (broken symmetry case), the steady state is unique and the average excitonic current does not depend on the mixing angle $\theta$. This dramatic variation of the molecule steady transport properties with the probe position is hence the smoking gun of an underlying molecular symmetry. 

Unfortunately, perfectly symmetric molecules are rarely feasible and realistic modelling calls for the addition of a small quenched disorder in the on-site energies. This conformational disorder adds up to the unavoidable environmental noise and other decoherence sources, from which it is difficult to isolate a typical molecule. Both noise sources (either environmental decoherence and/or conformational disorder), despite being weak, violate in principle the molecular symmetries, leading in all cases to a unique steady state in the long time limit and hence blurring the steady state signatures of the underlying (quasi-)symmetries.

As pointed out in Ref. \cite{thingna16a}, this weak violation suggests to search for \emph{dynamical} signatures of molecular symmetries, due to the expected separation of timescales between the different noise sources and the molecular symmetries. In particular, both noise sources discussed above are typically weak and hence affect the system dynamics at long times, while the underlying molecular symmetries are expected to affect dynamics on much shorter timescales.  Ref. \cite{macieszczak:prl16} describes a theory of metastability discussing the properties of Lindbladians when such separation of timescales is present.
%In Ref.  \cite{macieszczak:prl16} a theory of metastability is described discussing the properties of Lindbladians when such separation of timescales is present
%covers this phenomena 

\begin{figure}
\includegraphics[width=8.cm]{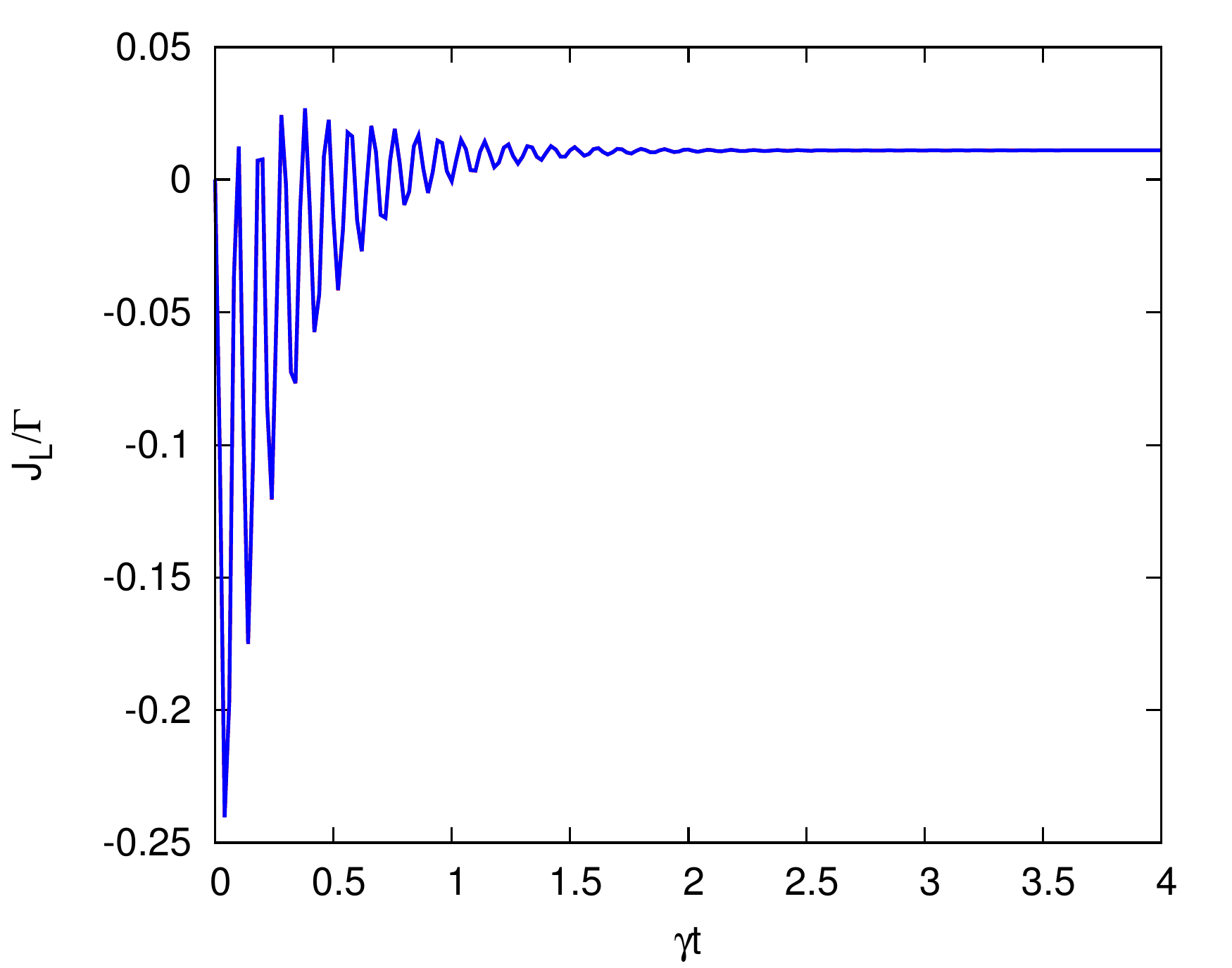}
\includegraphics[width=8.cm]{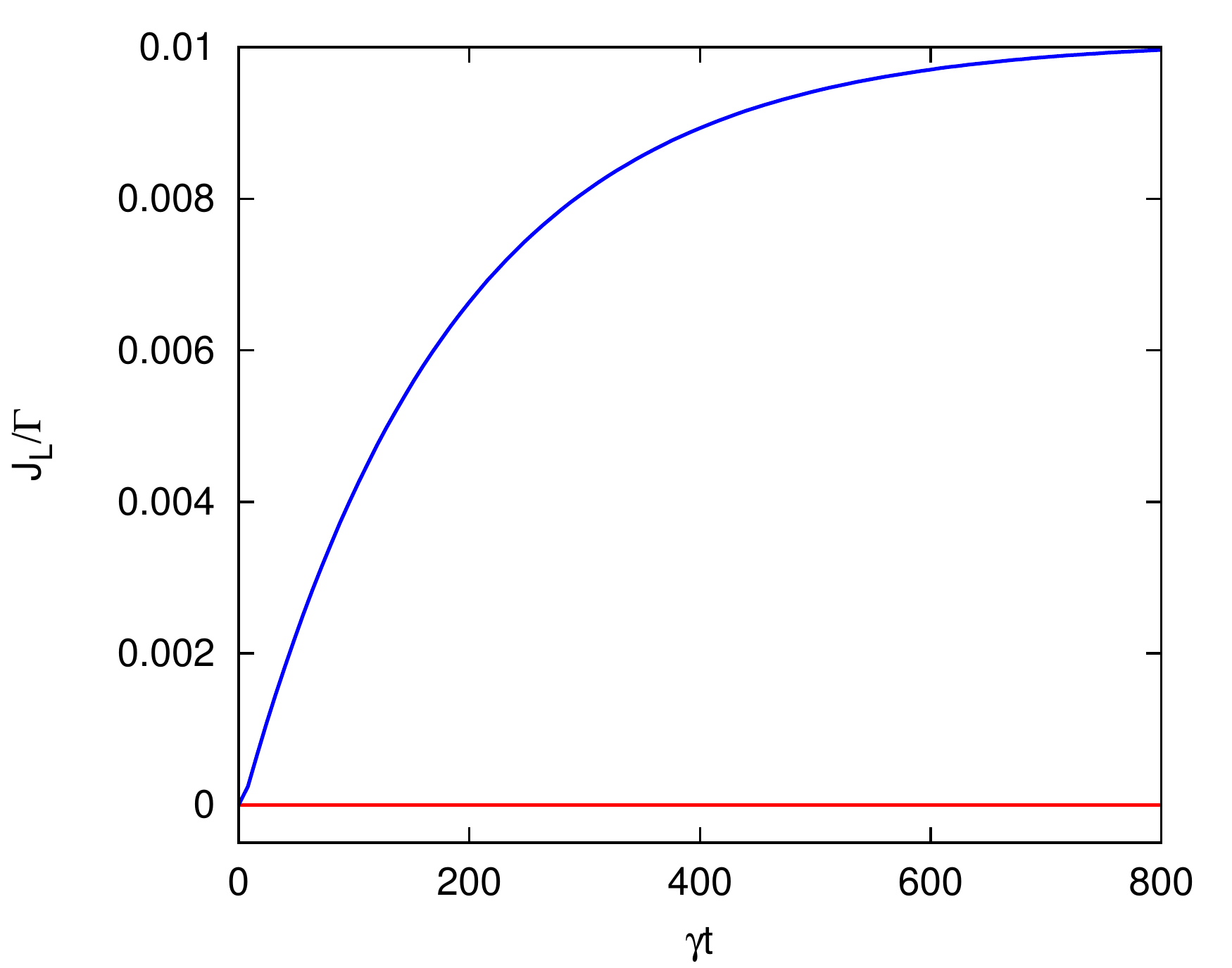}
\caption{Time evolution of the excitonic current across the four-level molecular system initialized in a totally symmetric state $\rho_+$ (left) or in an antisymmetric state $\rho_-$ (right), 
%under the exchange of molecular sites $2\leftrightarrow 4$, 
see Eq. (\ref{rhoini4}). Red lines correspond to the probe acting on site $1$ (molecular structure invariant under exchanges of sites 2 and 4)
%(symmetric case), 
while blue lines correspond to the probe acting on site 2 (broken $2\leftrightarrow 4$ exchange symmetry), see Fig. \ref{fig:4-level}. While red and blue lines overlap almost completely in the left panel (symmetric initial condition $\rho_+$; the overlap is so good that the two lines cannot be distinguished and only the top --blue-- curve is apparent), a clear difference in their time dependence appears for antisymmetric initial conditions $\rho_-$ (right panel). System parameters are $\epsilon=-142.2 \;\te{meV}, \; J=-9.35 \; \te{meV}$. Baths parameters are: $T_L=330\; \te{K}, \; T_R=270\; \te{K}$, and probe parameters are fixed to $T=300\; \te{K}, \Gamma=196, \te{GHz},\; \gamma=19.6\; \te{GHz}, \omega_0=78.55\; \te{THz}, \; \omega_D=1.96\; \te{THz}.$ See also Ref. \cite{thingna16a}. }
\label{fig:sec3_currents}
\end{figure}

But how does symmetry show up at the dynamical level? Let us focus first on molecule dynamics in the absence of configurational and environmental noise. Fig. \ref{fig:sec3_currents} represents the exciton current flowing through our 4-site molecular system as a function of time when the system is initialized in a symmetric state $\rho_+$ (left panel) or in an antisymmetric state $\rho_-$ (right panel). In both cases, the red lines represent the current time evolution when the probe is located at site 1 (symmetric case, see Fig. \ref{fig:4-level}.a) while blue lines correspond to the probe interferring with the molecule at site 2 (broken symmetry case, see Fig. \ref{fig:4-level}.b). For the totally symmetric initial state (left panel) there are clearly no dynamical signatures of molecular symmetries in the current evolution. Indeed the current dynamics does not depend on the underlying molecular symmetries in this case, as red and blue lines mostly overlap. On the other hand, a clear-cut dynamical signature of the underlying molecular symmetry emerges for antisymmetric initial condition (right panel in Fig. \ref{fig:sec3_currents}): the excitonic current is blocked when the symmetry is present (probe at site 1), but it increases steadily when the probe (now at site 2) breaks the molecule's internal symmetry. In this way, initializing the molecule in a dark state for bulk sites and monitoring the current time evolution when the probe is located at different sites provides a clear-cut method to detect underlying symmetries dynamically.

\begin{figure}
\bc
\includegraphics[width=16cm]{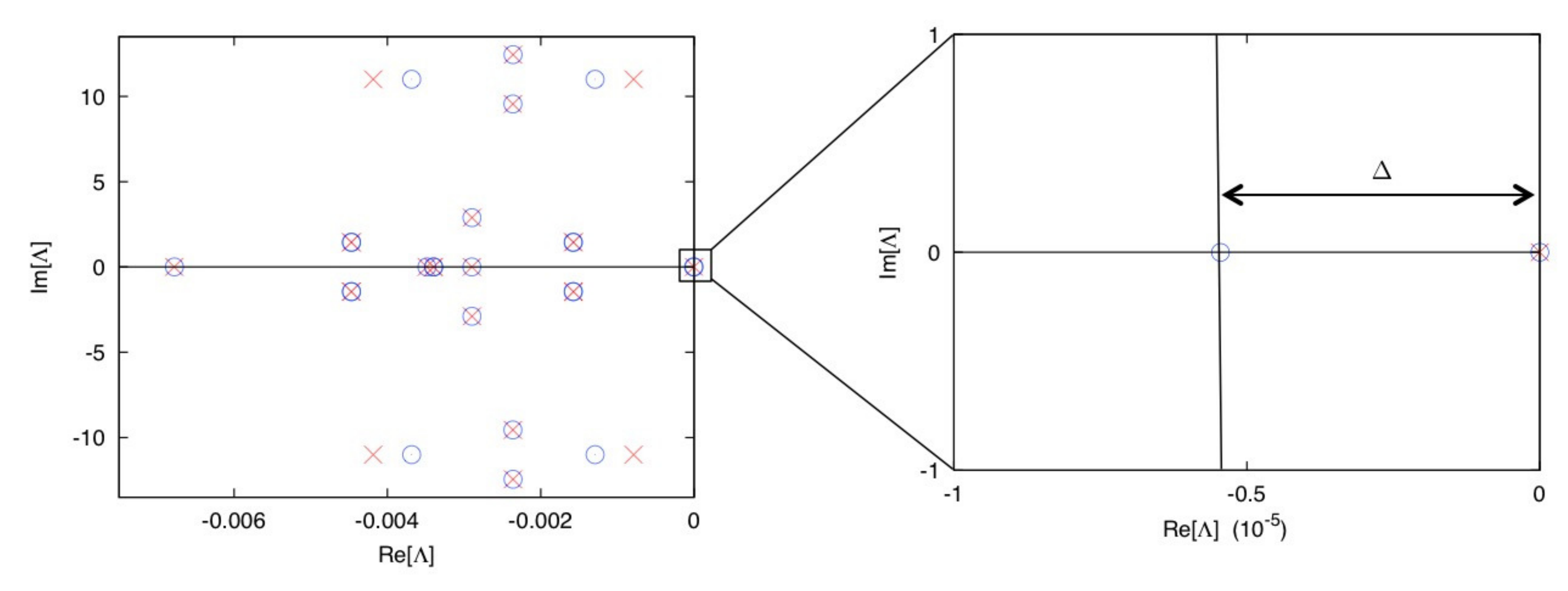}
\caption{Spectrum of the Lindblad-Redfield Liouvillian of Eq. \ref{eq:L_RL} with the probe acting on site $1$ (red crosses, see Fig. \ref{fig:4-level}.a) and site $2$ (blue circles, see Fig. \ref{fig:4-level}.b). When the probe is acting on site $1$ the system remains symmetric and there are two degenerate zero eigenvalues. When the probe is acting on site $2$ the symmetry is broken, degeneracy disappears and there is a gap $\Delta$ between the zero eigenvalue and the first decay mode. See also Ref. \cite{thingna16a}. }
\label{fig:sec3_eigen}
\ec
\end{figure}

To better understand the dynamical clues of the underlying molecular symmetry, we  use now the spectral approach introduced in Section \S\ref{sec:sym1}, and define $\phi_k, \, \hat{\phi}_k \in\cBH$ as the right and left eigenoperators of the Lindblad-Redfield Liouvillian $\mathcal{L}_{LR}$ defined in Eq. (\ref{eq:L_RL}), respectively, with (common) eigenvalue $\Lambda_k\in\mathbb{C}$. In particular
\be
\cL_{LR} \phi_k = \Lambda_k \phi_k \, , \qquad \hat{\phi}_k \cL_{LR} = \Lambda_k \hat{\phi}_k \, ,
\label{eq:eigM}
\ee
and any arbitrary density matrix $\rho(0)$ can be spectrally decomposed using this biorthogonal basis of $\cBH$. In this way, we may write the time-evolved molecule's density matrix, $\rho(t)=\exp(+t\cL_{LR}) \rho(0)$, as
\be
\rho(t)=\sum_k e^{+\Lambda_k t} \bbraket{\hat{\phi}_k}{\rho(0)}\; \phi_k \, ,
\label{eq:LF-evolM}
\ee
with the Hilbert-Schmidt inner product $\bbraket{\hat{\phi}_k}{\rho(0)}=\Tr ({\hat{\phi}_k}^\dagger \rho(0))$. 

The previous expression makes manifest that the system time evolution depends on two important factors, namely the spectrum of $\mathcal{L}_{LR}$ and the projections of the initial state $\rho(0)$ on the different eigenspaces of $\mathcal{L}_{LR}$. Fig. \ref{fig:sec3_eigen} shows the spectrum of $\mathcal{L}_{LR}$ for the 4-site toy molecule of interest when the probe acts on either site 1 ($\times$) or 2 ($\bigcirc$), see also Fig. \ref{fig:4-level}. This eigenspectrum is a particular instance of the general spectral structure discussed in Section \ref{sec:sym1}, see Fig. \ref{fig:spectrum} there and the ensuing discussion. When the probe acts on site 1, the molecule symmetry is preserved and there are two different steady states, both with $\Lambda_k=0$, one for each symmetry subspace (i.e. one symmetric and another antisymmetric under the exchange of atomic sites 2 and 4). Initializing the molecule in a dark bulk state, $\rho(0)=\rho_-$, then constraints the system into the antisymmetric steady state, which has zero current (red line in right panel of Fig. \ref{fig:sec3_currents}). On the other hand, when the probe acts on site 2 the molecular symmetry is broken, and one of the zero eigenvalues becomes a decay mode, defining a spectral gap $\Delta<0$ (see zoom in right panel of Fig. \ref{fig:sec3_eigen}). Interestingly, the (now unique) steady state is mostly symmetrical (i.e. its overlap with $\rho_+$ is high), while the decay mode is mostly antisymmetric. In this way, initializing the molecule in state $\rho(0)=\rho_-$ when the probe acts on site 2 we expect an exponential increase of the current at short times of the form $\sim\exp(+|\Delta| t)$, see Eq. (\ref{eq:LF-evolM}) above and blue line in Fig. \ref{fig:sec3_currents} (right), so $\tau_\text{R}=|\Delta|^{-1}$ defines the timescale to detect dynamically the transport effects of the probe-induced molecular symmetry breaking. This timescale (and hence the spectral gap $\Delta$) can be easily determined by following the short-time current evolution. Finally, when the molecular system is initialized in a symmetric state, $\rho(0)=\rho_+$, the current will converge quickly to its steady state value (note the timescale difference between left and right panels in Fig. \ref{fig:sec3_currents}), irrespective of the position of the probe since in both cases (probe at site 1 or 2) the initial state is very close to the final steady state. This explains the observation of Fig. \ref{fig:sec3_currents} (left).

In many experiments thermal (canonical) initial conditions for the density matrix are employed, $\rho(0)=Z^{-1}\exp(-\beta H^\text{4s})$, with $Z$ the associated partition function. Such initial condition strongly overlaps with the symmetric density matrx $\rho_+$, and hence we expect a similar behavior of the exciton current as a function of time when the molecule is initialized in such thermal state, see left panel in Fig. \ref{fig:sec3_currents}, namely no dynamical signature of the internal molecule's symmetry in the current evolution for this particular initial condition. We also mention that the interplay between the relaxation timescale $\tau_\text{R}$ and the probe characteristic timescales can be also studied in detail, see \cite{thingna16a}.

The previous dynamical signatures of molecular symmetries are robust even in the presence of weak conformational disorder and/or environmental noise. Indeed, if both noise sources are much weaker than the probe action (i.e. $\delta\ll \gamma$, with $\delta$ the generic noise strength and $\gamma$ the probe-molecule coupling constant, see Eq. (\ref{eq:gamma}) above), then one expects their effect on the spectrum of $\mathcal{L}_{LR}$ to be negligible at first order so the previous discussion remains valid in this case. This should remain true at least for intermediate times, since the timescale for dissipation to become relevant, $\tau_\text{dis}\sim \delta^{-1}$, is such that $\tau_\text{R} \ll \tau_\text{dis}$.

\subsection{A more realistic example: the benzene molecule}
\label{sec8:benz}

\begin{figure}
\centerline{\includegraphics[width=16cm]{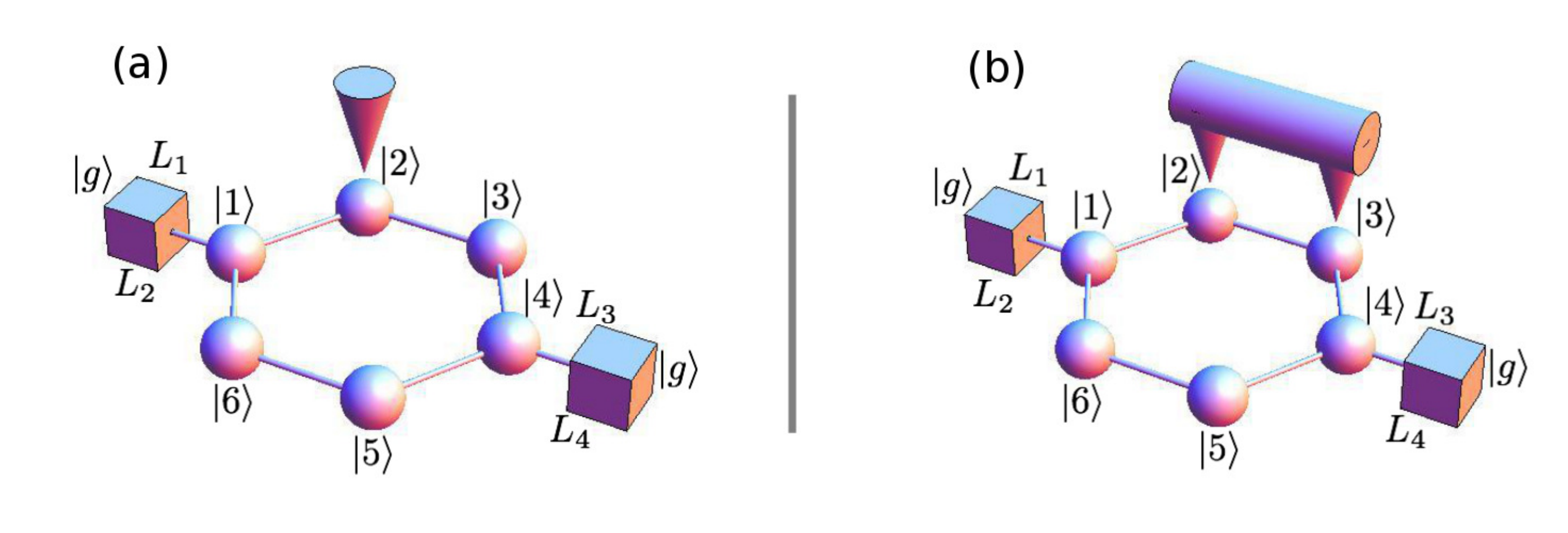}}
\caption{Four level system used in Ref. \cite{thingna16a} to study symmetry breaking. Left: System without probe. Right: System with a local probe acting on site $\ket{2}$.}
\label{fig:benzene}
\end{figure}

Next we focus on the dynamical detection of symmetries in more complex molecules. In particular, we briefly describe now the method described above as applied to the para-benzene molecule, a system with 6 different atomic sites (see Fig. \ref{fig:benzene}). The Huckel Hamiltonian for this molecule reads now
\be
H^{\text{6s}} = \epsilon \sum_{n=1}^6 \op{n}{n} + J \sum_{\la n,m\ra} \op{n}{m} \, , \label{Hb}
\ee
where the second sum runs over all pairs of nearest neighbors according to the interaction topology depicted in Fig. \ref{fig:benzene}, and the different parameters $\epsilon$ and $J$ have the same meaning as in Eq. (\ref{H4s}). Moreover, two (terminal) sites are weakly connected to thermal baths as in the case of the 4-site toy model. 

In the absence of a probe, the open para-benzene molecular ring exhibits a symmetry corresponding to the following unitary operator
\be
\pi = \exp \Big[i(\op{2}{6} + \op{6}{2}) \otimes (\op{3}{5} + \op{5}{3}) \Big] \, , \label{sym6}
\ee
which corresponds to exchanging molecular sites $2\leftrightarrow 6$ and $3\leftrightarrow 5$, or equivalently a $180^\text{o}$ rotation around the axis connecting the terminal atomic sites interacting with the baths (see Fig. \ref{fig:benzene}). This symmetry thus leads to multiple, coexisting steady states. For this particular case, three different steady states emerge, $\{\rho_k,\, k=1,2,3\}$. The first two steady states are pure states and contain no information on the molecule's coupling to the reservoirs. These two pure steady states can be written as $\rho_k=\op{\psi_k}{\psi_k}$, with $k=1,2$, such that
\ben
\ket{\psi_1} &=& \frac{1}{2}\big(\ket{5} + \ket{6} - \ket{2} - \ket{3} \big) \, , \\
\ket{\psi_2} &=& \frac{1}{2}\big(\ket{3} + \ket{6} - \ket{2} - \ket{5} \big) \, .
\een
The remaining steady state $\rho_3$ is non-trivial and contains nonequilibrium information associated to the coupling with thermal baths at different temperatures.

The presence of a probe interacting with the molecular ring breaks its internal symmetry and hence the degeneracy of the steady state. In particular, it can be shown \cite{thingna16a} that the two pure steady states become decay modes in the presence of a probe, while the (nontrivial) nonequilibrium steady state remains (all three slightly perturbed when compared to the probe-free case \cite{li14a}). Interestingly, and in contrast with the 4-site toy molecule discussed in the previous section, we may now break the molecule symmetry in different ways, i.e. with probes in different configurations, and the resulting decay modes reflect this richness, opening the door to the dynamical identification of all possible molecular symmetries.

For the para-bencene ring, when the probe acts on a single bulk atomic site (see left panel in Fig. \ref{fig:benzene}), it can be shown \cite{thingna16a} that the two former pure steady states, $\rho_{1,2}$ become decay modes with complex conjugate eigenvalues. As discussed in previous section, this gives rise to a exponential-like relaxation of the current as a function of time, with a characteristic time scale defined by the real part of these eigenvalues (which is the same for both).

The second possibility consists in introducing a double (non-local) probe, as depicted in the right panel of Fig. \ref{fig:benzene}. In this case one can demonstrate that the former steady states $\rho_{1,2}$ become again decay modes \cite{thingna16a}, but now with different real eigenvalues $\Lambda_2<\Lambda_1<0$. Both eigenvalues, being close enough to the steady state, define different timescales $\tau_{1,2}=|\Lambda_{1,2}|^{-1}$ which show up in the relaxation to the (unique) steady state. Indeed, demanding the initial state to be antisymmetric under the exchange of sites 2 and 5 (an state equivalent to $\rho_1$) and following the time evolution of the excitonic current, we find a double exponential relaxation with distinct timescales $\tau_1$ and $\tau_2$. The amplitudes of both exponentials have opposite sign, and hence the current exhibits a peak as a function of time which is the signature of the two unstable symmetry-related steady states. This shows how, by using a combination of probe configurations (local and non-local), we can detect the number of different steady states by measuring the current multi-exponential relaxation at long enough times. More details on this technique can be found in Ref. \cite{thingna16a}.

\newpage

\section{Other symmetry-mediated control mechanisms} 
\label{sec:other}

In this section we briefly review other approaches to control transport in open quantum systems using symmetry as a tool. These approaches are different from our control theory above in that the symmetries used as a resource to manipulate quantum transport are different from the strong symmetries that we introduce in \S\ref{sec:sym2}. In particular, we will review next how (weak) symmetries of the steady-state density matrix strongly constraint the transport properties of the system of interest \cite{popkov13a}. These weak symmetries can then be used to switch on or off currents in the system, even under large boundary gradients, by suitably adjusting the boundary pumping of excitations. We will study these effects in a particular model already discussed above, an open $1d$ XXZ spin chain. In a second section we will also study how quantum transport can be enhanced by breaking another symmetry inherent to the evolution of quantum systems, namely time-reversal symmetry \cite{zimboras13a}. We will demonstrate this effect in a model of \emph{chiral} quantum walk, relevant in the study of transport phenomena when magnetic fields are applied \cite{zimboras13a}.

\subsection{Manipulating transport in interacting qubit systems using weak symmetries of the density matrix} 
\label{sec:other1}

As in previous sections, we are interested here in studying transport in open quantum systems evolving in time according to a Lindblad-type master equation $\dot\rho=\cL \rho$, with $\cL$ the associated Lindbladian, see e.g. Eq. (\ref{eq:lindblad}). In particular, and following Ref. \cite{popkov13a}, we want to explore in this section the effect of \emph{weak symmetries} \cite{buca12a} on the Lindblad dynamics, and the constraints that these weak symmetries impose on the resulting nonequilibrium steady state (NESS). We say that a system exhibits a weak symmetry iff there exists a unitary operator $U\in\cBH$ such that
\be
\cL (U \rho U^\dagger) = U (\cL \rho) U^\dagger \, , \qquad \forall \rho\in\cBH \, .
\ee
The presence of such a weak symmetry ensures that the steady-state solution $\rho^{\scriptscriptstyle\text{NESS}}$ to the Lindblad equation, defined by the condition $\cL \rho^{\scriptscriptstyle\text{NESS}}=0$, remains invariant under the weak symmetry operator $U$, i.e.
\be
\rho^{\scriptscriptstyle\text{NESS}} = U \rho^{\scriptscriptstyle\text{NESS}} U^\dagger \, .
\ee
Note that, provided Evans theorem holds \cite{evans77a}, this steady state is unique despite the existence of a weak symmetry. Indeed, strong symmetries of the dynamics as those defined in Eq. (\ref{eq:sym}) are also weak symmetriers, but the reverse statement is not true in general. The existence of a weak symmetry also implies that the steady-state expectation value of any physical observable of interest $\eta\in\cBH$, defined as $\la \eta\ra=\Tr(\eta \rho^{\scriptscriptstyle\text{NESS}})$, obeys the following relation \cite{popkov13a}
\be
\la \eta\ra = \la U^\dagger \eta U\ra \, ,
\ee
where we have used the cyclic property of the trace. In particular, if the observable $\eta$ changes sign under the action of the symmetry operator $U$, so $U^\dagger \eta U = - \eta$, the previous relation forces $\la \eta \ra =0$. On the other hand, if $\eta$ remains invariant under $U$, so $U^\dagger \eta U = \eta$, the above symmetry relation does not constraint the expectation value $\la \eta \ra$. In this way, for a given quantum Hamiltonian, one can engineer the Lindblad jump operators driving the system out of equilibrium to build up weak symmetries of the full Lindbladian capable of switching off desired currents. This transport manipulation method is markedly different from the symmetry-based tools described in previous sections, for which the existence of multiple steady states or transport channels dynamically coexisting due to the existence of a strong symmetry was pivotal.

In what follows we describe a specific example of this symmetry-mediated mechanism for transport control. In particular, we will be interested again in a $1d$ XXZ Heisenberg spin chain, a relevant model in the study of energy transport in one-dimensional systems \cite{znidaric10a,manzano12a,asadian13a,manzano16b}. This model is described by the Hamiltonian
\be
H^\text{XXZ}=\sum_{i=1}^{L-1} \left(\sigma_i^x \sigma_{i+1}^x +  \sigma_i^y \sigma_{i+1}^y + \Delta~\sigma_i^z \sigma_{i+1}^z \right) \equiv \sum_{i=1}^{L-1} h_{i,i+1}\, ,
\label{eq:chain_xxz2}
\ee
with $\Delta$ a dimensionless constant and $\sigma_i^{x,y,z}$ the standard Pauli operators acting on site $i$. In order to demonstrate the effect of a weak symmetry on the dynamics, we now drive this spin chain out of equilibrium by coupling it to two baths acting at the ends of the chain. The left bath is then defined by the following jump operators \cite{popkov13a}
\ben
L_1=\sqrt{A} \pare{\sigma_1^y-i\sigma_1^z} \, ,\\
L_2= \sqrt{\alpha} \pare{\sigma_1^z+ i \sigma_1^x } \, ,
\een
while the right baths is represented by 
\ben
L_3=\pare{\sigma_L^y + i\sigma_L^z} \, ,\\
L_4= \sqrt{A\alpha} \pare{\sigma_L^z + i\sigma_L^x } \, ,
\een
where $A\ne1$ and $0\le \alpha \le 1$ are two different constants which characterize the strength of the pumping/extraction of excitations at both ends. For this type of reservoir coupling, the system exhibits weak symmetries as defined above for the extreme cases $\alpha=0,1$ \cite{popkov13a}. Indeed, for the case $\alpha=0$ the operator $\Omega_x=\pare{\sigma^x}^{\otimes L}$ satisfies the relation 
\be
\cL \pare{  \Omega_x \rho \Omega_x^{\dagger}} = \Omega_x  \cL \pare{  \rho } \Omega_x^{\dagger}.
\ee
The presence of this symmetry ensures that the solution to the equation $\cL \pare{  \rho^{\scriptscriptstyle\text{NESS}} }=0$ is invariant under the effect of $\Omega_x$ 
\be
\rho^{\scriptscriptstyle\text{NESS}}_{\pare{\alpha=0}}= \Omega_x \rho^{\scriptscriptstyle\text{NESS}}_{\pare{\alpha=0}} \Omega_x^{\dagger} \, , \qquad \forall \rho\in\cBH.
\ee
On the other hand, when $\alpha=1$ both the system dynamics and the steady state remain invariant under a more complicated symmetry operator  \cite{popkov13a}
\be
\rho^{\scriptscriptstyle\text{NESS}}_{\pare{\alpha=1}}=  \Omega_x U_{\text{rot}} R  \rho^{\scriptscriptstyle\text{NESS}}_{\pare{\alpha=1}} R U_\text{rot}^{\dagger} \Omega_x^{\dagger}.
\ee
Here $R$ is the right-left reflection operator, such that $R\pare{S_1 \otimes S_2 \otimes \cdots \otimes S_L }= \pare{S_L \otimes S_{L-1} \otimes \cdots \otimes S_1}R$, and $U_\text{rot}$ is a rotation in the $XY$ plane such that $U_\text{rot}^{\phantom{\dagger}} \sigma_i^x U_\text{rot}^{\dagger}=\sigma_i^y$ and $U_\text{rot}^{\phantom{\dagger}} \sigma_i^y U_\text{rot}^{\dagger}= -\sigma_i^x$. In any case, and despite the existence of these weak symmetries for particular values of $\alpha$, this system presents a unique fixed point of the dynamics because of the completeness of the algebra generated by the set of operators $\key{H,L_i}\;i=1,\dots,4$ (see Ref. \cite{prosen12a} for a detailed discussion about the uniqueness of steady states in the XXZ chain). 

\begin{figure}
\centerline{\includegraphics[width=9cm]{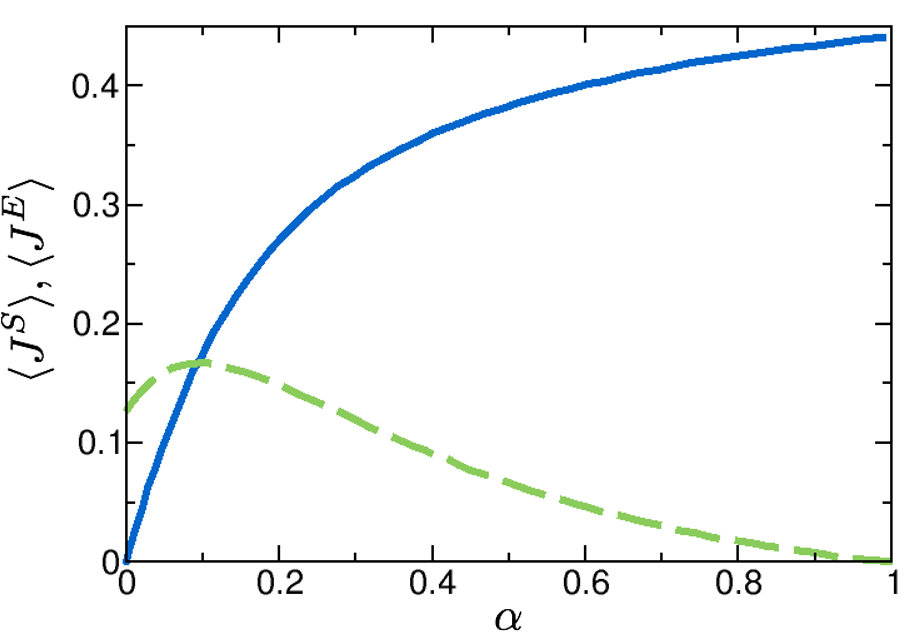}}
\caption{Fig. 3.a from Ref. \cite{popkov13a}. Average spin (solid line) and energy (dashed line) currents for the XXZ spin chain as a function of parameter $\alpha$. Note that the spin current is switched off for $\alpha=0$, while for $\alpha=1$ the energy current goes to 0.}
\label{fig:livi_currents}
\end{figure}

Bearing in mind the previous weak symmetries, we turn our attention to the system transport properties. These are fully characterized by the spin and energy current operators, defined respectively as
\be
J^S_{i,j} \equiv 2 \pare{\sigma_i^x \sigma_j^y - \sigma_j^y \sigma_i^x} \, ,
\label{eq:spin_chain}
\ee
and 
\be
J^E_{i} \equiv -\sigma_i^z J^S_{i-1,i+1} + \Delta \pare{ J^S_{i-1,i} \sigma_{i+1}^z + \sigma_{i-1}^z J^S_{i,i+1} } \, .
\label{eq:energy_chain}
\ee
These operators can be obtained starting from the continuity equations $\frac{d}{dt} \sigma_i^z = J_{i-1,i}^S - J_{i,i+1}^S $ and $\frac{d}{dt} h_{i,i+1} = J_i^E - J_{i+1}^E$, with $h_{i,i+1}$ the energy of bond $(i,i+1)$, see Eq. (\ref{eq:chain_xxz2}). It is straightforward to prove that in the steady state the currents across all bonds are equal and hence we can define the general current operators as $J^S\equiv J^S_{L,L-1}$ and $J^E\equiv J^E_{L}$. Next we study their steady state expectation value for the extreme cases $\alpha=0,1$ for which different weak symmetries emerge \cite{popkov13a}. In particular, when $\alpha=0$ it is found that 
\ben
\hspace{-2cm} \mean{J^S}{\pare{\alpha=0}}= \Tr \pare{\rho^{\scriptscriptstyle\text{NESS}}_{\pare{\alpha=0}} J^S }= -\Tr\pare{\Omega_x \rho^{\scriptscriptstyle\text{NESS}}_{\pare{\alpha=0}} \Omega_x^{\dagger} J^S } = - \mean{J^S}{\pare{\alpha=0}}=0\\
\hspace{-2cm} \mean{J^E}{\pare{\alpha=0}}= \Tr \pare{\rho^{\scriptscriptstyle\text{NESS}}_{\pare{\alpha=0}} J^E } = \Tr\pare{\Omega_x \rho^{\scriptscriptstyle\text{NESS}}_{\pare{\alpha=0}} \Omega_x^{\dagger} J^E } = \mean{J^E}{\pare{\alpha=0}} \, . 
\een
This means that for $\alpha=0$ the weak symmetry constraints the average value of the spin current to zero, switching it off, while there is no restriction on the energy current. On the other hand, when $\alpha=1$ one finds the opposite situation \cite{popkov13a}, leading to $\mean{J^E}{\pare{\alpha=1}}=0$ and no restriction for $\mean{J^S}{\pare{\alpha=1}}$. There are hence two symmetry-controlled regimes, one were the spin current is zero but not the energy flux, and another regime with zero energy current and non-zero spin flow. For arbitrary values of the control parameter $\pare{0<\alpha<1}$, the energy and spin currents can be calculated by solving numerically the Lindblad master equation for the steady state $\cL\pare{\rho^{\scriptscriptstyle\text{NESS}}}=0$ and calculating the expected values of the operators given by Eqs. (\ref{eq:spin_chain}) and (\ref{eq:energy_chain}), see Ref. \cite{popkov13a} for details. The measured values of the currents are shown in Fig. \ref{fig:livi_currents}, where a smooth crossover between the two extremal transport regimes is apparent. This shows how weak symmetries of the density matrix can be engineered to control transport in open quantum systems.

\subsection{Breaking time-reversal symmetry to enhance quantum transport} 
\label{sec:other2}

In Refs. \cite{zimboras13a,lu16a} a different approach for the use of symmetry (and its breaking) for quantum transport control is presented. In this case the symmetry broken is not represented by a weak or strong symmetry in Bu\v{c}a's and Prosen's terminology \cite{buca12a}. What is studied now is the effect of breaking the time-reversal symmetry (TRS) of the Hamiltonian. In particular, Zimbor\'as and collaborators \cite{zimboras13a} demonstrate this effect in
%the effect of time-reversal symmetry breaking on quantum transport by studying 
a model of continuous-time quantum walk broadly used in literature to investigate optimal transport.

A continuous quantum walk in a graph is defined by the Hamiltonian
\be
H^{QW}=\sum_{n,m} J_{n,m} \pare{ \op{n}{m} + \op{m}{n}},
\ee
where the summation runs over all the sites $n,m$ that are connected in the graph, and the weights $J_{n,m}$ defining the adjacency matrix are real-valued. This Hamiltonian exhibits time-reversal symmetry in the sense that the site-to-site transfer probability, i.e. the probability $P_{n\to m}(t)=\bra{m}\rho(t)\ket{m}$ to occupy site $m$ at time $t$ with initial condition $\rho(0)=\ket{n}\bra{n}$, remains invariant under the transformation $t\to -t$, i.e. $P_{n\to m}(t)=P_{n\to m}(-t)$. Time-reversal symmetry is a well known feature of quantum walks \cite{kempe03a,makmal14a}, and difficults the design of quantum systems with \emph{directional} control of transport. This problem has been addressed by different means, e.g. by increasing the Hilbert space of the walker \cite{szegedy04a}. We note that directional quantum walks have several applications in quantum information science \cite{paparo13a,paparo14a}.

A  way to break the time-reversal symmetry of a continuous quantum walk without increasing the Hilbert space of the system is to introduce a phase at each edge of the graph  \cite{zimboras13a}. This results in a continuous-time \emph{chiral} quantum walk (CQW) with Hamiltonian
\be
H^{CQW}= \sum_{n,m}  J_{n,m} \pare{e^{i\theta_{n,m}} \op{n}{m} +   e^{-i\theta_{n,m}} \op{m}{n}},
\label{eq:CQW}
\ee
where $\theta_{n,m}$ is the phase corresponding to the transition between sites $n$ and $m$. In the general case, each link can have a different phase, though this multiplicity is not necessary for time-revesal symmetry breaking and a single phase $\theta$ can already alter the behavior of the site-to-site transfer probability under the transformation $t\to -t$ (see \cite{zimboras13a} for a detailed discussion). 

The task in hand is to transfer a single-excitation between two specific sites of a graph, that we denote here as $I$ and $T$. The excitation is placed initially at site $I$ and it travels through the network with a dynamics which mix coherent quantum evolution given by the Hamiltonian (\ref{eq:CQW}) and incoherent stochastic jumps between different quantum states. All together, the CQW dynamics is described by a Lindblad equation
%The outgoing site is labeled as $T$. 
%The probability of finding the excitation at the outgoing site $T$ at time $t$ is given by $P_T(t)=\bra{T}\rho(t)\ket{T}$. Although, the transport properties of this system can also be analysed by introducing a trapping mechanism modelled by an incoherent channel modelled by a Lindblad-type equation
\be
\dot \rho(t) = \cL\rho(t)= -i\cor{H^{CQW},\rho} + \sum_k L_k^{\phantom{\dagger}} \rho L_k^{\dagger} - \frac{1}{2} \key{L_k^{\dagger}L_k^{\phantom{\dagger}}, \rho},
\ee
with Lindblad operators $L_k$, and the probability of finding the excitation at the outgoing site $T$ at time $t$ is given by $P_T(t)=\bra{T}\rho(t)\ket{T}$ with initial condition $\rho(0)=\op{I}{I}$.
%in the form $\op{I}{s}$. This channel connects one-way one of the sites of the system $\ket{s}$ with the trap site $\ket{T}$. 

\begin{figure}
\bc 
\includegraphics[scale=.28]{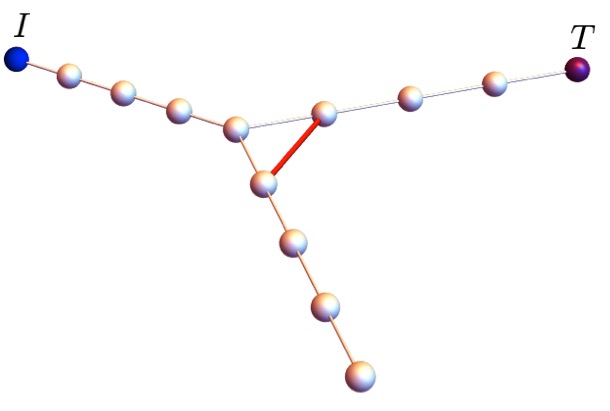}
\includegraphics[scale=.32]{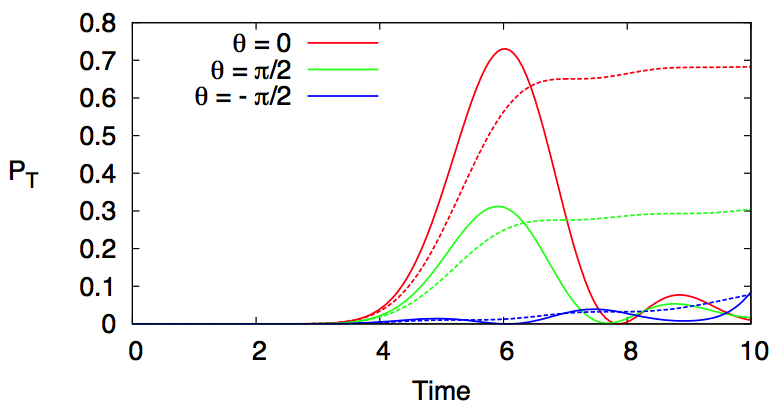}
\caption{Figure (1) from Ref. \cite{zimboras13a}. Left: Example of directional switch. The red line represents the only connection with a non-vanishing phase. A single excitation is initially placed at site $I$ (blue) and its transport can be enhanced/supressed in the direction of site $T$ (purple) by the introduction of this phase. Right: Probability of finding the excitation at site $T$ as a function of time for different values of $\theta$. Solid lines represent purely coherent evolution while dashed lines correspond to an open-system with an incoherent channel $L_k=\op{T}{s}$ between site $T$ and the previous one $s$. Results are mostly independent of the system length.}
\label{fig:CQW1}
\ec
\end{figure}

The CQW can be used to create a quantum switch between two edges of a ladder, see Fig. \ref{fig:CQW1}. The task to be performed in this case is to transfer one excitation from an initial site $I$ to a target site $T$ with the maximum possible efficiency. This is done by introducing a phase $\theta$ between the two first sites of each leg of the ladder, see Fig. \ref{fig:CQW1}. Without the phase, the two legs are indistinguishable and the probability of finding the excitation in one of them is equal than the other one for all times. The introduction of a phase $\theta$ changes this situation, promoting transport along one leg. 

The efficiency of this directional switch is studied from both the closed- and open-system perspectives. In Fig. \ref{fig:CQW1} the probability $P_T(t)$ of finding the excitation at site $T$
%, $P_T(t)=\bra{T}\rho(t)\ket{T}$, 
is displayed as a function of time. In the purely coherent evolution (solid lines) we appreciate an oscillatory behaviour. The height of the first maximum can be modulated by choosing a specific value of $\theta$. In particular, this maximum can be enhanced by $134\%$ or supressed by $91\%$ compared to the maximum in the time-symmetric case $(\theta=0)$ by changing $\theta$.
%The size of the maxima in the symmetric case $(\theta=0)$ can be enhanced by $134\%$ or diminished by $91\%$. 
In the open quantum framework the outgoing site $T$ is incoherently connected to the previous site, resulting in a different, saturating time-dependence for the transfer probability (dashed lines in Fig. \ref{fig:CQW1}). In this case the figure of merit is just the population of site $T$ at long times. By introducing an optimal phase $\theta$ this can be enhanced up to an $81.4\%$ with respect to the $\theta=0$ case \cite{zimboras13a}. Note that the dynamics of this system for any $\theta\ne 0$ is time-irreversible as $P_{T} (t) \neq P_{T} (-t)$ $\forall t>0$.
 
\begin{figure}
\bc
\begin{minipage}{10in}
%\centering
\hspace{-0.2cm}
\raisebox{0.3\height}{\includegraphics[scale=.2]{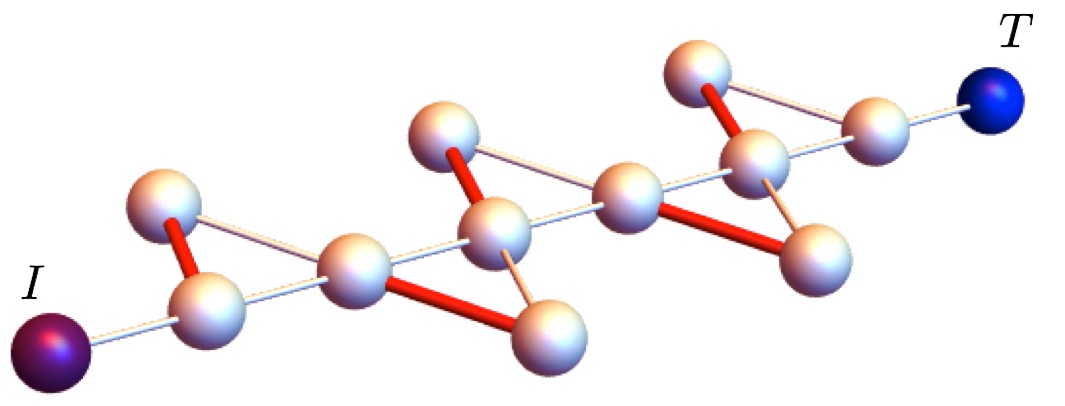}}
\hspace{0.15cm}
\includegraphics[scale=.3]{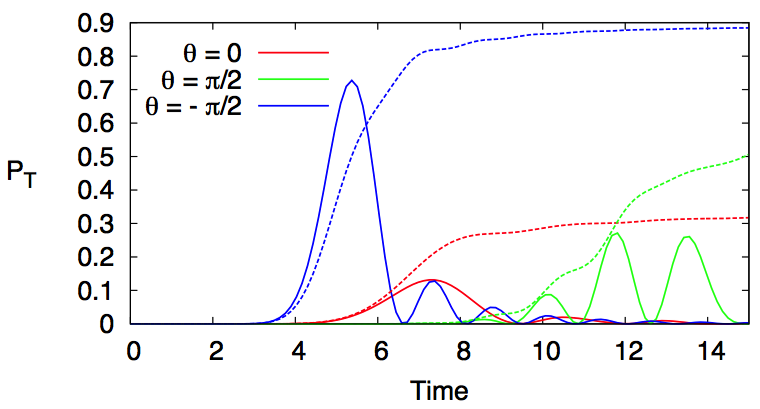}
\end{minipage}
\ec
\caption{Figure (2.a) from Ref. \cite{zimboras13a}. Left:  Sketch of the zig-zag chain that allows the enhancement of energy transfer by time-reversal symmetry breaking. Red links represent the edges with non-vanishing phase. Right: Probability of finding the excitation at site $T$ as a function of time for different phases. Solid lines represent purely coherent evolution while dashed lines correspond to open system dynamics with an incoherent channel between site $T$ and the previous site.}
\label{fig:CQW4}
\end{figure}
   
Chiral quantum walks can be also useful to enhance the transport in one dimension. Figure \ref{fig:CQW4} shows a zig-zag chain design also studied in Ref. \cite{zimboras13a} with this purpose. This design exploits the directional switching effect in order to enhance the quantum transport in a quasi-one dimensional system. In order to break the time-reversal symmetry, one phase is added to one of the connections of each triangle. The transport properties of the zig-zag chain are presented in Fig. \ref{fig:CQW4}, see also \cite{zimboras13a}. In this case there is only one possible target site, and the purpose is not to redirect the excitation in a specific direction, but to accelerate the energy transfer. To measure the speed of energy transfer Ref. \cite{zimboras13a} defines the {\it half-arrival time}, $\tau_{1/2}$, as the first time when the occupancy probability of the target reaches one half. One can also define the transport speed, $v_{1/2}$, as the inverse of $\tau_{1/2}$. The occupancy probabilities of the target site as a function of time are displayed in Fig. \ref{fig:CQW4} for different values of the phase $\theta$. Under purely coherent dynamics this probability presents again an oscillatory behaviour. In contrast with the previous case of the directional switch, in this case the time when the first maximum occurs depends on the value of the phase $\theta$. This time can be minimised by choosing $\theta=-\pi/2$, and the height of the maximum is also optimal for this value of the phase. When the incoherent trapping is included, the half arrival time is reduced from a non-chiral value $\tau_{1/2}(\theta=0)=38.1$ to $\tau_{1/2}(\theta=\pi/2)=5.2$ in the chiral case. This represents a 633\% enhancement of the transport speed \cite{zimboras13a}.

Remarkably, transport enhancement in chiral quantum walks has been experimentally studied in Ref. \cite{lu16a} using room-temperature liquid-state nuclear magnetic resonance (NRM) on a three qubit system. The simplest graph with the possibility of time-reversal symmetry breaking is a fully connected graph of three qubits. In the single excitation picture the system Hilbert space has dimension three, defined by the basis $\key{\ket{1},\ket{2},\ket{3}}$. The evolution used to study the effect of time-reversal symmetry breaking is discrete and based in two-qubit gates of the form
\be
U_{i,j} (\alpha,\theta) = \exp \pare{ \frac{\theta \pare{-i \cos \pare{\alpha} S_{i,j} + \sin \pare{ \alpha} A_{i,j} }}{2}}.
\label{eq:2qubit}
\ee
This operator is a combination of two different unitaries, one that preserves time-symmetry $S_{i,j} = \sigma^x_i \sigma^x_j + \sigma^y_i \sigma^y_j$ and another one $A_{i,j}= \sigma_i^x \sigma_j^y - \sigma_i^y \sigma_j^x$ that is time-antisymmetric, with $\sigma_i^{x,y}$ the standard $x,y$-Pauli matrices acting on qubit $i$. In this way, parameter $\alpha\in \left[ 0,2\pi\right)$ is a time-symmetry parameter leading to a completely time-symmetric dynamics only for $\alpha=\pi$. 

\begin{figure}
\bc
\includegraphics[scale=.15]{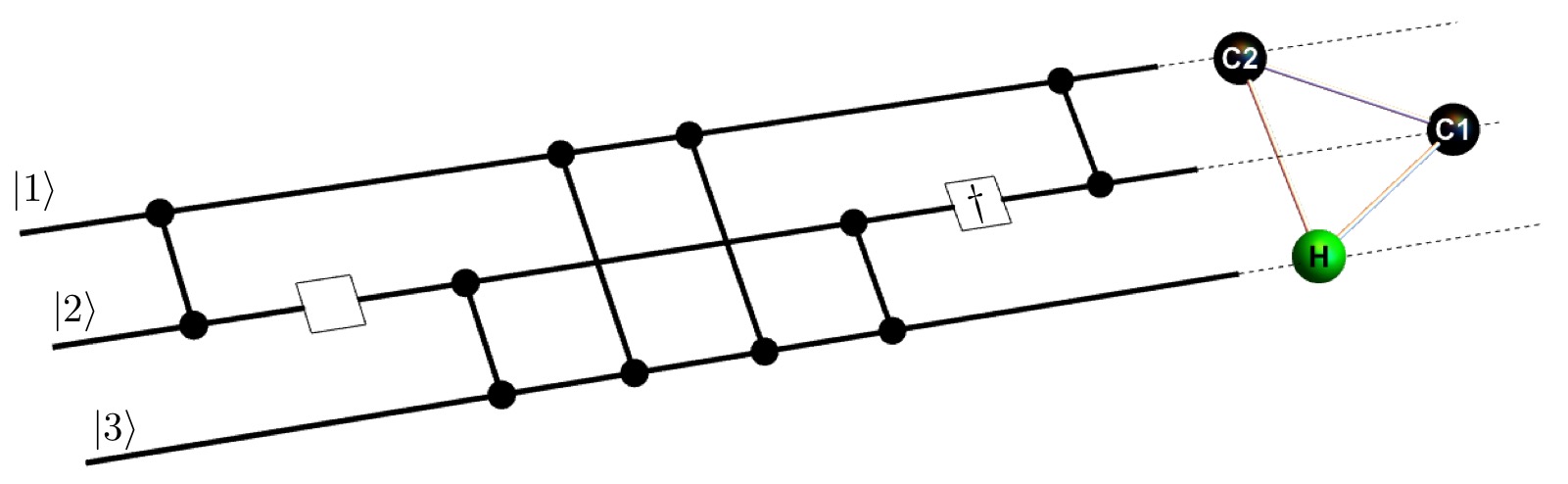}
\includegraphics[scale=.28]{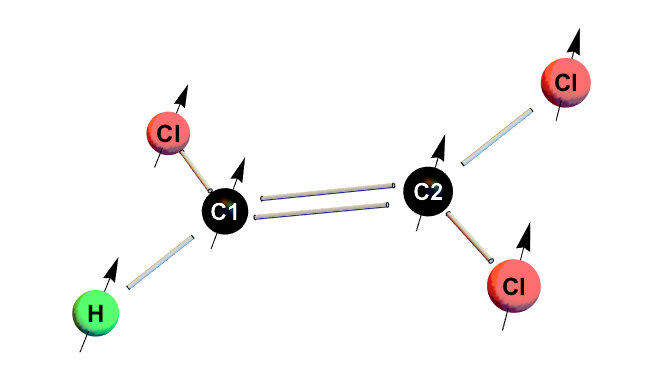}
\caption{Figs. 1.b-c and 2.a from Ref. \cite{lu16a}. Left: Quantum circuit representing the discrete evolution of the 3-qubit system. Lines connecting two qubits correspond to gates $U(0,\theta)$. The left blank box corresponds to the $z$-rotation $R^z(\alpha)$, and the right box corresponds to its adjoint. Right: Trichloroethylene molecule. Two $^{13}$C atoms (in black) and one $^1$H atom (green) form the 3-qubit system of the experiment.}
\label{fig:CQW5}
\ec
\end{figure}

An evolution like the one given by $S_{i,j}$ is naturally found in many quantum systems. The antisymmetric evolution $A_{i,j}$ is much more complicated to perform \cite{lu16a}. Luckily, the two-site gate (\ref{eq:2qubit}) can be decomposed as 
\be
U_{i,j} (\alpha,\theta) = R^z_j (\alpha) U_{i,j} (0,\theta)R^{z\dagger}_j (\alpha),
\ee
involving only symmetric unitaries combined with a local $z$-rotation $R^z_j(\alpha)= e^{-i(\alpha/2)\sigma_j^z}$. The simplest unitary evolution that allows time-reversal symmetry breaking is 
\be
U\pare{\alpha,\theta} = U_{1,2}U_{2,3}U_{3,1}U_{3,1} U_{2,3}U_{1,2} \, .
\ee
This is a palindromic circuit, meaning that it is composed for a series of gates that is the same from right to left than from left to right. Non-palindromic circuits are trivially non time symmetric and therefore they do not allow for time-reversal symmetry breaking. The circuit designed to perform this evolution is sketched in Fig \ref{fig:CQW5} \cite{lu16a}. 

\begin{figure}
\bc
\includegraphics[scale=.5]{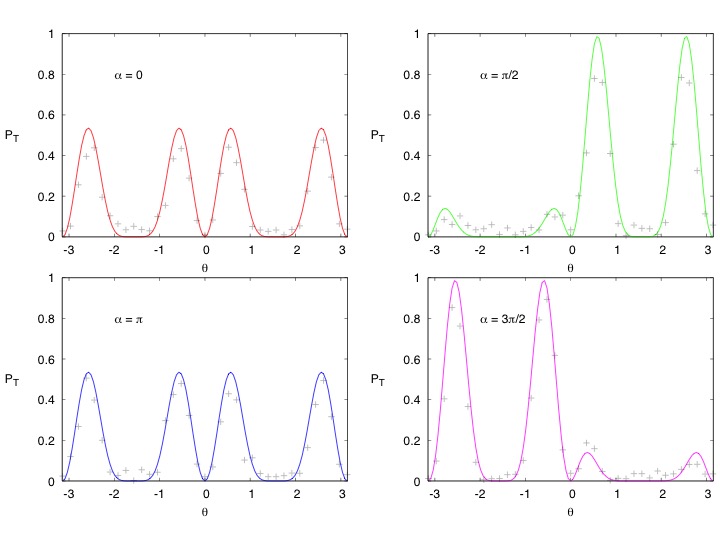}
\caption{Fig. 1.a from Ref. \cite{lu16a}. State transfer probability $\abs{\bra{3}U(\alpha,\theta)\ket{1}}^2$. Solid lines are theoretical predictions while dots represent experimental measures.}
\label{fig:CQW6}
\ec
\end{figure}

The experiment was performed using Nuclear Magnetic Resonance (NRM) in a molecular system of $^{13}$C-labeled trichloroethylene dissolved in deuterated chloroform \cite{lu16a}. The molecule structure is displayed in Fig. \ref{fig:CQW5}. The 3-qubit network is formed by the spins of the two $^{13}$C and the $^1$H atoms, and they are labeled as $C1$, $C2$, and $H$. The experiment used gradient ascent pulse engineering (GRAPE) pulses of different lengths to implement the two-body interactions \cite{ryan08a}. The run time of the circuit is 26 ms, much smaller than the decoherence time. The experiment was performed for $\alpha=0,\frac{\pi}{2},\pi,\frac{3\pi}{2}$, and $\theta$ was chosen to span $\cor{-\pi,\pi}$ in $\pi/18$ steps. Each experiment was performed 37 times. The results of the experiment, together with the theoretical prediction can be seen in Fig \ref{fig:CQW6}. In particular, for the time-reversal symmetric case $(\alpha=0)$, the maximum achievable probability of transporting the excitation is below 0.6. Moreover, in this case the probability $P_T$ is independent of the sign of $\theta$. The same behaviour is observed for $\alpha=\pi$. For the case of maximal time-asymmetry ($\alpha=\pi/2$) the transfer probability can go well beyond 0.6, approaching 1. The average error of the experimental data compared to the theoretical prediction is about $6.0\%$. This error is mainly due to decoherence and imperfection of the GRAPE pulses. 

In summary, these results show how time-reversal symmetry breaking can be used to enhance or suppress quantum transport.

\newpage

\section{Summary and conclusions} 
\label{sec:conclusions}

In this paper we have shown how to harness symmetry to control energy transport and fluctuations in quantum systems interacting with an environment. The natural framework to develop this program is the theory of open quantum systems and its extension to deal with the rare-event statistics. Consequently, we have introduced with some detail in the initial sections of this paper both the quantum master equation and the quantum jumps formalism, together with a brief description of full counting statistics for the current and the associated large deviation theory. Building on the ideas of Refs. \cite{buca12a,albert14a,albert16a}, we have studied how the interplay between dissipative processes and an underlying symmetry drives the quantum system of interest to a degenerate steady state which preserves part of the information of the initial state due to the lack of mixing between the different symmetry sectors. This opens the door to a complete control of transport properties (both the average current and its statistics) by tailoring this information via initial-state preparation techniques. We have reviewed two different examples involving both spin chains \cite{buca12a} and ladders \cite{znidaric13b} where these symmetries and their effect on transport properties become apparent.
%analysing their symmetries and how they affect their transport properties.

Interestingly, the degeneracy of the nonequilibrium steady state in the presence of a symmetry reflects a dynamical coexistence of different transport channels in the quantum system, each with a well defined and different current. We demonstrate that such coexistence stems from a general first-order-type dynamic phase transition (DPT) in the statistics of current fluctuations. This DPT, which appears as a non-analyticity in the cumulant generating function of the current or equivalently as a non-convex regime in the current large deviation function, separates two (fluctuating) transport phases characterized by a maximal and minimal current, respectively. Moreover, the time-reversibility of the microscopic dynamics results in the appearance of a \emph{twin} DPT for rare, reversed current fluctuations.

The previous results apply to general open quantum systems, provided that the few hypotheses leading to a description in terms of quantum master equations remain valid. With the aim of validating our results, and motivated by the problem of energy harvesting in (natural and artificial) photosynthetic complexes, we have studied transport and current fluctuations in open quantum networks. These models, which play an important role in many active areas of research\footnote[1]{These range from e.g. the understanding of quantum effects in photosynthesis to the functional role of dephasing in quantum transport or the interplay between network topology and quantum dynamics, to mention just a few.}, exhibit different exchange symmetries associated with their network topology, and as such are expected to display the general phenomenology unveiled in previous sections, something that we confirm in detailed numerical analyses. Furthermore, our symmetry-based approach to transport in quantum networks suggests design strategies to build quantum devices with controllable transport properties. In particular, based on our analysis of quantum networks, we introduce the concept of a \emph{symmetry-controlled quantum thermal switch}: this is a quantum qubit device with network architecture that, when coupled to heat reservoirs at different temperatures, is capable of modulating at will the heat current flowing from the hot to the cold bath by just playing with the initial state symmetry. This is essentially different from other transport control setups that have been recently introduced, e.g. by coupling lattice vibrons to internal states of trapped ions in crystal \cite{bermudez13a}. Finally, the symmetry-induced coexistence of two dynamical phases with different activity levels in a three-spin network model with collective dissipation has been reviewed with some detail \cite{pigeon15a}, as this system can in principle be engineered in hybrid electro/opto-mechanical settings.
%coexistence of two dynamical phases with different activity levels
%can in principle be engineered in hybrid electro/opto-mechanical settings

Furthermore, the recent possibility of creating coherent cavity networks with arbitrary connectivity \cite{kyoseva12a} has opened the door to new potential  experimental realizations of these symmetry-based quantum control ideas. Advancing along these lines, and based on our schematic model of symmetry-controlled switch, we describe a detailed experimental setup introduced in \cite{manzano16a} where symmetry-enabled quantum control can be realized in the lab. The system consists in three optical cavities coupled as a linear array, with both terminal cavities coupled in turn to reservoirs at different temperatures, and a central cavity doped with two identical $\Lambda$-atoms externally driven by laser fields. Note that an exchange of the two atoms leaves the system invariant and hence constitutes a symmetry in the language of previous sections. Under well-defined conditions, the excited states of both $\Lambda$-atoms can be adiabatically eliminated, and the two-atoms complex behaves as a four level system with symmetric and antisymmetric manifolds. The mixing between these two manifolds is controlled by the external laser field, so switching off the laser at the right moment leaves the diatomic system entrained in one of the symmetry subspaces, enabling the control of the photon current across the optical cavities. This phenomenon is fully confirmed in numerical investigations of the current statistics in this atomic switch, and the chances are that it can be realized in laboratory experiments using current technologies. Interestingly, the proposed device can be used also as a quantum memory to store maximally-entangled states \cite{thingna16a}, since the state of the two-atom system (in particular whether the system is in a dark state) can be measured without interferring with its dynamics by just monitoring the current flowing to the thermal baths.

The previous discussion revolves around the use of symmetry to control quantum transport. However, in an interesting U-turn, it is also possible to use transport as an indicator of hidden symmetries, and Section \ref{sec:transient} describes such an application to molecular systems. In particular, we describe a dynamical method first introduced in \cite{thingna16a} to detect molecular symmetries by analyzing the time evolution of the exciton current in molecules coupled to external reservoirs which impose a gradient. To detect the different symmetries, one needs to introduce a probe acting on the molecule and explore the dependence of the current evolution with the probe position and the symmetry of the initial state. The probe serves as a possible symmetry-breaking element, depending on its location in the molecular complex. Signatures of molecular symmetries are found in particular when initializing the molecule in an antisymmetric state and moving the probe from a symmetry-preserving site to a symmetry-breaking site: the exciton current is blocked in the former case by the dark state formed, while it increases steadily in time when the probe breaks the underlying molecular symmetry. These dynamical signatures of symmetry remain robust even in the presence of conformational disorder and environmental noise, provided that both noise sources are weaker than the molecule-probe interaction.

To end this review, we describe with some detail other symmetry-based mechanisms to control quantum transport. These methods differ from the control mechanisms described previously in that the symmetries used as a resource to manipulate quantum transport are different from the strong symmetries introduced in \S\ref{sec:sym2}. In particular, we review a method described in \cite{popkov13a} to manipulate spin and energy currents in driven spin chains using weak symmetries of the steady-state density matrix. We also describe the enhancement of quantum transport that emerges when time-reversal symmetry is broken, as may happen in chiral quantum walks \cite{zimboras13a}. 

In summary, we have analyzed the impact of symmetry in the transport properties of dissipative quantum systems, and how this interplay can be used to control or characterize the systems of interest. We expect the symmetry-based toolbox here described will trigger further advances in dissipative state engineering and dissipative quantum computation \cite{barreiro11a,diehl08a,kraus08a,lin13a,verstraete09a,pastawski11a}.

\ack
D.M. thanks J. Cao, E. Kyoseva and J. Thingna for many useful discussions and insights. Financial support from Spanish project FIS2017-84256-P and FIS2013-43201-P (MINECO) and Junta de Andalic\'{\i}a and EU Project TAHUB/II-148 (Program ANDALUC\'{I}A TALENT HUB 291780) is acknowledged. We also want to thank B. Buca and T Prosen for providing the data of Figs \ref{fig:buca_profiles} and \ref{fig:buca_spectrum}, M. Znidaric for providing the data of Fig \ref{fig:znidaric_fig1}, S. Pigeon for the data of Fig \ref{fig:3-qubits}, and M. Faccin and J. Biamonte for the data of Figs \ref{fig:CQW4},  \ref{fig:CQW5}, and \ref{fig:CQW6}.

\newpage

\section*{Appendix: Glossary}

Glossary of most used terms:

\begin{itemize}
\item $\cH$: Hilbert space
\item $D\equiv \text{dim}(\cH)$: dimension of Hilbert space
\item $\ket{\psi}\in\cH$: state vector of a system
\item $\bracket{\psi}{\varphi}$: inner product, $\ket{\psi},\ket{\varphi}\in \cH$
\item $\cBH$ space of linear bounded operators acting on the Hilbert space $\cH$
\item $H\in\cBH$: Hamiltonian
\item $\rho\in \cBH$: density matrix
\item $\bbraket{\sigma}{\rho} = \Tr(\sigma^\dagger \rho)$: Hilbert-Schmidt inner product, $\sigma,\rho \in \cBH$
\item $\id$: identity matrix
\item $\cL\in\cBBH$: generic Liouville-type superoperator (including Lindblad, Redfield, etc. superoperators) that determines the time evolution of a density matrix.
\item $\cor{A,B}=AB-BA$ commutator
\item $\key{A,B}=AB+BA$ anticommutator
\item $\Tr_A$: trace over the subspace $A$.
\item $\tO(t)= \text{e}^{\ii H t} \, O \, \text{e}^{-\ii H t}$: operator in the interaction (Dirac) picture, $O\in\cBH$
\item $\eta$: system-environment interaction strength
\item $V=\sum_l S_l\otimes E_l\in\cBHt$: system-environment interaction operator
\item $S_l\in \cBH$: interaction operator acting on the system
\item $E_l \in \cBH_E$: interaction operator acting on the environment
\item $\rho_{th}$: thermal density matrix
\item $L_i\in \cBH$: Lindblad operators
\item ${\cal D}\in\cBBH$: dissipator superoperator
\item $H^\text{XXZ}$ Hamiltonian of the $XXZ$-chain model.
\item $H^\text{Ld}$ Hamiltonian of the $XX$-$XXZ$ ladder model.
\item $\Lambda_k$: eigenvalues of Lindblad superoperator $\cL$.
\item $\phi_k$: Right eigenmatrix of $\cL$ 
\item $\hat{\phi}_k$: Left eigenmatrix of $\cL$ 
\item $U\in\cBH$: symmetry operator, such that $\cor{U,H}=0=\cor{U,S_l}$ $\forall l$
\item $\cU_{l,r}\in \cBBH$: right and left adjoint superoperators associated to the unitary operator $U$
\item $\rho_\alpha^{\scriptscriptstyle\text{NESS}}$: nonequilibrium steady-state (NESS) density matrix with symmetry index $\alpha$
\item $\Delta$: spectral gap in the spectrum of $\cL$
\item $\cW(t)\in\cBBH$: full propagator to time $t$ associated to master equation
\item $\chi$: quantum jumps trajectory
\item $\Pi_0(t)\in\cBBH$: current-free propagator
\item $\cW_\chi(t)\in\cBBH$: unnormalized propagator conditioned on trajectory $\chi$
\item $Q_\chi$: total current (or total number of quanta interchanged between the system and the selected incoherent channel) conditioned on trajectory $\chi$
\item $X_Q(t)=\{\chi:Q_\chi=Q\}$: set of all trajectories of duration $t$ with a fixed extensive current $Q$
\item $\rho_Q(t)$: current-resolved density-matrix, or reduced density matrix in the space of $Q$ quanta interchanged with an incoherent channel
\item $\rho_\lambda(t) = \sum_{Q=-\infty}^\infty \rho_Q(t) \text{e}^{-\lambda Q}$: Laplace transform of current-resolved density matrix
\item $\lambda$: counting field conjugated to the current
\item $\PP_t(Q)=\Tr \rho_Q(t)$: probability of observing a total current $Q$ after a time $t$
\item $Z_\lambda(t)=\Tr \rho_\lambda(t)$: moment generating function of the current probability distribution
\item $q=Q/t$: time-averaged current
\item $\la q \ra$: average current
\item $G(q)$: current large deviation function (LDF)
\item $\mu(\lambda)$: large deviation function associated to the counting field $\lambda$. Scaled cumulant generating function of the current distribution. Also Legendre transform of the current LDF, $\mu(\lambda)=\max_q[G(q)-\lambda q]$
\item $\cLl$: deformed or tilted Lindblad superoperator defining the evolution of $\rho_\lambda(t)$
\item $\omega_{\alpha\beta\nu}(\lambda)\in\cBH$: common right eigenfunction of $\cLl$ and $\cU_{l,r}$
\item $\hat{\omega}_{\alpha\beta\nu}(\lambda)\in\cBH$: common left eigenfunction of $\cLl$ and $\cU_{l,r}$
\item $\Omega_\alpha$: eigenphases of $\cU_{l,r}$
\item $\mu_\nu(\lambda)$: eigenvalue of $\cLl$ associated to eigenfunction $\omega_{\alpha\beta\nu}(\lambda)$
\item $\mu_0^{\alpha_0}(\lambda)$: is the eigenvalue of $\cLl$ with largest real part and symmetry index $\alpha_0$ among all symmetry diagonal eigenspaces with nonzero projection on the initial state $\rho(0)$
\item $\la q_\alpha\ra$: average current for NESS $\rho_\alpha^{\scriptscriptstyle\text{NESS}}$
\item $H^\text{net}$: Hamiltonian of the open quantum network model
\item $\sigma^{x,y,z}$: Pauli matrices
\item $\sigma^\pm$: raising and lowering operators 
\item $L_j^\text{deph}$: Lindblad operator associated to environment-induced dephasing
\item $H^{\text{SW}}$ Hamiltonian of the atomic-switch model.
\item $\Omega_{L,M}$: Rabi frequencies of laser fields
\item $H^\text{4s}$: H\"uckel Hamiltonian for the 4-site toy molecule
\item $H^\text{6s}$: H\"uckel Hamiltonian for benzene molecule
\item  $\mathcal{L}_{LR}$: Redfield-Lindblad Liouvillian superoperator for the molecule+probe system
\item $C(t)$: time-correlator describing the probe dynamics in the Redfield approach
\item $\omega_D$: Lorentz-Drude cutoff frequency
\end{itemize}

\section*{References}
%%\bibliographystyle{apsrev4-1}
%\bibliographystyle{unsrturl-v2}
%\bibliography{./referencias-BibDesk-v2-Dani}

%%%%%%%%%%%%%%%%%%%%%%%%%%%%%%%%%%%%%%%%%%%%%%%%%%%%%%%%%%%%%%
\end{document}